\documentclass[notitlepage,openbib,12pt]{article}
\usepackage{amsfonts}
\usepackage{amsmath}
\usepackage{amssymb}
\usepackage{graphicx}
\usepackage{color}
\usepackage[colorlinks,citecolor=blue]{hyperref}
\usepackage{natbib}
\usepackage{mathpazo}
\usepackage{amssymb}

\newtheorem{theo}{Theorem}

\newtheorem{assum}{Assumption}
\newtheorem{define}{Definition}

\newtheorem{lemma}{Lemma}

\makeindex
\input epsf
\input{tcilatex}
\begin{document}

\title{\textbf{Nash implementation by stochastic mechanisms: a simple full
characterization}\thanks{%
I thank Eddie Dekel, Jeff Ely and Roberto Serrano for insightful comments.}}
\author{Siyang Xiong\thanks{
Department of Economics, University of California, Riverside, United States,
siyang.xiong@ucr.edu}}
\maketitle

\begin{abstract}
We study Nash implementation by stochastic mechanisms, and provide a
surprisingly simple full characterization, which is in sharp contrast to the
classical, albeit complicated, full characterization (of Nash implementation
by deterministic mechanisms) in \cite{jmrr} and \cite{tomas}. Our current
understanding on the following four pairs of notions in Nash implementation
is limited: "mixed-Nash-implementation VS pure-Nash-implementation,"
"ordinal-approach VS cardinal-approach," "almost-full-characterization (in 
\cite{em}) VS full-characterization (in \cite{jmrr})," and
"Nash-implementation VS rationalizable-implementation." Our results build a
bridge connecting the two notions in each of the four pairs.
\end{abstract}

\newpage

\baselineskip= 19pt

\section{Introduction}

\label{sec:intro}

Mechanism design studies how to achieve a social goal in the presence of
decentralized decision making, and one important subfield is Nash
implementation.\footnote{%
There are two paradigms in mechanism design: full implementation (e.g., Nash
implementation) and partial implementation (e.g., auction design). The
former requires all solutions deliver the social goal, while the latter
requires one solution only. By adopting different solution concepts, we may
have different full implementation notions, e.g., Nash implementation (i.e.,
adopting Nash equilibria), and rationalizable implementation (i.e., adopting
rationalizability).} \cite{em2, em}\footnote{\cite{em2} is published as \cite%
{em}.} propose the famous notion of \emph{Maskin monotonicity}, and prove
that it is necessary for Nash implementation. Given an additional assumption
of \emph{no-veto power}, \cite{em2, em} further prove that Maskin
monotonicity is also sufficient.

The gap between necessity and sufficiency of Nash implementation is finally
eliminated by \cite{jmrr}, \cite{vd} and \cite{ds}, which provide necessary
and sufficient conditions. Throughout the paper, we focus on \cite{jmrr}.%
\footnote{\cite{ds} focuses on 2-agent environments, and we assume three or
more agents. Our results can be easily extended to 2-agent environments. 
\cite{vd}'s full characterization hinges on a domain assumption, while \cite%
{jmrr}'s does not, and that is why we focus on the latter only.} \cite{tomas}
provides algorithms to check the conditions in \cite{jmrr}.

Compared to the simple and intuitive Maskin monotonicity, the full
characterization in \cite{jmrr} is complicated and hard to interpret. It is
one of the most celebrated result, and yet, our understanding on it is still
limited.

Most papers in the literature use a canonical mechanism as in \cite{em} to
achieve Nash implementation. There are three cases in the canonical
mechanism: (1) consensus, (2) unilateral deviation, (3) multi-lateral
deviation. The difference between \cite{em} and \cite{jmrr} is how to
eliminate bad equilibria when case (2) (or case (3)) is triggered in the
canonical mechanism. \cite{em} solves this problem by an \emph{exogenous}
condition of no-veto power, which is \emph{essentially} equivalent to
requiring all equilibria in case (2) to be good. Instead, \cite{jmrr}
consider all possible equilibria in case (2), and identify \emph{endogenous }%
necessary conditions (of Nash implementation), and then embed them into the
canonical mechanism, which achieves Nash implementation.---Therefore, such
conditions are both necessary and sufficient.

From a normative view, the result in \cite{jmrr} is better than that in \cite%
{em}, because the former implies the latter. However, the advantage of the
Maskin approach is that the characterization is much simpler and more
intuitive. Furthermore, no-veto power is usually considered as a weak
condition. Thus, from a practical view, Maskin's characterization is usually
considered as an almost full characterization. Because of this, almost all
of the papers in the literature on full implementation follow the Maskin
approach, i.e., identify a Maskin-monotonicity-style necessary condition,
and prove that it is sufficient, given no-veto power, e.g., \cite{cmlr}, 
\cite{kt2012}.

In this paper, we study Nash implementation by stochastic mechanisms, and
provide a surprisingly simple full characterization. By taking full
advantage of the convexity structure of lotteries, we show that the
complicated Moore-Repullo-style full characterization collapses into a
Maskin-monotonicity-style condition. That is, not only does our simple full
characterization have a form similar to \cite{em}, but also it has an
interpretation parallel to \cite{jmrr}. In this sense, we build a bridge
connecting \cite{em} and \cite{jmrr}.\footnote{%
The outcome space in \cite{jmrr} is not convex, while it is convex in our
environment, and our full characterization hinges crucially on this
convexity assumption. As a result, the full characterization in \cite{jmrr}
does not imply ours, even though we share the same conceptual idea. See more
discussion in Section \ref{sec:connection}.}

More importantly, our results show a conceptual advantage of the
Moore-Repullo approach, which is not shared by the Maskin approach: the full
characterization rigorously pin down the logical relation between different
notions. For example, the notion of mixed-Nash implementation in \cite{cmlr}
does not require existence of pure-strategy equilibria, but the notion in 
\cite{em} does. Both \cite{cmlr} and \cite{cksx2022} argue that this is a
significant difference in their setups. However, how much impact does this
difference induce? With only almost full characterization in both \cite{em}
and \cite{cmlr}, we cannot find an answer for this question. Given the full
characterization in our paper, we prove that this difference alone actually
does not induce any impact (Theorem \ref{thm:pure:SCC}).

In a second example, \cite{bmt} try to compare rationalizable implementation
to mixed-Nash implementation. \cite{bmt} observe that the necessary
condition of the former is stronger than the necessary condition of the
latter, which, clearly, sheds limited light on their rigorous relation,
generally. Only with no-veto power, we can conclude that the former is
stronger than the latter.---This may be misleading. First, as illustrated in 
\cite{bmt} and \cite{xiong2022c}, no-veto power has no role in
rationalizable implementation, and hence, there is not much justification to
impose it, when we compare the two implementation notions. Second, no-veto
power is not the reason that rationalizable implementation is stronger than
mixed-Nash implementation, because with our full characterization, we are
able to prove that the former always implies the latter with or without
no-veto power (Theorem \ref{theorem:rationalizable}). The intuition is that
mixed-Nash implementation is fully characterized by \emph{a condition on
agents' modified lower-contour sets} (Theorem \ref{theorem:full:mix:SCC-A}),
which is also shared by rationalizable implementation. The difference
between mixed-Nash equilibria and rationalizability is whether agents have
correct beliefs in the corresponding solutions, and this difference has no
impact on the identified condition on agents' modified lower-contour sets.

\cite{obochet} and \cite{bo2} are the first two papers which propose to use
stochastic mechanisms to achieve Nash implementation. Their results are
orthogonal to ours, because of two differences. First, they impose weaker
assumptions on agents' preference on lotteries, and in this sense, their
results are stronger. Second, they impose exogenous assumptions on agents'
preference on deterministic outcomes, which immediately makes no-veto power 
\emph{vacuously} true on non-degenerate lotteries.\footnote{%
That is, they do not impose no-veto power on deterministic outcomes, but
(implicitly) impose no-veto power on non-degenerate lotteries.} Allowing for
non-degenerate lotteries only in case (2) and case (3) in the canonical
mechanism, \cite{obochet} and \cite{bo2} establish an almost full
characterization as \cite{em} does i.e., \cite{obochet} and \cite{bo2} take
the Maskin approach. In contrast, we establish our full characterization
without any assumption on agents' preference on deterministic outcomes, and
in this sense, our results are stronger. More importantly, we take the
Moore-Repullo approach, i.e., no-veto power may fail for both degenerate and
no-degenerate lotteries, and in spite of this, we find necessary and
sufficient conditions. To the best of our knowledge, we are the first paper
after \cite{jmrr} and \cite{tomas}, which takes the Moore-Repullo approach%
\footnote{%
Here is a difference between the Maskin approach and the Moore-Repullo
approach: there is an exogenously given subset of outcomes that satisfies
no-veto power in the former, but such an exogenous subset does not exist in
the latter.} and establishes a simple full characterization for Nash
implementation.

The remainder of the paper proceeds as follows: we describe the model in
Section \ref{sec:model} and fully characterize Nash implementation for
social choice functions in Section \ref{sec:mix:stochastic}; we compare
pure-Nash and mixed-Nash implementation in Section \ref{sec:pure}; we
compare the ordinal and the cardinal approaches in Section \ref{sec:ordinal}%
; we focus on social choice correspondences in Section \ref%
{sec:extension:SCC:A} and study ordinal and rationalizable implementation in
Sections \ref{sec:ordinal:full} and \ref{sec:rationalizable}, respectively;
we establish connection to \cite{jmrr} and \cite{tomas} in Section \ref%
{sec:connection} and Section \ref{sec:conclude} concludes.

\section{Model}

\label{sec:model}

\subsection{Environment}

\label{sec:environment}

We take a cardinal approach, and a (cardinal) model consists of 
\begin{equation}
\left\langle \mathcal{I}=\left\{ 1,..,I\right\} \text{, \ }\Theta \text{, \ }%
Z\text{, }f:\Theta \longrightarrow Z\text{, }Y\equiv \triangle \left(
Z\right) \text{, }\left( u_{i}^{\theta }:Z\longrightarrow 
\mathbb{R}
\right) _{\left( i,\theta \right) \in \mathcal{I}\times \Theta
}\right\rangle \text{,}  \label{yyr2}
\end{equation}%
where $\mathcal{I}$ is a finite set of $I$ agents with $I\geq 3$, $\Theta $
a finite or countably-infinite set of possible states, $Z$ a finite set of
pure social outcomes, $f$ a social choice function (hereafter, SCF)\footnote{%
For simplicity, we focus on social choice functions first. We will introduce
social choice correspondences in Section \ref{sec:pure}, and we extend our
results in Section \ref{sec:extension:SCC:A}.} which maps each state in $%
\Theta $ to an outcome in $Z$, $Y$ the set of all possible lotteries on $Z$, 
$u_{i}$ the Bernoulli utility function of agent $i$ at state $\theta $.
Throughout the paper, we assume that agents have expected utility, i.e.,%
\begin{equation*}
U_{i}^{\theta }\left( y\right) =\dsum\limits_{z\in Z}y_{z}u_{i}^{\theta
}\left( z\right) \text{, }\forall y\in Y\text{,}
\end{equation*}%
where $y_{z}$ denotes the probability of $z$ under $y$, and $U_{i}^{\theta
}\left( y\right) $ is agent $i$'s expected utility of $y$ at state $\theta $%
. Without loss of generality, we impose the following assumption throughout
the paper.

\begin{assum}[non-triviality]
\label{assm:non-trivial} $\left\vert f\left( \Theta \right) \right\vert \geq
2$.\footnote{%
If $\left\vert f\left( \Theta \right) \right\vert =1$, e.g., $f\left( \Theta
\right) =\left\{ z\right\} $ for some $z\in Z$, the implementation problem
can be solved trivially, i.e., we implement $z$ at every state.}
\end{assum}

For each $z\in Z$, we regard $z$ as a degenerate lottery in $Y$. With abuse
of notation, we write $z\in Z\subset Y$. Throughout the paper, we use $-i$
to denote $\mathcal{I}\backslash \left\{ i\right\} $. For any $\left( \alpha
,i,\theta \right) \in Y\times \mathcal{I}\times \Theta $, define%
\begin{eqnarray*}
\mathcal{L}_{i}^{Y}\left( \alpha ,\theta \right) &\equiv &\left\{ y\in
Y:U_{i}^{\theta }\left( \alpha \right) \geq U_{i}^{\theta }\left( y\right)
\right\} \text{,} \\
\mathcal{L}_{i}^{Z}\left( \alpha ,\theta \right) &\equiv &\left\{ z\in
Z:U_{i}^{\theta }\left( \alpha \right) \geq U_{i}^{\theta }\left( z\right)
\right\} \text{.}
\end{eqnarray*}

For any finite or countably-infinite set $E$, we use $\triangle \left(
E\right) $ to denote the set of probabilities on $E$. For any $\mu \in
\triangle \left( E\right) $, we let SUPP$\left[ \mu \right] $ denote the
support of $\mu $, i.e.,%
\begin{equation*}
\text{SUPP}\left[ \mu \right] \equiv \left\{ x\in E:\mu \left( x\right)
>0\right\} \text{.}
\end{equation*}%
Furthermore, define%
\begin{equation*}
\triangle ^{\circ }\left( E\right) \equiv \left\{ \mu \in \triangle \left(
E\right) :\text{SUPP}\left[ \mu \right] =E\right\} \text{.}
\end{equation*}%
For any finite set $E$, we use UNIF$\left( E\right) $ to denote the uniform
distribution on $E$, and use $\left\vert E\right\vert $ to denote the number
of elements in $E$.

\subsection{Mechanisms and Nash implementation}

\label{sec:model:pure}

A mechanism is a tuple $\mathcal{M}=\left\langle M\equiv \times _{i\in 
\mathcal{I}}M_{i}\text{, \ }g:M\longrightarrow Y\right\rangle $, where each $%
M_{i}$ is a countable set, and it denotes the set of strategies for agent $i$
in $\mathcal{M}$. A profile $\left( m_{i}\right) _{i\in \mathcal{I}}\in
\times _{i\in \mathcal{I}}M_{i}$ is a pure-strategy Nash equilibrium in $%
\mathcal{M}$ at state $\theta $ if and only if%
\begin{equation*}
U_{i}^{\theta }\left[ g\left( m_{i},m_{-i}\right) \right] \geq U_{i}^{\theta
}\left[ g\left( m_{i}^{\prime },m_{-i}\right) \right] \text{, }\forall i\in 
\mathcal{I}\text{, }\forall m_{i}^{\prime }\in M_{i}\text{.}
\end{equation*}%
Let $PNE^{\left( \mathcal{M},\text{ }\theta \right) }$ denote the set of
pure-strategy Nash equilibria in $\mathcal{M}$ at $\theta $. Furthermore, a
profile $\lambda \equiv \left( \lambda _{i}\right) _{i\in \mathcal{I}}\in
\times _{i\in \mathcal{I}}\triangle \left( M_{i}\right) $ is a
mixed-strategy Nash equilibrium in $\mathcal{M}$ at $\theta $ if and only if%
\begin{eqnarray*}
&&\Sigma _{m\in M}\left[ \lambda _{i}\left( m_{i}\right) \times \Pi _{j\in 
\mathcal{I}\diagdown \left\{ i\right\} }\lambda _{j}\left( m_{j}\right)
\times U_{i}^{\theta }\left[ g\left( m_{i},m_{-i}\right) \right] \right] \\
&\geq &\Sigma _{m\in M}\left[ \lambda _{i}^{\prime }\left( m_{i}\right)
\times \Pi _{j\in \mathcal{I}\diagdown \left\{ i\right\} }\lambda _{j}\left(
m_{j}\right) \times U_{i}^{\theta }\left[ g\left( m_{i},m_{-i}\right) \right]
\right] \text{, }\forall i\in \mathcal{I}\text{, }\forall \lambda
_{i}^{\prime }\in \triangle \left( M_{i}\right) \text{,}
\end{eqnarray*}%
where $\lambda _{j}\left( m_{j}\right) $ is the probability that $\lambda
_{j}$ assigns to $m_{j}$. Let $MNE^{\left( \mathcal{M},\text{ }\theta
\right) }$ denote the set of mixed-strategy Nash equilibria in $\mathcal{M}$
at state $\theta $.

For any mechanism $\mathcal{M}=\left\langle M\text{, \ }g:M\longrightarrow
Y\right\rangle $, and any $\lambda \equiv \left( \lambda _{i}\right) _{i\in 
\mathcal{I}}\in \times _{i\in \mathcal{I}}\triangle \left( M_{i}\right) $,
we use $g\left( \lambda \right) $ to denote the lottery induced by $\lambda $%
.

\begin{define}[mixed-Nash-implemenation]
\label{def:implementation:mixed}An SCF $f:\Theta \longrightarrow Z$ is
mixed-Nash-implemented by a mechanism $\mathcal{M}=\left\langle M\text{, \ }%
g:M\longrightarrow Y\right\rangle $ if%
\begin{equation*}
\dbigcup\limits_{\lambda \in MNE^{\left( \mathcal{M},\text{ }\theta \right)
}}\text{SUPP}\left( g\left[ \lambda \right] \right) =\left\{ f\left( \theta
\right) \right\} \text{, }\forall \theta \in \Theta \text{.}
\end{equation*}%
$f$ is mixed-Nash-implementable if there exists a mechanism that
mixed-Nash-implements $f$.
\end{define}

\begin{define}[pure-Nash-implemenation]
\label{def:implementation:pure}An SCF $f:\Theta \longrightarrow Z$ is
pure-Nash-implemented by a mechanism $\mathcal{M}=\left\langle M\text{, \ }%
g:M\longrightarrow Y\right\rangle $ if%
\begin{equation*}
\dbigcup\limits_{\lambda \in PNE^{\left( \mathcal{M},\text{ }\theta \right)
}}\text{SUPP}\left( g\left[ \lambda \right] \right) =\left\{ f\left( \theta
\right) \right\} \text{, }\forall \theta \in \Theta \text{.}
\end{equation*}%
$f$ is pure-Nash-implementable if there exists a mechanism that
pure-Nash-implements $f$.
\end{define}

\section{Mixed-Nash-implementation: a full characterization}

\label{sec:mix:stochastic}

In this section, we focus on mixed-Nash-implementation,\footnote{%
We will show that mixed-Nash-implementation is equivalent to
pure-Nash-implementation in Theorem \ref{theorem:full:equivalence:double}.}
and provide a surprisingly simple full characterization. As a benchmark, we
first list the theorem of \cite{em} in Section \ref%
{sec:mix:stochastic:Maskin}. We present the full characterization in Section %
\ref{sec:mix:stochastic:full}, which also contains the necessity part of the
proof. The sufficiency part of the proof is more complicated, which is
provided in Section \ref{sec:mix:stochastic:sufficiency}.

\subsection{Maskin's theorem}

\label{sec:mix:stochastic:Maskin}

By applying the ideas in \cite{em} to our environment with stochastic
mechanisms, we adapt Maskin monotonicity as follows.\footnote{%
With stochastic mechanisms, there are two ways to define Maskin monotonicity
(or related monotnicity conditions): (i) it is defined on the outcome space
of $Y$ (e.g., \cite{bmt} and \cite{cksx2022}) and (ii) it is defined on the
outcome space of $Z$ (e.g., \cite{cmlr}). We follow the tradition of the
former.}

\begin{define}[Maskin monotonicity]
Maskin monotonicity holds if%
\begin{equation*}
\left[ 
\begin{array}{c}
\mathcal{L}_{i}^{Y}\left( f\left( \theta \right) ,\theta \right) \subset 
\mathcal{L}_{i}^{Y}\left( f\left( \theta \right) ,\theta ^{\prime }\right) 
\text{, } \\ 
\forall i\in \mathcal{I}%
\end{array}%
\right] \text{ }\Longrightarrow \text{ }f\left( \theta \right) =f\left(
\theta ^{\prime }\right) \text{, }\forall \left( \theta ,\theta ^{\prime
}\right) \in \Theta \times \Theta \text{.}
\end{equation*}
\end{define}

\begin{define}[no-veto power]
No-veto power holds if for any $\left( a,\theta \right) \in Z\times \Theta $%
, we have%
\begin{equation*}
\left\vert \left\{ i\in \mathcal{I}:a\in \arg \max_{z\in Z}u_{i}^{\theta
}\left( z\right) \right\} \right\vert \geq \left\vert \mathcal{I}\right\vert
-1\Longrightarrow f\left( \theta \right) =a\text{.}
\end{equation*}
\end{define}

\begin{theo}[\protect\cite{em}]
\label{theorem:Maskin:sufficient}Maskin monotonicity holds if $f$ is
pure-Nash implementable. Furthermore, $f$ is pure-Nash implementable if
Maskin monotonicity and no-veto power hold.
\end{theo}

\subsection{A simple full characterization}

\label{sec:mix:stochastic:full}

The following easy-to-check notion plays a critical role in our full
characterization.

\begin{define}[$i$-max set]
\label{def:i-max}For any $\left( i,\theta \right) \in \mathcal{I}\times
\Theta $, a set $E\in 2^{Z}\diagdown \left\{ \varnothing \right\} $ is an $i$%
-$\theta $-max set if%
\begin{equation*}
E\subset \arg \max_{z\in E}u_{i}^{\theta }\left( z\right) \text{ and }%
E\subset \arg \max_{z\in Z}u_{j}^{\theta }\left( z\right) \text{, }\forall
j\in \mathcal{I}\diagdown \left\{ i\right\} \text{.}
\end{equation*}%
Furthermore, $E$ is an $i$-max set if $E$ is an $i$-$\theta $-max set for
some $\theta \in \Theta $.
\end{define}

This immediately leads to the following lemma, which sheds light on
mechanisms that mixed-Nash-implement $f$. The proof is relegated to Appendix %
\ref{sec:lem:mixed:deviation:SCF}.

\begin{lemma}
\label{lem:mixed:deviation:SCF}Suppose that an SCF $f$ is mixed-Nash
implemented by $\mathcal{M}=\left\langle M\text{, \ }g:M\longrightarrow
Y\right\rangle $. For any $\left( i,\theta \right) \in \mathcal{I}\times
\Theta $ and any $\lambda \in MNE^{\left( \mathcal{M},\text{ }\theta \right)
}$, we have%
\begin{equation}
\left( 
\begin{array}{c}
f\left( \theta \right) \in \arg \min_{z\in Z}u_{i}^{\theta }\left( z\right) 
\text{,} \\ 
\text{and }\mathcal{L}_{i}^{Z}\left( f\left( \theta \right) ,\theta \right) 
\text{ is an }i\text{-max set }%
\end{array}%
\right) \Longrightarrow \dbigcup\limits_{m_{i}\in M_{i}}\text{SUPP}\left[
g\left( m_{i},\lambda _{-i}\right) \right] =\left\{ f\left( \theta \right)
\right\} \text{.}  \label{iit1}
\end{equation}
\end{lemma}

$\dbigcup\limits_{m_{i}\in M_{i}}$SUPP$\left[ g\left( m_{i},\lambda
_{-i}\right) \right] $ is the set of outcomes that can be induced with
positive probability by $i$'s unilateral deviation from $\lambda $. Lemma %
\ref{lem:mixed:deviation:SCF} says that any unilateral deviation of $i$ from 
$\lambda $ must induce $f\left( \theta \right) $ if the condition on the
left-hand side of "$\Longrightarrow $" in (\ref{iit1})\ holds. In light of
Lemma \ref{lem:mixed:deviation:SCF}, we refine lower-counter sets as
follows. For each $\left( i,\theta ,a\right) \in \mathcal{I}\times \Theta
\times Z$,

\begin{equation}
\widehat{\mathcal{L}}_{i}^{Y}\left( a,\theta \right) \equiv \left\{ 
\begin{tabular}{ll}
$\left\{ a\right\} \text{,}$ & if $a=f\left( \theta \right) \in \arg
\min_{z\in Z}u_{i}^{\theta }\left( z\right) \text{ }$ and $\mathcal{L}%
_{i}^{Z}\left( f\left( \theta \right) ,\theta \right) \text{ is an }i\text{%
-max set}$, \\ 
&  \\ 
$\mathcal{L}_{i}^{Y}\left( a,\theta \right) \text{,}$ & otherwise%
\end{tabular}%
\right. \text{.}  \label{yjj8}
\end{equation}

\begin{define}[$\protect\widehat{\mathcal{L}}^{Y}$-monotonicity]
\label{def:L-monotonicity-SCF}$\widehat{\mathcal{L}}^{Y}$-monotonicity holds
if%
\begin{equation*}
\left[ 
\begin{array}{c}
\widehat{\mathcal{L}}_{i}^{Y}\left( f\left( \theta \right) ,\theta \right)
\subset \mathcal{L}_{i}^{Y}\left( f\left( \theta \right) ,\theta ^{\prime
}\right) \text{, } \\ 
\forall i\in \mathcal{I}%
\end{array}%
\right] \text{ }\Longrightarrow \text{ }f\left( \theta \right) =f\left(
\theta ^{\prime }\right) \text{, }\forall \left( \theta ,\theta ^{\prime
}\right) \in \Theta \times \Theta \text{.}
\end{equation*}
\end{define}

This leads to a simple full characterization of mixed-Nash-implementation.

\begin{theo}
\label{theorem:full:mix}An SCF $f:\Theta \longrightarrow Z$ is
mixed-Nash-implementable if and only if $\widehat{\mathcal{L}}^{Y}$%
-monotonicity holds.
\end{theo}

\noindent \textbf{Proof of the "only if" part of Theorem \ref%
{theorem:full:mix}:} Suppose that $f$ is mixed-Nash-implemented by $\mathcal{%
M}=\left\langle M\text{, \ }g:M\longrightarrow Y\right\rangle $. Fix any $%
\left( \theta ,\theta ^{\prime }\right) \in \Theta \times \Theta $ such that%
\begin{equation}
\left[ 
\begin{array}{c}
\widehat{\mathcal{L}}_{i}^{Y}\left( f\left( \theta \right) ,\theta \right)
\subset \mathcal{L}_{i}^{Y}\left( f\left( \theta \right) ,\theta ^{\prime
}\right) \text{, } \\ 
\forall i\in \mathcal{I}%
\end{array}%
\right] \text{,}  \label{hrr1}
\end{equation}%
and we aim to show $f\left( \theta \right) =f\left( \theta ^{\prime }\right) 
$, i.e., $\widehat{\mathcal{L}}^{Y}$-monotonicity holds.

We prove $f\left( \theta \right) =f\left( \theta ^{\prime }\right) $ by
contradiction. Suppose $f\left( \theta \right) \neq f\left( \theta ^{\prime
}\right) $. Pick any $\lambda \in MNE^{\left( \mathcal{M},\text{ }\theta
\right) }$, and we have SUPP$\left[ g\left( \lambda \right) \right] =\left\{
f\left( \theta \right) \right\} $ because $f$ is implemented by $\mathcal{M}$%
. Since $f\left( \theta \right) \neq f\left( \theta ^{\prime }\right) $, we
have $\lambda \notin MNE^{\left( \mathcal{M},\text{ }\theta ^{\prime
}\right) }$. As a result, there exists an agent $j$ who has a profitable
deviation $m_{j}\in M_{j}$ from $\lambda $ at $\theta ^{\prime }$, i.e.,%
\begin{equation}
\exists j\in \mathcal{I}\text{, }\exists m_{j}\in M_{j}\text{, }g\left(
m_{j},\lambda _{-j}\right) \in \mathcal{L}_{j}^{Y}\left( f\left( \theta
\right) ,\theta \right) \diagdown \mathcal{L}_{j}^{Y}\left( f\left( \theta
\right) ,\theta ^{\prime }\right) \text{,}  \label{ttat1}
\end{equation}%
where $g\left( m_{j},\lambda _{-j}\right) \in \mathcal{L}_{j}^{Y}\left(
f\left( \theta \right) ,\theta \right) $ and $g\left( m_{j},\lambda
_{-j}\right) \notin \mathcal{L}_{j}^{Y}\left( f\left( \theta \right) ,\theta
^{\prime }\right) $ follow from $\lambda \in MNE^{\left( \mathcal{M},\text{ }%
\theta \right) }$ and $\lambda \notin MNE^{\left( \mathcal{M},\text{ }\theta
^{\prime }\right) }$, respectively.

In particular, $g\left( m_{j},\lambda _{-j}\right) \notin \mathcal{L}%
_{j}^{Y}\left( f\left( \theta \right) ,\theta ^{\prime }\right) $ implies $%
g\left( m_{j},\lambda _{-j}\right) \neq f\left( \theta \right) $, and hence,
Lemma \ref{lem:mixed:deviation:SCF} implies failure of the following
condition:%
\begin{equation*}
\left( 
\begin{array}{c}
f\left( \theta \right) \in \arg \min_{z\in Z}u_{j}^{\theta }\left( z\right) 
\text{,} \\ 
\text{and }\mathcal{L}_{j}^{Z}\left( f\left( \theta \right) ,\theta \right) 
\text{ is an }j\text{-max set }%
\end{array}%
\right) \text{,}
\end{equation*}%
which, together with (\ref{yjj8}), further implies%
\begin{equation}
\widehat{\mathcal{L}}_{j}^{Y}\left( f\left( \theta \right) ,\theta \right) =%
\mathcal{L}_{j}^{Y}\left( f\left( \theta \right) ,\theta \right) \text{.}
\label{ttat2}
\end{equation}%
(\ref{ttat1}) and (\ref{ttat2}) imply%
\begin{equation*}
g\left( m_{j},\lambda _{-j}\right) \in \widehat{\mathcal{L}}_{j}^{Y}\left(
f\left( \theta \right) ,\theta \right) \diagdown \mathcal{L}_{j}^{Y}\left(
f\left( \theta \right) ,\theta ^{\prime }\right) \text{,}
\end{equation*}%
contradicting (\ref{hrr1}).$\blacksquare $

\subsection{Sufficiency of $\protect\widehat{\mathcal{L}}^{Y}$-monotonicity}

\label{sec:mix:stochastic:sufficiency}

\subsubsection{Preliminary construction}

In order to build our canonical mechanism to implement $f$, we need to take
three preliminary constructions. First, for each $\left( i,\theta \right)
\in \mathcal{I}\times \Theta $, we define%
\begin{equation}
\widehat{\Gamma }_{i}\left( \theta \right) \equiv \dbigcup\limits_{y\in 
\widehat{\mathcal{L}}_{i}^{Y}\left( f\left( \theta \right) ,\theta \right) }%
\text{SUPP}\left[ y\right] =\left\{ z\in Z:%
\begin{tabular}{l}
$\exists y\in \widehat{\mathcal{L}}_{i}^{Y}\left( f\left( \theta \right)
,\theta \right) $, \\ 
$z\in $SUPP$\left[ y\right] $%
\end{tabular}%
\right\} \text{,}  \label{kkl2}
\end{equation}%
i.e., $\widehat{\Gamma }_{i}\left( \theta \right) $\ is the set of outcomes
that can be induced with positive probability by lotteries in $\widehat{%
\mathcal{L}}_{i}^{Y}\left( f\left( \theta \right) ,\theta \right) $. This
leads to the following lemma, and the proof is relegated to Appendix \ref%
{sec:lem:no-veto:generalized}.

\begin{lemma}
\label{lem:no-veto:generalized}Suppose that $\widehat{\mathcal{L}}^{Y}$%
-monotonicity holds. Then, $Z$ is not an $i$-max set for any $i\in \mathcal{I%
}$ and%
\begin{equation*}
\left[ \widehat{\Gamma }_{j}\left( \theta \right) \text{ is a }j\text{-}%
\theta ^{\prime }\text{-max set}\right] \Longrightarrow \widehat{\Gamma }%
_{j}\left( \theta \right) =\left\{ f\left( \theta ^{\prime }\right) \right\} 
\text{, }\forall \left( j,\theta ,\theta ^{\prime }\right) \in \mathcal{I}%
\times \Theta \times \Theta \text{.}
\end{equation*}
\end{lemma}

Second, for each $\left( \theta ,j\right) \in \Theta \times \mathcal{I}$,
fix any function $\phi _{j}^{\theta }:\Theta \longrightarrow Y$ such that 
\begin{equation}
\phi _{j}^{\theta }\left( \theta ^{\prime }\right) \in \left( \arg
\max_{y\in \widehat{\mathcal{L}}_{j}^{Y}\left( f\left( \theta \right)
,\theta \right) }U_{j}^{\theta ^{\prime }}\left[ y\right] \right) \text{, }%
\forall \theta ^{\prime }\in \Theta \text{,}  \label{yuy1a}
\end{equation}%
Suppose that the true state is $\theta ^{\prime }\in \Theta $. In a
canonical mechanism that implements $f$, when agent $j$ unilaterally
deviates from "all agents reporting $\theta $," we let $j$ choose any
lottery in $\widehat{\mathcal{L}}_{j}^{Y}\left( f\left( \theta \right)
,\theta \right) $, in order to ensure that truthful reporting is always a
Nash equilibrium. Thus, $\phi _{j}^{\theta }\left( \theta ^{\prime }\right) $
in (\ref{yuy1a}) is an optimal lottery in $\widehat{\mathcal{L}}%
_{j}^{Y}\left( f\left( \theta \right) ,\theta \right) $ for $j$ at $\theta
^{\prime }\in \Theta $.

Furthermore, by (\ref{kkl2}), we have%
\begin{equation}
\phi _{j}^{\theta }\left( \theta ^{\prime }\right) \in \left( \arg
\max_{y\in \widehat{\mathcal{L}}_{j}^{Y}\left( f\left( \theta \right)
,\theta \right) }U_{j}^{\theta ^{\prime }}\left[ y\right] \right) \cap
\triangle \left( \widehat{\Gamma }_{j}\left( \theta \right) \right) \text{, }%
\forall \theta ^{\prime }\in \Theta \text{.}  \label{yuy1}
\end{equation}

Finally, the following lemma completes our third construction, and the proof
is relegated to Appendix \ref{sec:lem:if}.

\begin{lemma}
\label{lem:if}For each $\left( \theta ,j\right) \in \Theta \times \mathcal{I}
$, there exist%
\begin{equation*}
\varepsilon _{j}^{\theta }>0\text{ and }y_{j}^{\theta }\in \widehat{\mathcal{%
L}}_{j}^{Y}\left( f\left( \theta \right) ,\theta \right) \text{,}
\end{equation*}%
such that%
\begin{equation}
\left[ \varepsilon _{j}^{\theta }\times y+\left( 1-\varepsilon _{j}^{\theta
}\right) \times y_{j}^{\theta }\right] \in \widehat{\mathcal{L}}%
_{j}^{Y}\left( f\left( \theta \right) ,\theta \right) \text{, }\forall y\in
\triangle \left( \widehat{\Gamma }_{j}\left( \theta \right) \right) \text{.}
\label{yuy2}
\end{equation}
\end{lemma}

\subsubsection{A canonical mechanism}

\label{sec:canonic}

Let $\mathbb{N}$ denote the set of positive integers. We use the mechanism $%
\mathcal{M}^{\ast }=\left\langle \times _{i\in \mathcal{I}}M_{i}\text{, \ }%
g:M\longrightarrow Y\right\rangle $ defined below to implement $f$. In
particular, we have 
\begin{equation}
M_{i}=\left\{ \left( \theta _{i},k_{i}^{2},k_{i}^{3},\gamma
_{i},b_{i}\right) \in \Theta \times \mathbb{N}\times \mathbb{N}\times \left(
Z\right) ^{\left[ 2^{Z}\diagdown \left\{ \varnothing \right\} \right]
}\times Z:%
\begin{tabular}{l}
$\gamma _{i}\left( E\right) \in E$, \\ 
$\forall E\in \left[ 2^{Z}\diagdown \left\{ \varnothing \right\} \right] $%
\end{tabular}%
\right\} \text{, }\forall i\in \mathcal{I}\text{,}  \label{yuy2a}
\end{equation}%
and $g\left[ m=\left( m_{i}\right) _{i\in \mathcal{I}}\right] $ is defined
in three cases.

\begin{description}
\item[Case (1): consensus] if there exists $\theta \in \Theta $ such that%
\begin{equation*}
\left( \theta _{i},k_{i}^{2}\right) =\left( \theta ,1\right) \text{, }%
\forall i\in \mathcal{I}\text{,}
\end{equation*}%
then $g\left[ m\right] =f\left( \theta \right) $;

\item[Case (2), unilateral deviation: ] if there exists $\left( \theta
,j\right) \in \Theta \times \mathcal{I}$ such that%
\begin{equation*}
\left( \theta _{i},k_{i}^{2}\right) =\left( \theta ,1\right) \text{ if and
only if }i\in \mathcal{I}\diagdown \left\{ j\right\} \text{,}
\end{equation*}%
then 
\begin{eqnarray}
g\left[ m\right] &=&\left( 1-\frac{1}{k_{j}^{2}}\right) \times \phi
_{j}^{\theta }\left( \theta _{j}\right)  \label{uue1} \\
&&+\frac{1}{k_{j}^{2}}\times \left( 
\begin{tabular}{l}
$\varepsilon _{j}^{\theta }\times \left[ \left( 1-\frac{1}{k_{j}^{3}}\right)
\times \gamma _{j}\left( \widehat{\Gamma }_{j}\left( \theta \right) \right) +%
\frac{1}{k_{j}^{3}}\times \text{UNIF}\left( \widehat{\Gamma }_{j}\left(
\theta \right) \right) \right] $ \\ 
$+\left( 1-\varepsilon _{j}^{\theta }\right) \times y_{j}^{\theta }$%
\end{tabular}%
\right) \text{,}  \notag
\end{eqnarray}%
where $\left( \varepsilon _{j}^{\theta },y_{j}^{\theta }\right) $ are chosen
for each $\left( \theta ,j\right) \in \Theta \times \mathcal{I}$ according
to Lemma \ref{lem:if}. Note that $\gamma _{j}\left( \widehat{\Gamma }%
_{j}\left( \theta \right) \right) \in \widehat{\Gamma }_{j}\left( \theta
\right) $ by (\ref{yuy2a}) and UNIF$\left( \widehat{\Gamma }_{j}\left(
\theta \right) \right) \in \triangle \left( \widehat{\Gamma }_{j}\left(
\theta \right) \right) $;

\item[Case (3), multi-lateral deviation: ] otherwise, 
\begin{equation}
g\left[ m\right] =\left( 1-\frac{1}{k_{j^{\ast }}^{2}}\right) \times
b_{j^{\ast }}+\frac{1}{k_{j^{\ast }}^{2}}\times \text{UNIF}\left( Z\right) 
\text{,}  \label{uue2}
\end{equation}%
where $j^{\ast }=\max \left( \arg \max_{i\in \mathcal{I}}k_{i}^{2}\right) $,
i.e., $j^{\ast }$ is the largest-numbered agent who submits the highest
number on the second dimension of the message.
\end{description}

Each agent $i$ uses $k_{i}^{2}$ to show intention to be a whistle-blower,
i.e., $i$ \emph{voluntarily} challenges agents $-i$'s reports if and only if 
$k_{i}^{2}>1$. In Case (1), agents reach a consensus, i.e., all agents
report the same state $\theta $, and choose not to challenge voluntarily (or
equivalently, $k_{i}^{2}=1$ for every $i\in \mathcal{I}$). In this case, $%
f\left( \theta \right) $ is assigned by $g$.

Case (2) is triggered if any agent $j$ unilaterally deviates from Case (1)
(in the first two dimensions of $j$'s message): either $j$ challenges
voluntarily (i.e., $k_{j}^{2}>1$), or $j$ challenges involuntarily (i.e., $%
k_{j}^{2}=1$ and $j$ reports a different state $\theta _{j}\left( \neq
\theta \right) $ in the first dimension). In this case, $g$ assigns the
compound lottery in (\ref{uue1}), which is determined by the state $\theta $
being agreed upon by $-j$, and by $\left( \theta
_{j},k_{j}^{2},k_{j}^{3},\gamma _{j}\right) $ of $j$'s message. By (\ref%
{yuy1a}) and (\ref{yuy2}), the compound lottery in (\ref{uue1}) is an
element in $\widehat{\mathcal{L}}_{j}^{Y}\left( f\left( \theta \right)
,\theta \right) $, which ensures that truth-reporting is always a Nash
equilibrium at each state $\theta \in \Theta $. Note that $k_{j}^{2}$ and $%
k_{j}^{3}$ determine the probabilities in the compound lottery in (\ref{uue1}%
). Furthermore, $\left( \theta ,\theta _{j}\right) $ determines $j$'s
challenge scheme $\phi _{j}^{\theta }\left( \theta _{j}\right) $ in (\ref%
{uue1})---revealing the true state is always $j$'s best challenge scheme due
to (\ref{yuy1a}). Finally, $\theta $ determines the set $\widehat{\Gamma }%
_{j}\left( \theta \right) $, and $j$ is entitled to pick an optimal outcome
in $\widehat{\Gamma }_{j}\left( \theta \right) $ via $\gamma _{j}$ (i.e.,
the fourth dimensions in $j$'s message): the picked outcome $\gamma
_{j}\left( \widehat{\Gamma }_{j}\left( \theta \right) \right) $ occurs with
probability $\frac{1}{k_{j}^{2}}\times \varepsilon _{j}^{\theta }\times
\left( 1-\frac{1}{k_{j}^{3}}\right) $, and the outcome UNIF$\left( \widehat{%
\Gamma }_{j}\left( \theta \right) \right) $ occurs with probability $\frac{1%
}{k_{j}^{2}}\times \varepsilon _{j}^{\theta }\times \frac{1}{k_{j}^{3}}$%
.---The higher $k_{j}^{3}$, the more probability is shifted from "UNIF$%
\left( \widehat{\Gamma }_{j}\left( \theta \right) \right) $" to "$\gamma
_{j}\left( \widehat{\Gamma }_{j}\left( \theta \right) \right) $."

Case (3) includes all other scenarios, and as usual, agents compete in an
integer game. The agent $j^{\ast }$ who reports the highest integer in the
second dimension wins, and $j^{\ast }$ is entitled to pick an optimal
outcome $b_{j^{\ast }}$ in $Z$. In this case, $g$ assigns the compound
lottery in (\ref{uue2}): $b_{j^{\ast }}$ occurs with probability $\left( 1-%
\frac{1}{k_{j^{\ast }}^{2}}\right) $ and UNIF$\left( Z\right) $ occurs with
probability $\frac{1}{k_{j^{\ast }}^{2}}$.---The higher $k_{j^{\ast }}^{2}$,
the more probability is shift from "UNIF$\left( Z\right) $" to "$b_{j^{\ast
}}$."

The following lemma substantially simplifies the analysis of mixed-strategy
Nash equilibria in $\mathcal{M}^{\ast }$, and the proof is relegated to
Appendix \ref{sec:lem:mixed:canonical:pure}.

\begin{lemma}
\label{lem:mixed:canonical:pure}Consider the canonical mechanism $\mathcal{M}%
^{\ast }$ above. For any $\theta \in \Theta $ and any $\lambda \in
MNE^{\left( \mathcal{M}^{\ast },\text{ }\theta \right) }$, we have SUPP$%
\left[ \lambda \right] \subset PNE^{\left( \mathcal{M}^{\ast },\text{ }%
\theta \right) }$.
\end{lemma}

Lemma \ref{lem:mixed:canonical:pure} says that every pure-strategy profile
on the support of a mixed-strategy Nash equilibrium in $\mathcal{M}^{\ast }$
at $\theta $ must also be a pure-strategy Nash equilibrium at $\theta $. As
a result, it suffers no loss of generality to focus on pure-strategy Nash
equilibria.

\subsubsection{Proof of "if" part of Theorem \textbf{\protect\ref%
{theorem:full:mix}}}

\label{sec:canonic:proof}

Suppose that $\widehat{\mathcal{L}}^{Y}$-monotonicity holds. Fix any true
state $\theta ^{\ast }\in \Theta $. We aim to prove%
\begin{equation*}
\dbigcup\limits_{\lambda \in MNE^{\left( \mathcal{M}^{\ast },\text{ }\theta
^{\ast }\right) }}\text{SUPP}\left( g\left[ \lambda \right] \right) =\left\{
f\left( \theta ^{\ast }\right) \right\} \text{.}
\end{equation*}%
First, truth revealing is a Nash equilibrium, i.e., any pure strategy profile%
\begin{equation*}
m^{\ast }=\left( \theta _{i}=\theta ^{\ast },k_{i}^{1}=1,k_{i}^{2},\gamma
_{i},b_{i}\right) _{i\in \mathcal{I}}
\end{equation*}%
is a Nash equilibrium, which triggers Case (1) and $g\left[ m^{\ast }\right]
=f\left( \theta ^{\ast }\right) $. Any unilateral deviation $\overline{m}%
_{j}=\left( \overline{\theta _{j}},\overline{k_{j}^{2}},\overline{k_{j}^{3}},%
\overline{\gamma _{j}},\overline{b_{j}}\right) $ of agent $j\in \mathcal{I}$
would either still trigger Case (1) and induce $f\left( \theta ^{\ast
}\right) $, or trigger Case (2) and induce%
\begin{eqnarray*}
g\left[ \overline{m}_{j},m_{-j}^{\ast }\right] &=&\left( 1-\frac{1}{%
\overline{k_{j}^{2}}}\right) \times \phi _{j}^{\theta ^{\ast }}\left( 
\overline{\theta _{j}}\right) \\
&&+\frac{1}{\overline{k_{j}^{2}}}\times \left( 
\begin{tabular}{l}
$\varepsilon _{j}^{\theta ^{\ast }}\times \left[ \left( 1-\frac{1}{\overline{%
k_{j}^{3}}}\right) \gamma _{j}\left( \widehat{\Gamma }_{j}\left( \theta
^{\ast }\right) \right) +\frac{1}{\overline{k_{j}^{3}}}\times \text{UNIF}%
\left( \widehat{\Gamma }_{j}\left( \theta ^{\ast }\right) \right) \right] $
\\ 
$+\left( 1-\varepsilon _{j}^{\theta ^{\ast }}\right) \times y_{j}^{\theta
^{\ast }}$%
\end{tabular}%
\right) \text{,}
\end{eqnarray*}%
and by (\ref{yuy1}) and (\ref{yuy2}), we have%
\begin{equation*}
g\left[ \overline{m}_{j},m_{-j}^{\ast }\right] \in \widehat{\mathcal{L}}%
_{j}^{Y}\left( f\left( \theta ^{\ast }\right) ,\theta ^{\ast }\right)
\subset \mathcal{L}_{j}^{Y}\left( f\left( \theta ^{\ast }\right) ,\theta
^{\ast }\right) \text{, }\forall \overline{m}_{j}\in M_{j}\text{.}
\end{equation*}%
Therefore, any $\overline{m}_{j}\in M_{j}$ is not a profitable deviation and 
$m^{\ast }$ is a Nash equilibrium which induces $g\left[ m^{\ast }\right]
=f\left( \theta ^{\ast }\right) $.

Second, by Lemma \ref{lem:mixed:canonical:pure}, it suffers no loss of
generality to focus on pure-strategy equilibria. Fix any%
\begin{equation*}
\widetilde{m}=\left( \widetilde{\theta _{i}},\widetilde{k_{i}^{2}},%
\widetilde{k_{i}^{3}},\widetilde{\gamma _{i}},\widetilde{b_{i}}\right)
_{i\in \mathcal{I}}\in PNE^{\left( \mathcal{M}^{\ast },\text{ }\theta ^{\ast
}\right) }\text{,}
\end{equation*}%
and we aim to prove $g\left[ \widetilde{m}\right] =f\left( \theta ^{\ast
}\right) $. We first prove that $\widetilde{m}$ does not trigger Case (3).
Suppose otherwise, i.e., 
\begin{equation*}
g\left[ \widetilde{m}\right] =\left( 1-\frac{1}{\widetilde{k_{j^{\ast }}^{2}}%
}\right) \times \widetilde{b_{j^{\ast }}}+\frac{1}{\widetilde{k_{j^{\ast
}}^{2}}}\times \text{UNIF}\left( Z\right) \text{,}
\end{equation*}%
where $j^{\ast }=\max \left( \arg \max_{i\in \mathcal{I}}\widetilde{k_{i}^{2}%
}\right) $. By Lemma \ref{lem:no-veto:generalized}, $Z$ is not an $i$-max
set for any $i\in \mathcal{I}$, and thus, 
\begin{equation*}
\exists j\in \mathcal{I}\text{, }\min_{z\in Z}u_{j}^{\theta ^{\ast }}\left(
z\right) <\max_{z\in Z}u_{j}^{\theta ^{\ast }}\left( z\right) \text{,}
\end{equation*}%
and as a result, 
\begin{equation*}
U_{j}^{\theta ^{\ast }}\left( \text{UNIF}\left( Z\right) \right) <\max_{z\in
Z}u_{j}^{\theta ^{\ast }}\left( z\right) \text{,}
\end{equation*}%
which further implies that agent $j$ finds it strictly profitable to deviate
to%
\begin{equation*}
m_{j}=\left( \widetilde{\theta _{j}},k_{j}^{2},\widetilde{k_{j}^{3}},%
\widetilde{\gamma _{j}},b_{j}\right) \text{ with }b_{j}\in \arg \max_{z\in
Z}u_{j}^{\theta ^{\ast }}\left( z\right) \text{, for sufficiently large }%
k_{j}^{2}\text{,}
\end{equation*}%
contradicting $\widetilde{m}\in PNE^{\left( \mathcal{M}^{\ast },\text{ }%
\theta ^{\ast }\right) }$.

Thus, $\widetilde{m}$ must trigger either Case (1) or Case (2). Suppose $%
\widetilde{m}$ triggers Case (1), i.e.,%
\begin{equation*}
\widetilde{m}=\left( \widetilde{\theta _{i}}=\widetilde{\theta },\widetilde{%
k_{i}^{2}}=1,\widetilde{k_{i}^{3}},\widetilde{\gamma _{i}},\widetilde{b_{i}}%
\right) _{i\in \mathcal{I}}\text{ for some }\widetilde{\theta }\in \Theta 
\text{,}
\end{equation*}%
and $g\left[ \widetilde{m}\right] =f\left( \widetilde{\theta }\right) $. We
prove $g\left[ \widetilde{m}\right] =f\left( \theta ^{\ast }\right) $ by
contradiction. Suppose $f\left( \widetilde{\theta }\right) \neq f\left(
\theta ^{\ast }\right) $. By $\widehat{\mathcal{L}}^{Y}$-monotonicity, there
exists $j\in \mathcal{I}$ such that%
\begin{equation*}
\exists y^{\ast }\in \widehat{\mathcal{L}}_{j}^{Y}\left[ f\left( \widetilde{%
\theta }\right) ,\widetilde{\theta }\right] \diagdown \mathcal{L}_{j}^{Y}%
\left[ f\left( \widetilde{\theta }\right) ,\theta ^{\ast }\right] \text{,}
\end{equation*}%
which, together with (\ref{yuy1}), implies%
\begin{equation*}
U_{j}^{\theta ^{\ast }}\left[ \phi _{j}^{\widetilde{\theta }}\left( \theta
^{\ast }\right) \right] \geq U_{j}^{\theta ^{\ast }}\left[ y^{\ast }\right]
>U_{j}^{\theta ^{\ast }}\left[ f\left( \widetilde{\theta }\right) \right] 
\text{.}
\end{equation*}%
Therefore, it is strictly profitable for agent $j$ to deviate to%
\begin{equation*}
m_{j}=\left( \theta ^{\ast },k_{j}^{2},\widetilde{k_{j}^{3}},\widetilde{%
\gamma _{j}},\widetilde{b_{j}}\right) \text{ for sufficiently large }%
k_{j}^{2}\text{,}
\end{equation*}%
contradicting $\widetilde{m}\in PNE^{\left( \mathcal{M}^{\ast },\text{ }%
\theta ^{\ast }\right) }$.

Finally, suppose $\widetilde{m}$ triggers Case (2), i.e., there exists $j\in 
\mathcal{I}$ such that 
\begin{equation*}
\exists \widetilde{\theta }\in \Theta \text{, }\widetilde{m}_{i}=\left( 
\widetilde{\theta _{i}}=\widetilde{\theta },\widetilde{k_{i}^{2}}=1,%
\widetilde{k_{i}^{3}},\widetilde{\gamma _{i}},\widetilde{b_{i}}\right) \text{
, }\forall i\in \mathcal{I\diagdown }\left\{ j\right\} \text{,}
\end{equation*}%
and%
\begin{eqnarray}
g\left[ \widetilde{m}\right] &=&\left( 1-\frac{1}{\widetilde{k_{j}^{2}}}%
\right) \times \phi _{j}^{\widetilde{\theta }}\left( \widetilde{\theta _{j}}%
\right)  \label{yuy4} \\
&&+\frac{1}{\widetilde{k_{j}^{2}}}\times \left( 
\begin{tabular}{l}
$\varepsilon _{j}^{\widetilde{\theta }}\times \left[ \left( 1-\frac{1}{%
\widetilde{k_{j}^{3}}}\right) \times \gamma _{j}\left( \widehat{\Gamma }%
_{j}\left( \widetilde{\theta }\right) \right) +\frac{1}{\widetilde{k_{j}^{3}}%
}\times \text{UNIF}\left( \widehat{\Gamma }_{j}\left( \widetilde{\theta }%
\right) \right) \right] $ \\ 
$+\left( 1-\varepsilon _{j}^{\widetilde{\theta }}\right) \times y_{j}^{%
\widetilde{\theta }}$%
\end{tabular}%
\right) \text{.}  \notag
\end{eqnarray}%
We now prove $g\left[ \widetilde{m}\right] =f\left( \theta ^{\ast }\right) $%
. By (\ref{yuy1}) and (\ref{yuy2}), we have%
\begin{equation}
g\left[ \widetilde{m}\right] \in \triangle \left[ \widehat{\Gamma }%
_{j}\left( \widetilde{\theta }\right) \right] \text{.}  \label{uiu1}
\end{equation}%
Since every $i\in \mathcal{I\diagdown }\left\{ j\right\} $ can deviate to
trigger Case (3), and dictate her top outcome in $Z$ with arbitrarily high
probability, $\widetilde{m}\in PNE^{\left( \mathcal{M}^{\ast },\text{ }%
\theta ^{\ast }\right) }$ implies%
\begin{equation}
\widehat{\Gamma }_{j}\left( \widetilde{\theta }\right) \subset \arg
\max_{z\in Z}u_{i}^{\theta ^{\ast }}\left( z\right) \text{, }\forall i\in 
\mathcal{I}\diagdown \left\{ j\right\} \text{.}  \label{uiu2}
\end{equation}%
Inside the the compound lottery $g\left[ \widetilde{m}\right] $ in (\ref%
{yuy4}), conditional on an event with probability $\frac{1}{\widetilde{%
k_{j}^{2}}}\times \varepsilon _{j}^{\widetilde{\theta }}$, we have the
compound lottery 
\begin{equation*}
\left[ \left( 1-\frac{1}{\widetilde{k_{j}^{3}}}\right) \times \widetilde{%
\gamma _{j}}\left( \widehat{\Gamma }_{j}\left( \widetilde{\theta }\right)
\right) +\frac{1}{\widetilde{k_{j}^{3}}}\times \text{UNIF}\left( \widehat{%
\Gamma }_{j}\left( \widetilde{\theta }\right) \right) \right] \text{,}
\end{equation*}%
and hence, agent $j$ can always deviate to%
\begin{equation*}
m_{j}=\left( \widetilde{\theta _{j}},\widetilde{k_{j}^{2}},k_{j}^{3},\gamma
_{j},\widetilde{b_{j}}\right) _{i\in \mathcal{I\diagdown }\left\{ j\right\} }%
\text{ with }\gamma _{j}\left( \widehat{\Gamma }_{j}\left( \widetilde{\theta 
}\right) \right) \in \arg \max_{z\in \widehat{\Gamma }_{j}\left( \widetilde{%
\theta }\right) }u_{j}^{\theta ^{\ast }}\left( z\right)
\end{equation*}%
for sufficiently large $k_{j}^{3}$. Thus, $\widetilde{m}\in PNE^{\left( 
\mathcal{M}^{\ast },\text{ }\theta ^{\ast }\right) }$ implies%
\begin{equation}
\widehat{\Gamma }_{j}\left( \widetilde{\theta }\right) \subset \arg
\max_{z\in \widehat{\Gamma }_{j}\left( \widetilde{\theta }\right)
}u_{j}^{\theta ^{\ast }}\left( z\right) \text{.}  \label{uiu3}
\end{equation}%
(\ref{uiu2}) and (\ref{uiu3}) imply that $\widehat{\Gamma }_{j}\left( 
\widetilde{\theta }\right) $ is a $j$-$\theta ^{\ast }$-max set, which
together Lemma \ref{lem:no-veto:generalized}, further implies%
\begin{equation}
\widehat{\Gamma }_{j}\left( \widetilde{\theta }\right) =\left\{ f\left(
\theta ^{\ast }\right) \right\} \text{.}  \label{uiu3a}
\end{equation}%
(\ref{uiu1}) and (\ref{uiu3a}) imply $g\left[ \widetilde{m}\right] =f\left(
\theta ^{\ast }\right) $.$\blacksquare $

\section{Discussion: implementation-in-PNE Vs implementation-in-MNE}

\label{sec:pure}

The current literature has limited understanding on the difference between
implementation in pure Nash equilibria and in mixed Nash equilibria.
Compared to \cite{em}, \cite{cmlr} argue that mixed-Nash implementation
substantially expand the scope of implementation.%
\begin{equation}
\text{What drives such difference?}  \label{ggh1}
\end{equation}%
Both \cite{cmlr} and \cite{cksx2022} argue that a significant difference
between \cite{em} and \cite{cmlr} is whether we require existence of pure
Nash equilibria in mixed-Nash implementation. That is, \cite{em} actually
adopts the notion of double implementation defined as follows.

\begin{define}[double-Nash-implemenation, \protect\cite{em}]
\label{def:implementation:double}An SCF $f:\Theta \longrightarrow Z$ is
double-Nash-implementable if there exists a mechanism $\mathcal{M}%
=\left\langle M\text{, \ }g:M\longrightarrow Y\right\rangle $ such that%
\begin{equation*}
\dbigcup\limits_{\lambda \in MNE^{\left( \mathcal{M},\text{ }\theta \right)
}}\text{SUPP}\left( g\left[ \lambda \right] \right)
=\dbigcup\limits_{\lambda \in PNE^{\left( \mathcal{M},\text{ }\theta \right)
}}\text{SUPP}\left( g\left[ \lambda \right] \right) =\left\{ f\left( \theta
\right) \right\} \text{, }\forall \theta \in \Theta \text{.}
\end{equation*}
\end{define}

The following theorem says that this difference does not answer the question
in (\ref{ggh1}).

\begin{theo}
\label{theorem:full:equivalence:double}Consider any SCF $f:\Theta
\longrightarrow Z$. The following statements are equivalent.

(i) $f$ is pure-Nash-implementable;

(ii) $f$ is mixed-Nash-implementable;

(iii) $f$ is double-Nash-implementable;

(iv) $\widehat{\mathcal{L}}^{Y}$-monotonicity holds.
\end{theo}

The proof in Section \ref{sec:canonic:proof} shows sufficiency of $\widehat{%
\mathcal{L}}^{Y}$-monotonicity for all of (i), (ii) and (iii) in Theorem \ref%
{theorem:full:equivalence:double}, while the necessity of $\widehat{\mathcal{%
L}}^{Y}$-monotonicity is described by (a slightly modified version of) the
proof for the "only if" part of Theorem \ref{theorem:full:mix} in Section %
\ref{sec:mix:stochastic:full}.\footnote{%
We omit the proof Theorem \ref{theorem:full:equivalence:double}, because it
is implied by Theorem \ref{thm:pure:SCC}, which is proved in Appendix \ref%
{sec:thm:pure:SCC}.}

If we focus on SCFs, Theorem \ref{theorem:full:equivalence:double} implies
that the analysis in \cite{cmlr} remains the same if we replace mixed-Nash
implementation in their setup with pure-Nash-implementation or
double-Nash-implementation. However, \cite{cmlr} considers social choice
correspondences (hereafter, SCC) besides SCFs, and hence, Theorem \ref%
{theorem:full:equivalence:double} does not provide a full answer for the
question in (\ref{ggh1}). An SCC is a set-valued function $F:\Theta
\longrightarrow 2^{Z}\diagdown \left\{ \varnothing \right\} $, and we thus
extend our definitions to SCCs.\footnote{%
We will consider six different definitions of mixed-Nash-implementation, and
we call them versions A, B, C, D, E and F.}

\begin{define}[mixed-Nash-A-implemenation, \protect\cite{cmlr}]
\label{sec:mixed:implementation:SCC:A}An SCC $F$ is
mixed-Nash-A-implementable if there exists a mechanism $\mathcal{M}%
=\left\langle M\text{, \ }g:M\longrightarrow Y\right\rangle $ such that%
\begin{equation*}
\dbigcup\limits_{\lambda \in MNE^{\left( \mathcal{M},\text{ }\theta \right)
}}\text{SUPP}\left( g\left[ \lambda \right] \right) =F\left( \theta \right) 
\text{, }\forall \theta \in \Theta \text{.}
\end{equation*}
\end{define}

\begin{define}
\label{sec:pure:implementation:SCC:A}An SCC $F$ is pure-Nash-implementable
if there exists a mechanism $\mathcal{M}=\left\langle M\text{, \ }%
g:M\longrightarrow Y\right\rangle $ such that%
\begin{equation*}
\dbigcup\limits_{\lambda \in PNE^{\left( \mathcal{M},\text{ }\theta \right)
}}\text{SUPP}\left( g\left[ \lambda \right] \right) =F\left( \theta \right)
,\forall \theta \in \Theta \text{.}
\end{equation*}
\end{define}

\begin{define}[mixed-Nash-B-implemenation]
\label{sec:mixed:implementation:SCC:B}An SCC $F$ is
mixed-Nash-B-implementable if there exists a mechanism $\mathcal{M}%
=\left\langle M\text{, \ }g:M\longrightarrow Y\right\rangle $ such that%
\begin{equation*}
\dbigcup\limits_{\lambda \in MNE^{\left( \mathcal{M},\text{ }\theta \right)
}}\text{SUPP}\left( g\left[ \lambda \right] \right)
=\dbigcup\limits_{\lambda \in PNE^{\left( \mathcal{M},\text{ }\theta \right)
}}\text{SUPP}\left( g\left[ \lambda \right] \right) =F\left( \theta \right)
,\forall \theta \in \Theta \text{.}
\end{equation*}
\end{define}

\begin{theo}
\label{thm:pure:SCC}Consider any SCC $F:\Theta \longrightarrow
2^{Z}\diagdown \left\{ \varnothing \right\} $. The following statements are
equivalent.

(i) $F$ is pure-Nash-implementable;

(ii) $F$ is mixed-Nash-A-implementable;

(iii) $F$ is mixed-Nash-B-implementable.
\end{theo}

Theorem \ref{theorem:full:mix:SCC-A} will provide a full characterization of
mixed-Nash-A-implemention, which will be used to prove Theorem \ref%
{thm:pure:SCC}. We will prove Theorem \ref{thm:pure:SCC} in Appendix \ref%
{sec:thm:pure:SCC},\footnote{\cite{cmlr} observe that
pure-Nash-implementation and mixed-Nash-A-implementation share the same
necessary condition (i.e., set-monotonicity), but do not provide their
relationship. Only with our full characterization, we are able to prove
their equivalence.} after we prove Theorem \ref{theorem:full:mix:SCC-A} in
Appendix \ref{sec:theorem:full:mix:SCC-A}.

Compared to \cite{em}, \cite{cmlr} introduce three new ingredients to the
model: (I) a new class of mechanisms (i.e., stochastic mechanisms), (II) new
solutions (i.e., mixed-strategy Nash equilibria, or pure-strategy Nash
equilibria, or both) and (III) how to interpret "implementing $F\left(
\theta \right) $" (see more discussion in Section \ \ref{sec:extension:SCC:4}%
). Given SCFs, (III) is the same in both \cite{em} and \cite{cmlr}, and
Theorem \ref{theorem:full:equivalence:double} shows that (II) is also the
same in the two papers, which immediately leads to an answer for the
question in (\ref{ggh1}):\ the difference is solely driven by the new class
of mechanisms, i.e., (I). Given SCCs, Theorem \ref{thm:pure:SCC} shows that
(II) is still the same in the two papers, and hence, the difference must be
driven by (I) and (III).\footnote{%
Given SCCs, (III) is not the same in the two papers. Maskin's notion
coresponds to mixed-Nash-D-implemention in Definition \ref%
{def:implementation:mixed:SCC:D}.}

\section{Discussion: cardinal approach Vs ordinal approach}

\label{sec:ordinal}

We take a cardinal approach in this paper, i.e., agents have cardinal
utility functions. However, an ordinal approach is usually adopted in the
literature of implementation (e.g., \cite{cmlr}). We show that our cardinal
approach is more general than the ordinal approach.

Throughout this section, we fix an ordinal model in \cite{cmlr}, which
consists of 
\begin{equation}
\left\langle \mathcal{I}=\left\{ 1,..,I\right\} \text{, \ }\Theta ^{\ast }%
\text{, \ }Z\text{, }f:\Theta ^{\ast }\longrightarrow Z\text{, }Y\equiv
\triangle \left( Z\right) \text{, }\left( \succeq _{i}^{\theta }\right)
_{\left( i,\theta \right) \in \mathcal{I}\times \Theta ^{\ast
}}\right\rangle \text{,}  \label{yyr1}
\end{equation}%
where each ordinal state $\theta \in \Theta ^{\ast }$ determines a profile
of preferences $\left( \succeq _{i}^{\theta }\right) _{i\in \mathcal{I}}$ on 
$Z$. This ordinal model differs from our cardinal model on two aspects only.
First, agents have ordinal preference (i.e., $\left( \succeq _{i}^{\theta
}\right) _{\left( i,\theta \right) \in \mathcal{I}\times \Theta ^{\ast }}$),
compared to the cardinal utility functions (i.e., $\left( u_{i}^{\theta
}:Z\longrightarrow 
\mathbb{R}
\right) _{\left( i,\theta \right) \in \mathcal{I}\times \Theta }$) in in
Section \ref{sec:environment}. Second, the ordinal state set, denoted by $%
\Theta ^{\ast }$, is finite, while the cardinal state set, denoted by $%
\Theta $ in Section \ref{sec:environment}, is either finite or
countably-infinite.

For each ordinal state $\theta \in \Theta ^{\ast }$, we say $u^{\theta
}\equiv \left( u_{i}^{\theta }:Z\longrightarrow 
\mathbb{R}
\right) _{i\in \mathcal{I}}$ \ is a cardinal representation of $\succeq
^{\theta }\equiv \left( \succeq _{i}^{\theta }\right) _{i\in \mathcal{I}}$
if and only if%
\begin{equation*}
z\succeq _{i}^{\theta }z^{\prime }\Longleftrightarrow u_{i}^{\theta }\left(
z\right) \geq u_{i}^{\theta }\left( z^{\prime }\right) \text{, }\forall
\left( z,z^{\prime },i\right) \in Z\times Z\times \mathcal{I}\text{.}
\end{equation*}%
Each $\left( u_{i}^{\theta }:Z\longrightarrow 
\mathbb{R}
\right) _{i\in \mathcal{I}}$ is called a cardinal state. That is, each
ordinal state $\theta \in \Theta ^{\ast }$ can be represented by a set of
cardinal states defined as follows.%
\begin{equation*}
\Omega ^{\left[ \succeq ^{\theta },\text{ }%
\mathbb{R}
\right] }\equiv \left\{ \left( u_{i}^{\theta }:Z\longrightarrow 
\mathbb{R}
\right) _{i\in \mathcal{I}}:%
\begin{tabular}{l}
$z\succeq _{i}^{\theta }z^{\prime }\Longleftrightarrow u_{i}^{\theta }\left(
z\right) \geq u_{i}^{\theta }\left( z^{\prime }\right) \text{,}$ \\ 
$\forall \left( z,z^{\prime },i\right) \in Z\times Z\times \mathcal{I}\text{.%
}$%
\end{tabular}%
\right\} \subset \left( \left( 
\mathbb{R}
\right) ^{Z}\right) ^{\mathcal{I}}\text{.}
\end{equation*}%
Mixed-Nash ordinal-implementation in \cite{cmlr} requires that $f$ be
implemented by a mechanism under any cardinal representation.

\begin{define}[mixed-Nash-ordinal-implemenation, \protect\cite{cmlr}]
$f\ $is mixed-Nash-ordinally-implementable if there exists a mechanism $%
\mathcal{M}=\left\langle M\text{, \ }g:M\longrightarrow Y\right\rangle $
such that%
\begin{equation*}
\dbigcup\limits_{\lambda \in MNE^{\left( \mathcal{M},\text{ }u^{\theta
}\right) }}\text{SUPP}\left( g\left[ \lambda \right] \right) =\left\{
f\left( \theta \right) \right\} \text{, }\forall \theta \in \Theta ^{\ast }%
\text{, }\forall u^{\theta }\in \Omega ^{\left[ \succeq ^{\theta },\text{ }%
\mathbb{R}
\right] }\text{.}
\end{equation*}
\end{define}

Since $\Omega ^{\left[ \succeq ^{\theta },\text{ }%
\mathbb{R}
\right] }$ is uncountably infinite, our results do not apply directly.
However, we may consider cardinal utility functions with rational values
only.%
\begin{equation*}
\Omega ^{\left[ \succeq ^{\theta },\text{ }\mathbb{Q}\right] }\equiv \left\{
\left( u_{i}^{\theta }:Z\longrightarrow \mathbb{Q}\right) _{i\in \mathcal{I}%
}:%
\begin{tabular}{l}
$z\succeq _{i}^{\theta }z^{\prime }\Longleftrightarrow u_{i}^{\theta }\left(
z\right) \geq u_{i}^{\theta }\left( z^{\prime }\right) \text{,}$ \\ 
$\forall \left( z,z^{\prime },i\right) \in Z\times Z\times \mathcal{I}\text{.%
}$%
\end{tabular}%
\right\} \subset \left( \left( \mathbb{Q}\right) ^{Z}\right) ^{\mathcal{I}}%
\text{.}
\end{equation*}%
Clearly, $\Omega ^{\left[ \succeq ^{\theta },\text{ }\mathbb{Q}\right]
}\subset \Omega ^{\left[ \succeq ^{\theta },\text{ }%
\mathbb{R}
\right] }$, and $\Omega ^{\left[ \succeq ^{\theta },\text{ }\mathbb{Q}\right]
}$ is countably infinite.

\begin{theo}
\label{thm:mixed:ordinal}$f\ $is mixed-Nash-ordinally-implementable (or
equivalently, $f\ $is mixed-Nash-A-implementable with $\Theta =\cup _{\theta
\in \Theta ^{\ast }}\Omega ^{\left[ \succeq ^{\theta },\text{ }%
\mathbb{R}
\right] }$) if and only if $f\ $is mixed-Nash-A-implementable with $\Theta
=\cup _{\theta \in \Theta ^{\ast }}\Omega ^{\left[ \succeq ^{\theta },\text{ 
}\mathbb{Q}\right] }$.
\end{theo}

The "only if" part of Theorem \ref{thm:mixed:ordinal} is implied by $\Omega
^{\left[ \succeq ^{\theta },\text{ }\mathbb{Q}\right] }\subset \Omega ^{%
\left[ \succeq ^{\theta },\text{ }%
\mathbb{R}
\right] }$, and the "if" part is immediately implied by the following lemma.

\begin{lemma}
\label{lem:ordinal}For any $\theta \in \Theta ^{\ast }$ and any $\overline{%
u^{\theta }}\in \Omega ^{\left[ \succeq ^{\theta },\text{ }%
\mathbb{R}
\right] }$, there exists $\left( \widehat{u^{\theta }},\widetilde{u^{\theta }%
}\right) \in \Omega ^{\left[ \succeq ^{\theta },\text{ }\mathbb{Q}\right]
}\times \Omega ^{\left[ \succeq ^{\theta },\text{ }\mathbb{Q}\right] }$,
such that%
\begin{equation}
\mathcal{L}_{i}^{Y}\left( z,\widehat{u^{\theta }}\right) \subset \mathcal{L}%
_{i}^{Y}\left( z,\overline{u^{\theta }}\right) \subset \mathcal{L}%
_{i}^{Y}\left( z,\widetilde{u^{\theta }}\right) \text{, }\forall \left(
i,z\right) \in \mathcal{I}\times Y\text{.}  \label{ggi6}
\end{equation}
\end{lemma}

The proof of Lemma \ref{lem:ordinal} is relegated to Appendix \ref%
{sec:lem:ordinal}.

\noindent \textbf{Proof of the "if" part of Theorem \ref{thm:mixed:ordinal}:}
We use the canonical mechanism $\mathcal{M}^{\ast }=\left\langle \times
_{i\in \mathcal{I}}M_{i}\text{, \ }g:M\longrightarrow Z\right\rangle $ (with 
$\Theta =\cup _{\theta \in \Theta ^{\ast }}\Omega ^{\left[ \succeq ^{\theta
},\text{ }\mathbb{Q}\right] }$) in Section \ref{sec:canonic} to implement $f$%
, i.e.,%
\begin{equation}
\dbigcup\limits_{\lambda \in MNE^{\left( \mathcal{M}^{\ast },\text{ }%
u^{\theta }\right) }}\text{SUPP}\left( g\left[ \lambda \right] \right)
=\left\{ f\left( \theta \right) \right\} \text{, }\forall \theta \in \Theta
^{\ast }\text{, }\forall u^{\theta }\in \Omega ^{\left[ \succeq ^{\theta },%
\text{ }\mathbb{Q}\right] }\text{.}  \label{ggi7}
\end{equation}%
Fix any $\theta \in \Theta ^{\ast }$ and pick any $\overline{u^{\theta }}\in
\Omega ^{\left[ \succeq ^{\theta },\text{ }%
\mathbb{R}
\right] }$. By Lemma \ref{lem:ordinal}, there exists $\left( \widehat{%
u^{\theta }},\widetilde{u^{\theta }}\right) \in \Omega ^{\left[ \succeq
^{\theta },\text{ }\mathbb{Q}\right] }\times \Omega ^{\left[ \succeq
^{\theta },\text{ }\mathbb{Q}\right] }$, such that (\ref{ggi6})\ holds. In
particular, $\times _{i\in \mathcal{I}}\mathcal{L}_{i}^{Y}\left( f\left(
\theta \right) ,\widehat{u^{\theta }}\right) \subset \times _{i\in \mathcal{I%
}}\mathcal{L}_{i}^{Y}\left( f\left( \theta \right) ,\overline{u^{\theta }}%
\right) $ immediately implies%
\begin{equation}
MNE^{\left( \mathcal{M},\text{ }\widehat{u^{\theta }}\right) }\subset
MNE^{\left( \mathcal{M},\text{ }\overline{u^{\theta }}\right) }\text{.}
\label{ggi8}
\end{equation}%
Lemma \ref{lem:no-veto:generalized} implies that $Z$ is not an $i$-max set
for any $i\in \mathcal{I}$. As a result, no equilibrium exists when Case (3)
occurs. In fact, as the proof in Section \ref{sec:canonic:proof} shows that,
at state $\overline{u^{\theta }}\in \Omega ^{\left[ \succeq ^{\theta },\text{
}%
\mathbb{R}
\right] }$, an equilibrium exists in $\mathcal{M}^{\ast }$ only when either
Case (1) occurs, or Case (2) in which agents $-i$ report $\theta ^{\prime }$
with $\widehat{\mathcal{L}}_{i}^{Y}\left( f\left( \theta ^{\prime }\right)
,\theta ^{\prime }\right) =\left\{ f\left( \theta ^{\prime }\right) \right\} 
$ occurs. That is, in both cases, we have%
\begin{equation*}
g\left[ \lambda \right] \in Z\text{, }\forall \lambda \in MNE^{\left( 
\mathcal{M}^{\ast },\text{ }\overline{u^{\theta }}\right) }\text{,}
\end{equation*}%
which, together with $\times _{i\in \mathcal{I}}\mathcal{L}_{i}^{Y}\left( z,%
\overline{u^{\theta }}\right) \subset \times _{i\in \mathcal{I}}\mathcal{L}%
_{i}^{Y}\left( z,\widetilde{u^{\theta }}\right) $ for every $z\in Z$, implies%
\begin{equation}
MNE^{\left( \mathcal{M},\text{ }\overline{u^{\theta }}\right) }\subset
MNE^{\left( \mathcal{M},\text{ }\widetilde{u^{\theta }}\right) }\text{.}
\label{ggi8a}
\end{equation}%
(\ref{ggi7}), (\ref{ggi8})\ and (\ref{ggi8a}) imply $\dbigcup\limits_{%
\lambda \in MNE^{\left( \mathcal{M},\text{ }\overline{u^{\theta }}\right) }}$%
SUPP$\left( g\left[ \lambda \right] \right) =\left\{ f\left( \theta \right)
\right\} $.$\blacksquare $

Theorem \ref{thm:mixed:ordinal} extends to SCCs (see e.g., Theorem \ref%
{theorem:MR:iff}).

\section{Extension to social choice correspondences}

\label{sec:extension:SCC:A}

\subsection{Four additional definitions}

\label{sec:extension:SCC:4}

Given any solution concept, what does it mean that an SCC $F:\Theta
\longrightarrow 2^{Z}\diagdown \left\{ \varnothing \right\} $ is implemented
in the solution? There are two views in the literature. The first view is
that, at each state $\theta \in \Theta $, each solution must induce a
deterministic outcome and $F\left( \theta \right) $ is the set of all such
deterministic outcomes.--This view is adopted in \cite{ks}. The second view
is that, at each state $\theta \in \Theta $, $F\left( \theta \right) $ is
the set of outcomes that can be induced with positive probability by some
solution--This view is adopted in \cite{cmlr} and \cite{rjain}. Furthermore,
we may or may not require existence of pure Nash equilibria, when we define
mixed-Nash-implementation. Definitions \ref{sec:mixed:implementation:SCC:A}
and \ref{sec:mixed:implementation:SCC:B} follow the second view, while the
former does not require existence of pure Nash equilibria, and the latter
does. Besides these two definitions, we can define four alternative versions
of mixed-Nash-implementation, with different combination of requirements.
For any mechanism $\mathcal{M}=\left\langle M\text{, \ }g:M\longrightarrow
Y\right\rangle $, define%
\begin{equation*}
\Phi ^{\mathcal{M}}\equiv \left\{ \left( \lambda _{i}\right) _{i\in \mathcal{%
I}}\in \times _{i\in \mathcal{I}}\triangle \left( M_{i}\right) :\left\vert 
\text{SUPP}\left( g\left[ \left( \lambda _{i}\right) _{i\in \mathcal{I}}%
\right] \right) \right\vert =1\right\} \text{,}
\end{equation*}%
i.e., $\Phi ^{\mathcal{M}}$ is the set of mixed strategy profiles that
induces a unique deterministic outcome.

\begin{define}[mixed-Nash-C-implemenation]
\label{def:implementation:mixed:SCC:C}An SCC $F:\Theta \longrightarrow
2^{Z}\diagdown \left\{ \varnothing \right\} $ is mixed-Nash-C-implementable
if there exists a mechanism $\mathcal{M}=\left\langle M\text{, \ }%
g:M\longrightarrow Y\right\rangle $ such that%
\begin{equation*}
\dbigcup\limits_{\lambda \in MNE^{\left( \mathcal{M},\text{ }\theta \right)
}}\text{SUPP}\left( g\left[ \lambda \right] \right) =g\left( MNE^{\left( 
\mathcal{M},\text{ }\theta \right) }\cap \Phi ^{\mathcal{M}}\right) =F\left(
\theta \right) \text{, }\forall \theta \in \Theta \text{.}
\end{equation*}
\end{define}

\begin{define}[mixed-Nash-D-implemenation, \protect\cite{em}]
\label{def:implementation:mixed:SCC:D}An SCC $F:\Theta \longrightarrow
2^{Z}\diagdown \left\{ \varnothing \right\} $ is mixed-Nash-D-implementable
if there exists a mechanism $\mathcal{M}=\left\langle M\text{, \ }%
g:M\longrightarrow Y\right\rangle $ such that%
\begin{equation*}
\dbigcup\limits_{\lambda \in MNE^{\left( \mathcal{M},\text{ }\theta \right)
}}\text{SUPP}\left( g\left[ \lambda \right] \right) =g\left( PNE^{\mathcal{M}%
,\text{ }\theta }\right) =F\left( \theta \right) \text{, }\forall \theta \in
\Theta \text{.}
\end{equation*}
\end{define}

\begin{define}[mixed-Nash-E-implemenation]
\label{def:implementation:mixed:SCC:E}An SCC $F:\Theta \longrightarrow
2^{Z}\diagdown \left\{ \varnothing \right\} $ is mixed-Nash-E-implementable
if there exists a mechanism $\mathcal{M}=\left\langle M\text{, \ }%
g:M\longrightarrow Y\right\rangle $ such that%
\begin{gather*}
\dbigcup\limits_{\lambda \in MNE^{\left( \mathcal{M},\text{ }\theta \right)
}}\text{SUPP}\left( g\left[ \lambda \right] \right) =F\left( \theta \right) 
\text{, }\forall \theta \in \Theta \text{,} \\
\text{and }MNE^{\mathcal{M},\text{ }\theta }\subset \Phi ^{\mathcal{M}}\text{%
.}
\end{gather*}
\end{define}

\begin{define}[mixed-Nash-F-implemenation]
\label{def:implementation:double:SCC:F}An SCC $F:\Theta \longrightarrow
2^{Z}\diagdown \left\{ \varnothing \right\} $ is mixed-Nash-F-implementable
if there exists a mechanism $\mathcal{M}=\left\langle M\text{, \ }%
g:M\longrightarrow Y\right\rangle $ such that%
\begin{gather*}
\dbigcup\limits_{\lambda \in MNE^{\left( \mathcal{M},\text{ }\theta \right)
}}\text{SUPP}\left( g\left[ \lambda \right] \right)
=\dbigcup\limits_{\lambda \in PNE^{\left( \mathcal{M},\text{ }\theta \right)
}}\text{SUPP}\left( g\left[ \lambda \right] \right) =F\left( \theta \right) 
\text{, }\forall \theta \in \Theta \text{,} \\
\text{and }MNE^{\left( \mathcal{M},\text{ }\theta \right) }\subset \Phi ^{%
\mathcal{M}}\text{.}
\end{gather*}
\end{define}

\subsection{Version E and version F: full characterization}

It is straightforward to extend Theorem \ref{theorem:full:mix} to
mixed-Nash-E-implementation and Mixed-Nash-F-implementation. Define

\begin{equation}
\widehat{\mathcal{L}}_{i}^{Y}\left( a,\theta \right) \equiv \left\{ 
\begin{tabular}{ll}
$\left\{ a\right\} \text{,}$ & if $a\in F\left( \theta \right) \cap \arg
\min_{z\in Z}u_{i}^{\theta }\left( z\right) $ and $\mathcal{L}_{i}^{Z}\left(
a,\theta \right) \text{ is an }i\text{-max set}$, \\ 
&  \\ 
$\mathcal{L}_{i}^{Y}\left( a,\theta \right) \text{,}$ & otherwise.%
\end{tabular}%
\right.  \label{nnt1}
\end{equation}

\begin{define}[$\protect\widehat{\mathcal{L}}^{Y}$-monotonicity]
\label{def:L-monotonicity-SCC}$\widehat{\mathcal{L}}^{Y}$-monotonicity holds
for an SCC $F$ if%
\begin{equation*}
\left[ 
\begin{array}{c}
a\in F\left( \theta \right) \text{,} \\ 
\widehat{\mathcal{L}}_{i}^{Y}\left( a,\theta \right) \subset \mathcal{L}%
_{i}^{Y}\left( a,\theta ^{\prime }\right) \text{, }\forall i\in \mathcal{I}%
\end{array}%
\right] \text{ }\Longrightarrow a\in F\left( \theta ^{\prime }\right) \text{%
, }\forall \left( \theta ,\theta ^{\prime },a\right) \in \Theta \times
\Theta \times Z\text{.}
\end{equation*}
\end{define}

In the degenerate case that $F$ is a social choice function, $\widehat{%
\mathcal{L}}_{i}^{Y}\left( a,\theta \right) $ in (\ref{nnt1}) becomes $%
\widehat{\mathcal{L}}_{i}^{Y}\left( a,\theta \right) $ in (\ref{yjj8}), and $%
\widehat{\mathcal{L}}^{Y}$-monotonicity in Definition \ref%
{def:L-monotonicity-SCC} becomes $\widehat{\mathcal{L}}^{Y}$-monotonicity in
Definition \ref{def:L-monotonicity-SCF}. Hence, we use the same notation.

\begin{theo}
\label{theorem:full:mix:SCC-E-F}Given an SCC $F:\Theta \longrightarrow
2^{Z}\diagdown \left\{ \varnothing \right\} $, the following three
statements are equivalent.

(i) $F$ is mixed-Nash-E-implementable;

(ii) $F$ is mixed-Nash-F-implementable;

(iii) $Z$ is not an $i$-max set for any $i\in \mathcal{I}$ and $\widehat{%
\mathcal{L}}^{Y}$-monotonicity holds for $F$.
\end{theo}

It is worth noting that we need the requirement of "$Z$ is not an $i$-max
set for any $i\in \mathcal{I}$" in (iii) of Theorem \ref%
{theorem:full:mix:SCC-E-F}. We do not need this requirement in Theorem \ref%
{theorem:full:mix}, because it is implied by $\widehat{\mathcal{L}}^{Y}$%
-monotonicity when $F$ is a degenerate SCF (see Lemma \ref%
{lem:no-veto:generalized}).

The proof of Theorem \ref{theorem:full:mix:SCC-E-F} is almost the same as
that of Theorem \ref{theorem:full:mix}, and the detailed proof is relegated
to \cite{sx2022b}.

\subsection{Version A and version B: full characterization}

Define%
\begin{equation}
Z^{\ast }\equiv \left\{ 
\begin{tabular}{ll}
$\cup _{\theta \in \Theta }F\left( \theta \right) $, & if $Z$ is an $i$-max
set for some $i\in \mathcal{I}$, \\ 
&  \\ 
$Z\text{,}$ & if $Z$ is not an $i$-max set for any $i\in \mathcal{I}$.%
\end{tabular}%
\right.  \label{tth2}
\end{equation}

\begin{lemma}
\label{lem:Z-A}Suppose that an SCC $F$ is mixed-Nash-A-implemented by $%
\mathcal{M}=\left\langle M\text{, \ }g:M\longrightarrow Y\right\rangle $.
Then, we have $g\left( M\right) \subset \triangle \left( Z^{\ast }\right) $.
\end{lemma}

Lemma \ref{lem:Z-A} says that only lotteries in $\triangle \left( Z^{\ast
}\right) $ can be used by a mechanism which mixed-Nash-A-implements an SCC,
and the proof is relegated to Appendix \ref{sec:lem:Z-A}. The implication of
Lemma \ref{lem:Z-A} is that, in order to achieve
mixed-Nash-A-implementation, we should delete $Z\diagdown Z^{\ast }$ from
our model.

In order to accommodate the new implementation notion, we need to further
adapt the notion of $i$-max set as follow.

\begin{define}[$i$-$Z^{\ast }$-$\protect\theta $-max set and $i$-$Z^{\ast }$%
-max set ]
\label{def:i-Z*-max}For any $\left( i,\theta \right) \in \mathcal{I}\times
\Theta $, a set $E\in 2^{Z^{\ast }}\diagdown \left\{ \varnothing \right\} $
is an $i$-$Z^{\ast }$-$\theta $-max set if%
\begin{equation*}
E\subset \arg \max_{z\in E}u_{i}^{\theta }\left( z\right) \text{ and }%
E\subset \arg \max_{z\in Z^{\ast }}u_{j}^{\theta }\left( z\right) \text{, }%
\forall j\in \mathcal{I}\diagdown \left\{ i\right\} \text{.}
\end{equation*}%
Furthermore, $E\in 2^{Z^{\ast }}\diagdown \left\{ \varnothing \right\} $ is
an $i$-$Z^{\ast }$-max set if%
\begin{equation}
\Lambda ^{i\text{-}Z^{\ast }\text{-}\Theta }\left( E\right) \equiv \left\{
\theta \in \Theta :E\text{ is an }i\text{-}Z^{\ast }\text{-}\theta \text{%
-max set}\right\} \neq \varnothing \text{.}  \label{grr1}
\end{equation}
\end{define}

For each $E\in 2^{Z}\diagdown \left\{ \varnothing \right\} $, define%
\begin{equation*}
\mathcal{L}_{i}^{Z}\left( E,\theta \right) \equiv \cap _{z\in E}\mathcal{L}%
_{i}^{Z}\left( z,\theta \right) \text{.}
\end{equation*}%
For each $\left( i,\theta \right) \in \mathcal{I}\times \Theta $, define%
\begin{equation}
\Theta _{i}^{\theta }\equiv \left\{ \theta ^{\prime }\in \Theta :F\left(
\theta \right) \text{ is an }i\text{-}Z^{\ast }\text{-}\theta ^{\prime }%
\text{-max set and }F\left( \theta \right) \subset F\left( \theta ^{\prime
}\right) \right\} \text{,}  \label{dtta}
\end{equation}

\begin{equation}
\Xi _{i}\left( \theta \right) \equiv \left\{ K\in 2^{\Theta _{i}^{\theta
}}\diagdown \left\{ \varnothing \right\} :\Theta _{i}^{\theta }\cap \left[
\Lambda ^{i\text{-}Z^{\ast }\text{-}\Theta }\left( Z^{\ast }\cap \mathcal{L}%
_{i}^{Z}\left( F\left( \theta \right) ,\theta \right) \cap \left[
\dbigcap\limits_{\theta ^{\prime }\in K}F\left( \theta ^{\prime }\right) %
\right] \right) \right] =K\right\} \text{,}  \label{ddttaa}
\end{equation}%
where $\Lambda ^{i\text{-}Z^{\ast }\text{-}\Theta }\left( \cdot \right) $ is
defined in (\ref{grr1}).

It is worthy of noting that we may replace the definition of $\Theta
_{i}^{\theta }$ in (\ref{dtta}) with $\Theta _{i}^{\theta }\equiv \Theta $,
and use it define $\Xi _{i}\left( \theta \right) $ and $\widehat{\mathcal{L}}%
_{i}^{Y\text{-}A\text{-}B}\left( \text{UNIF}\left[ F\left( \theta \right) %
\right] ,\theta \right) $ in (\ref{ddttaa}) and (\ref{ddtt}), respectively.
With this modification, our full characterization (i.e., Theorem \ref%
{theorem:full:mix:SCC-A}) still holds. However, since $\Theta _{i}^{\theta }$
is a much smaller set than $\Theta $, our definition of $\Theta _{i}^{\theta
}$ in (\ref{dtta}) is much more computationally efficient, i.e., we need to
check much fewer sets in (\ref{ddttaa}) and (\ref{ddtt}).

The full characterization is established by a monotonicity condition which
is defined on modified lower-contour sets. For each $\theta \in \Theta $,
define 
\begin{eqnarray}
&&\widehat{\mathcal{L}}_{i}^{Y\text{-}A\text{-}B}\left( \text{UNIF}\left[
F\left( \theta \right) \right] ,\theta \right)  \label{ddtt} \\
&\equiv &\left\{ 
\begin{tabular}{ll}
$\triangle \left[ Z^{\ast }\cap \mathcal{L}_{i}^{Z}\left( F\left( \theta
\right) ,\theta \right) \cap \left( \dbigcup\limits_{K\in \Xi _{i}\left(
\theta \right) }\dbigcap\limits_{\theta ^{\prime }\in K}F\left( \theta
^{\prime }\right) \right) \right] $, & if $\left( 
\begin{tabular}{l}
$F\left( \theta \right) \subset \arg \min_{z\in Z^{\ast }}u_{i}^{\theta
}\left( z\right) $, \\ 
$\Xi _{i}\left( \theta \right) \neq \varnothing $, \\ 
and $Z^{\ast }\cap \mathcal{L}_{i}^{Z}\left( F\left( \theta \right) ,\theta
\right) $ \\ 
is an $i\text{-}Z^{\ast }\text{-max set}$%
\end{tabular}%
\right) $, \\ 
&  \\ 
$\left[ \triangle \left( Z^{\ast }\right) \right] \cap \mathcal{L}%
_{i}^{Y}\left( \text{UNIF}\left[ F\left( \theta \right) \right] ,\theta
\right) \text{,}$ & otherwise%
\end{tabular}%
\right. \text{.}  \notag
\end{eqnarray}

\begin{define}[$\protect\widehat{\mathcal{L}}^{Y\text{-}A\text{-}B}$%
-uniform-monotonicity]
\label{defin:A-B}$\widehat{\mathcal{L}}^{Y\text{-}A\text{-}B}$%
-uniform-monotonicity holds for an SCC $F$ if%
\begin{equation*}
\left[ 
\begin{array}{c}
\widehat{\mathcal{L}}_{i}^{Y\text{-}A\text{-}B}\left( \text{UNIF}\left[
F\left( \theta \right) \right] ,\theta \right) \subset \mathcal{L}%
_{i}^{Y}\left( \text{UNIF}\left[ F\left( \theta \right) \right] ,\theta
^{\prime }\right) \text{, } \\ 
\forall i\in \mathcal{I}%
\end{array}%
\right] \text{ }\Longrightarrow F\left( \theta \right) \subset F\left(
\theta ^{\prime }\right) \text{, }\forall \left( \theta ,\theta ^{\prime
}\right) \in \Theta \times \Theta \text{.}
\end{equation*}
\end{define}

\begin{theo}
\label{theorem:full:mix:SCC-A}Given an SCC $F:\Theta \longrightarrow
2^{Z}\diagdown \left\{ \varnothing \right\} $, the following three
statements are equivalent.

(i) $F$ is mixed-Nash-A-implementable;

(ii) $F$ is mixed-Nash-B-implementable;

(iii) $\widehat{\mathcal{L}}^{Y\text{-}A\text{-}B}$-uniform-monotonicity
holds for $F$.
\end{theo}

The necessity part of $\widehat{\mathcal{L}}^{Y\text{-}A\text{-}B}$%
-uniform-monotonicity in Theorem \ref{theorem:full:mix:SCC-A} is implied by
the following lemma.

\begin{lemma}
\label{lem:mixed:deviation:SCC}Suppose that an SCC $F$ is
mixed-Nash-A-implemented by $\mathcal{M}=\left\langle M\text{, \ }%
g:M\longrightarrow Y\right\rangle $. For any $\left( i,\theta \right) \in 
\mathcal{I}\times \Theta $ and any $\lambda \in MNE^{\left( \mathcal{M},%
\text{ }\theta \right) }$, we have%
\begin{multline*}
\left( 
\begin{array}{c}
F\left( \theta \right) \subset \arg \min_{z\in Z^{\ast }}u_{i}^{\theta
}\left( z\right) \text{,} \\ 
\Xi _{i}\left( \theta \right) \neq \varnothing \text{ and} \\ 
\text{and }Z^{\ast }\cap \mathcal{L}_{i}^{Z}\left( F\left( \theta \right)
,\theta \right) \text{ is an }i\text{-}Z^{\ast }\text{-max set}%
\end{array}%
\right) \\
\Longrightarrow \dbigcup\limits_{m_{i}\in M_{i}}\text{SUPP}\left[ g\left(
m_{i},\lambda _{-i}\right) \right] \subset \left[ Z^{\ast }\cap \mathcal{L}%
_{i}^{Z}\left( F\left( \theta \right) ,\theta \right) \cap \left(
\dbigcup\limits_{E\in \Xi _{i}\left( \theta \right) }\dbigcap\limits_{\theta
^{\prime }\in E}F\left( \theta ^{\prime }\right) \right) \right] \text{.}
\end{multline*}
\end{lemma}

Like Lemma \ref{lem:mixed:deviation:SCF} for SCFs, Lemma \ref%
{lem:mixed:deviation:SCC} is the counterpart for SCCs, and the proof of
Lemma \ref{lem:mixed:deviation:SCC} is relegated to Appendix \ref%
{sec:lem:mixed:deviation:SCC}. The sufficiency part of $\widehat{\mathcal{L}}%
^{Y\text{-}A\text{-}B}$-uniform-monotonicity in Theorem \ref%
{theorem:full:mix:SCC-A} is implied by the following lemma.

\begin{lemma}
\label{lem:no-veto:generalized:SCC}Suppose that $\widehat{\mathcal{L}}^{Y%
\text{-}A\text{-}B}$-monotonicity holds. We have%
\begin{equation*}
\left[ Z^{\ast }\text{ is a }j\text{-}Z^{\ast }\text{-}\theta ^{\prime }%
\text{-max set}\right] \Longrightarrow Z^{\ast }\subset F\left( \theta
^{\prime }\right) \text{, }\forall \left( j,\theta ^{\prime }\right) \in 
\mathcal{I}\times \Theta \text{,}
\end{equation*}%
\begin{equation*}
\text{and }\left[ \widehat{\Gamma }_{j}^{A\text{-}B}\left( \theta \right) 
\text{ is a }j\text{-}Z^{\ast }\text{-}\theta ^{\prime }\text{-max set}%
\right] \Longrightarrow \widehat{\Gamma }_{j}^{A\text{-}B}\left( \theta
\right) \subset F\left( \theta ^{\prime }\right) \text{, }\forall \left(
j,\theta ,\theta ^{\prime }\right) \in \mathcal{I}\times \Theta \times
\Theta \text{,}
\end{equation*}%
\begin{equation}
\text{where }\widehat{\Gamma }_{j}^{A\text{-}B}\left( \theta \right) \equiv
\dbigcup\limits_{y\in \widehat{\mathcal{L}}_{j}^{Y\text{-}A\text{-}B}\left( 
\text{UNIF}\left[ F\left( \theta \right) \right] ,\theta \right) }\text{SUPP}%
\left[ y\right] \text{.}  \label{ddtt4}
\end{equation}
\end{lemma}

Like Lemma \ref{lem:no-veto:generalized} for SCFs, Lemma \ref%
{lem:no-veto:generalized:SCC} is the counterpart for SCCs, and the proof of
Lemma is relegated to Appendix \ref{sec:lem:no-veto:generalized:SCC}.

The detailed proof of Theorem \ref{theorem:full:mix:SCC-A} is relegated to
Appendix \ref{sec:theorem:full:mix:SCC-A}.

\subsection{Version C and version D: full characterization}

\label{sec:extension:SCC:E}

For each $\left( i,\theta \right) \in \mathcal{I}\times \Theta $, define%
\begin{equation*}
\Theta _{i}^{\theta \text{-}C\text{-}D}\equiv \left\{ \theta ^{\prime }\in
\Theta :%
\begin{tabular}{l}
$F\left( \theta \right) \cap \arg \min_{z\in Z^{\ast }}u_{i}^{\theta }\left(
z\right) \text{ is an }i\text{-}Z^{\ast }\text{-}\theta ^{\prime }\text{-max
set,}$ \\ 
$\text{and }F\left( \theta \right) \cap \arg \min_{z\in Z^{\ast
}}u_{i}^{\theta }\left( z\right) \subset F\left( \theta ^{\prime }\right) $%
\end{tabular}%
\right\} \text{,}
\end{equation*}

\begin{equation*}
\Xi _{i}^{C\text{-}D}\left( \theta \right) \equiv \left\{ K\in 2^{\Theta
_{i}^{\theta \text{-}C\text{-}D}}\diagdown \left\{ \varnothing \right\}
:\Theta _{i}^{\theta \text{-}C\text{-}D}\cap \left[ \Lambda ^{i\text{-}%
Z^{\ast }\text{-}\Theta }\left( Z^{\ast }\cap \mathcal{L}_{i}^{Z}\left(
F\left( \theta \right) ,\theta \right) \cap \left[ \dbigcap\limits_{\theta
^{\prime }\in K}F\left( \theta ^{\prime }\right) \right] \right) \right]
=K\right\} \text{.}
\end{equation*}%
For each $\left( i,\theta ,a\right) \in \mathcal{I}\times \Theta \times Z$,
define 
\begin{eqnarray*}
&&\widehat{\mathcal{L}}_{i}^{Y\text{-}C\text{-}D}\left( a,\theta \right) \\
&\equiv &\left\{ 
\begin{tabular}{ll}
$\triangle \left[ Z^{\ast }\cap \mathcal{L}_{i}^{Z}\left( F\left( \theta
\right) ,\theta \right) \cap \left( \dbigcup\limits_{K\in \Xi _{i}^{C\text{-}%
D}\left( \theta \right) }\dbigcap\limits_{\theta ^{\prime }\in K}F\left(
\theta ^{\prime }\right) \right) \right] \text{,}$ & if $\left( 
\begin{tabular}{l}
$a\in F\left( \theta \right) \cap \arg \min_{z\in Z^{\ast }}u_{i}^{\theta
}\left( z\right) $, \\ 
$\Xi _{i}^{C\text{-}D}\left( \theta \right) \neq \varnothing $, \\ 
and $Z^{\ast }\cap \mathcal{L}_{i}^{Z}\left( F\left( \theta \right) ,\theta
\right) $ \\ 
is an $i\text{-}Z^{\ast }\text{-max set}$%
\end{tabular}%
\right) $, \\ 
&  \\ 
$\triangle \left( Z^{\ast }\right) \cap \mathcal{L}_{i}^{Y}\left( a,\theta
\right) \text{,}$ & otherwise%
\end{tabular}%
\right. \text{.}
\end{eqnarray*}

\begin{define}[$\protect\widehat{\mathcal{L}}^{Y\text{-}C\text{-}D}$%
-Maskin-monotonicity]
$\widehat{\mathcal{L}}^{Y\text{-}C\text{-}D}$-Maskin-monotonicity holds for
an SCC $F$ if%
\begin{equation*}
\left[ 
\begin{array}{c}
a\in F\left( \theta \right) \text{,} \\ 
\widehat{\mathcal{L}}_{i}^{Y\text{-}C\text{-}D}\left( a,\theta \right)
\subset \mathcal{L}_{i}^{Y}\left( a,\theta \right) \text{, }\forall i\in 
\mathcal{I}%
\end{array}%
\right] \text{ }\Longrightarrow a\in F\left( \theta ^{\prime }\right) \text{%
, }\forall \left( \theta ,\theta ^{\prime },a\right) \in \Theta \times
\Theta \times Z\text{.}
\end{equation*}
\end{define}

\begin{theo}
\label{theorem:full:mix:SCC-C-D}Given an SCC $F:\Theta \longrightarrow
2^{Z}\diagdown \left\{ \varnothing \right\} $, the following three
statements are equivalent.

(i) $F$ is mixed-Nash-C-implementable;

(ii) $F$ is mixed-Nash-D-implementable;

(iii) $\widehat{\mathcal{L}}^{Y\text{-}C\text{-}D}$-Maskin-monotonicity
holds for $F$.
\end{theo}

The proof of Theorem \ref{theorem:full:mix:SCC-C-D} is similar to that of
Theorem \ref{theorem:full:mix:SCC-A}, and it is relegated to \cite{sx2022b}.

\section{Ordinal implementation: full characterization}

\label{sec:ordinal:full}

Throughout this section, we fix an ordinal model%
\begin{equation*}
\left\langle \mathcal{I}=\left\{ 1,..,I\right\} \text{, \ }\Theta ^{\ast }%
\text{, \ }Z\text{, }F:\Theta ^{\ast }\longrightarrow 2^{Z}\diagdown \left\{
\varnothing \right\} \text{, }Y\equiv \triangle \left( Z\right) \text{, }%
\left( \succeq _{i}^{\theta }\right) _{\left( i,\theta \right) \in \mathcal{I%
}\times \Theta ^{\ast }}\right\rangle \text{,}
\end{equation*}%
and show that it is straightforward to derive full characterization of
ordinal mixed-Nash implementation \emph{\`{a} la} \cite{cmlr}. For any $%
\left( a,i,\theta \right) \in Z\times \mathcal{I}\times \Theta ^{\ast }$,
consider%
\begin{eqnarray*}
\mathcal{L}_{i}^{Z}\left( a,\theta \right) &\equiv &\left\{ z\in Z:a\succeq
_{i}^{\theta }z\right\} \text{,} \\
S\mathcal{L}_{i}^{Z}\left( a,\theta \right) &\equiv &\left\{ z\in Z:a\succ
_{i}^{\theta }z\right\} \text{.}
\end{eqnarray*}

\begin{define}[set-monotonicity, \protect\cite{cmlr}]
\label{def:set-monotone}An SCC $F$ is set-monotonic if for any $\left(
\theta ,\theta ^{\prime }\right) \in \Theta ^{\ast }\times \Theta ^{\ast }$,
we have $F\left( \theta \right) \subset F\left( \theta ^{\prime }\right) $
whenever for any $i\in \mathcal{I}$, one of the following two condition
holds: either (1) $Z\subset \mathcal{L}_{i}^{Z}\left( F\left( \theta \right)
,\theta ^{\prime }\right) $ or (2) for any $a\in F\left( \theta \right) $,
both $\mathcal{L}_{i}^{Z}\left( a,\theta \right) \subset \mathcal{L}%
_{i}^{Z}\left( a,\theta ^{\prime }\right) $ and $\mathcal{SL}_{i}^{Z}\left(
a,\theta \right) \subset \mathcal{SL}_{i}^{Z}\left( a,\theta ^{\prime
}\right) $ hold.
\end{define}

\cite{cmlr} prove that set-monotonicity is necessary for ordinal
mixed-Nash-A-implementation.

\begin{theo}[\protect\cite{cmlr}]
\label{theorem:MR:necessary}Set-monotonicity holds if $F$ is
mixed-Nash-A-implementable on $\Theta =\cup _{\theta \in \Theta ^{\ast
}}\Omega ^{\left[ \succeq ^{\theta },\text{ }%
\mathbb{R}
\right] }$ (i.e., $F$ is ordinally-mixed-Nash-implementable \emph{\`{a} la} 
\cite{cmlr}).
\end{theo}

It is easy to show that $\mathcal{L}^{Y}$-uniform-monotonicity defined below
is necessary condition for mixed-Nash-A-implementation.\footnote{%
See Lemma \ref{lem:mixed-lottery:lower-contour} and the discussion in
Section \ref{sec:connection:A}.}

\begin{define}[$\mathcal{L}^{Y}$-uniform-monotonicity]
\label{defin:uniform}$\mathcal{L}^{Y}$-uniform-monotonicity holds for an SCC 
$F$ if%
\begin{gather*}
\left[ 
\begin{array}{c}
\mathcal{L}_{i}^{Y}\left( \text{UNIF}\left[ F\left( \theta \right) \right]
,\theta \right) \subset \mathcal{L}_{i}^{Y}\left( \text{UNIF}\left[ F\left(
\theta \right) \right] ,\theta ^{\prime }\right) \text{, } \\ 
\forall i\in \mathcal{I}%
\end{array}%
\right] \text{ }\Longrightarrow F\left( \theta \right) \subset F\left(
\theta ^{\prime }\right) \text{, }\forall \left( \theta ,\theta ^{\prime
}\right) \in \Theta \times \Theta \text{,} \\
\text{where }\Theta =\cup _{\theta \in \Theta ^{\ast }}\Omega ^{\left[
\succeq ^{\theta },\text{ }%
\mathbb{R}
\right] }\text{.}
\end{gather*}
\end{define}

\begin{lemma}[\protect\cite{cmlr}, Proposition 1]
\label{lem:set-monotone}The following statements are equivalent.

(i) set-monotonicity holds;

(ii) $\mathcal{L}^{Y}$-uniform-monotonicity holds for $F$ on $\Theta =\cup
_{\theta \in \Theta ^{\ast }}\Omega ^{\left[ \succeq ^{\theta },\text{ }%
\mathbb{R}
\right] }$;

(iii) $\mathcal{L}^{Y}$-uniform-monotonicity holds for $F$ on $\Theta =\cup
_{\theta \in \Theta ^{\ast }}\Omega ^{\left[ \succeq ^{\theta },\text{ }%
\mathbb{Q}\right] }$.
\end{lemma}

(i) being equivalent to (ii)\ in Lemma \ref{lem:set-monotone} is Proposition
1 in \cite{cmlr}, which provides its proof. The same argument shows that (i)
is equivalent (iii).

Given Lemma \ref{lem:set-monotone}, the following theorem shows that Theorem %
\ref{theorem:MR:necessary} is immediately implied by Theorem \ref%
{theorem:full:mix:SCC-A}, because $\mathcal{L}^{Y}$-uniform-monotonicity is
immediately implied by $\widehat{\mathcal{L}}^{Y\text{-}A\text{-}B}$%
-uniform-monotonicity. Theorem \ref{thm:mixed:ordinal:A} is implied by
Theorem \ref{theorem:MR:iff} below.

\begin{theo}
\label{thm:mixed:ordinal:A}An SCC $F\ $is mixed-Nash-A-implementable on $%
\Theta =\cup _{\theta \in \Theta ^{\ast }}\Omega ^{\left[ \succeq ^{\theta },%
\text{ }\mathbb{Q}\right] }$ if and only if $F\ $is
mixed-Nash-A-implementable on $\Theta ^{\prime }=\cup _{\theta \in \Theta
^{\ast }}\Omega ^{\left[ \succeq ^{\theta },\text{ }%
\mathbb{R}
\right] }$.
\end{theo}

In light of Theorem \ref{theorem:full:mix:SCC-A}, define%
\begin{gather*}
\forall \left( i,a,\theta \right) \in Z\times \Theta \text{,} \\
\widehat{\mathcal{L}}_{i}^{Z^{\ast }\text{-}A\text{-}B}\left( a,\theta
\right) \equiv \left\{ 
\begin{tabular}{ll}
$\left[ Z^{\ast }\cap \mathcal{L}_{i}^{Z}\left( F\left( \theta \right)
,\theta \right) \cap \left( \dbigcup\limits_{K\in \Xi _{i}\left( \theta
\right) }\dbigcap\limits_{\theta ^{\prime }\in K}F\left( \theta ^{\prime
}\right) \right) \right] $, & if $\left( 
\begin{tabular}{l}
$F\left( \theta \right) \subset \arg \min_{z\in Z^{\ast }}u_{i}^{\theta
}\left( z\right) $, \\ 
$\Xi _{i}\left( \theta \right) \neq \varnothing $, \\ 
and $Z^{\ast }\cap \mathcal{L}_{i}^{Z}\left( F\left( \theta \right) ,\theta
\right) $ \\ 
is an $i\text{-}Z^{\ast }\text{-max set}$%
\end{tabular}%
\right) $, \\ 
&  \\ 
$Z^{\ast }\cap \mathcal{L}_{i}^{Z}\left( a,\theta \right) \text{,}$ & 
otherwise%
\end{tabular}%
\right. \text{,} \\
\\
\widehat{S\mathcal{L}}_{i}^{Z^{\ast }\text{-}A\text{-}B}\left( a,\theta
\right) \equiv \widehat{\mathcal{L}}_{i}^{Z^{\ast }\text{-}A\text{-}B}\left(
a,\theta \right) \cap S\mathcal{L}_{i}^{Z}\left( a,\theta \right) \text{.}
\end{gather*}

\begin{define}[strong set-monotonicity]
\label{def:set-monotone:weak}An SCC $F$ is strongly set-monotonic if for any 
$\left( \theta ,\theta ^{\prime }\right) \in \Theta ^{\ast }\times \Theta
^{\ast }$, we have $F\left( \theta \right) \subset F\left( \theta ^{\prime
}\right) $ whenever for any $i\in \mathcal{I}$, one of the following two
condition holds: either (1) $Z^{\ast }\subset \mathcal{L}_{i}^{Z}\left(
F\left( \theta \right) ,\theta ^{\prime }\right) $ or (2) for any $a\in
F\left( \theta \right) $, both $\widehat{\mathcal{L}}_{i}^{Z^{\ast }\text{-}A%
\text{-}B}\left( a,\theta \right) \subset \mathcal{L}_{i}^{Z}\left( a,\theta
^{\prime }\right) $ and $\widehat{S\mathcal{L}}_{i}^{Z^{\ast }\text{-}A\text{%
-}B}\left( a,\theta \right) \subset \mathcal{SL}_{i}^{Z}\left( a,\theta
^{\prime }\right) $ hold.
\end{define}

Using a similar argument as in the proof of Lemma \ref{lem:set-monotone} (or
equivalently, Proposition 1 in \cite{cmlr}), it is straightforward to show
the following lemma.

\begin{lemma}
\label{lem:set-monotone:strong}The following statements are equivalent.

(i) strong set-monotonicity holds;

(ii) $\widehat{\mathcal{L}}^{Y\text{-}A\text{-}B}$-uniform-monotonicity
holds for $F$ on $\Theta =\cup _{\theta \in \Theta ^{\ast }}\Omega ^{\left[
\succeq ^{\theta },\text{ }%
\mathbb{R}
\right] }$;

(iii) $\widehat{\mathcal{L}}^{Y\text{-}A\text{-}B}$-uniform-monotonicity
holds for $F$ on $\Theta =\cup _{\theta \in \Theta ^{\ast }}\Omega ^{\left[
\succeq ^{\theta },\text{ }\mathbb{Q}\right] }$.
\end{lemma}

This immediately leads to the following full characterization, and the proof
is relegated to Appendix \ref{sec:theorem:MR:iff}.

\begin{theo}
\label{theorem:MR:iff}The following statements are equivalent.

(i) strong set-monotonicity holds;

(ii) $F$ is mixed-Nash-A-implementable on $\Theta =\cup _{\theta \in \Theta
^{\ast }}\Omega ^{\left[ \succeq ^{\theta },\text{ }\mathbb{Q}\right] }$;

(iii) $F$ is mixed-Nash-A-implementable on $\Theta =\cup _{\theta \in \Theta
^{\ast }}\Omega ^{\left[ \succeq ^{\theta },\text{ }%
\mathbb{R}
\right] }$.
\end{theo}

A prominent class of preferences discussed in \cite{cmlr} is the single-top
preferences. Given single-top preferences, it is straightforward to show 
\begin{eqnarray*}
Z &=&Z^{\ast }\text{,} \\
\mathcal{L}_{i}^{Y}\left( \text{UNIF}\left[ F\left( \theta \right) \right]
,\theta \right) &\equiv &\widehat{\mathcal{L}}_{i}^{Y\text{-}A\text{-}%
B}\left( \text{UNIF}\left[ F\left( \theta \right) \right] ,\theta \right) 
\text{,} \\
\text{and set-monotonicity} &\Longleftrightarrow &\text{strong
set-monotonicity.}
\end{eqnarray*}%
As a result, Theorem \ref{theorem:MR:iff} implies that set-monotonicity
fully characterizes ordinally-mixed-Nash-implementable \emph{\`{a} la} \cite%
{cmlr}.

Similarly, we can easy derive full characterization of ordinal
implementation for the other 5 versions of mixed-Nash implementation of SCCs.

\section{Compared to rationalizable implementation}

\label{sec:rationalizable}

Given a mechanism $\mathcal{M}=\left\langle M\equiv \times _{i\in \mathcal{I}%
}M_{i}\text{, \ }g:M\longrightarrow Y\right\rangle $, define $\mathcal{S}%
_{i}\equiv 2^{M_{i}}$ and $\mathcal{S}=\times _{i\in \mathcal{I}}\mathcal{S}%
_{i}$ for each $i\in \mathcal{I}$. For each state $\theta \in \Theta $,
consider an operator $b^{\mathcal{M},\text{ }\theta }:\mathcal{S}%
\longrightarrow \mathcal{S}$ with $b^{\mathcal{M},\text{ }\theta }\equiv %
\left[ b_{i}^{\mathcal{M},\text{ }\theta }:\mathcal{S}\longrightarrow 
\mathcal{S}_{i}\right] _{i\in \mathcal{I}}$, where each $b_{i}^{\mathcal{M},%
\text{ }\theta }$ is defined as follows. For every $S\in \mathcal{S}$, 
\begin{equation*}
b_{i}^{\mathcal{M},\text{ }\theta }\left( S\right) =\left\{ m_{i}\in M_{i}:%
\begin{tabular}{c}
$\exists \lambda _{-i}\in \triangle \left( M_{-i}\right) $ such that \\ 
(1) $\lambda _{-i}\left( m_{-i}\right) >0$ implies $m_{-i}\in S_{-i}$, and
\\ 
(2) $m_{i}\in \arg \max_{m_{i}^{\prime }\in M_{i}}\Sigma _{m_{-i}\in
M_{-i}}\lambda _{-i}\left( m_{-i}\right) u_{i}\left( g\left( m_{i}^{\prime
},m_{-i}\right) ,\theta \right) $%
\end{tabular}%
\right\} \text{.}
\end{equation*}%
Clearly, $\mathcal{S}$ is a lattice with the order of "set inclusion," and $%
b^{\mathcal{M},\text{ }\theta }$ is monotonically increasing.\footnote{%
That is, $S\subset S^{\prime }$ implies $b^{\mathcal{M},\text{ }\theta
}\left( S\right) \subset b^{\mathcal{M},\text{ }\theta }\left( S^{\prime
}\right) $.} Thus, Tarski's fixed point theorem implies existence of a
largest fixed point of $b^{\mathcal{M},\text{ }\theta }$, and we denote it
by $S^{\mathcal{M},\text{ }\theta }\equiv \left( S_{i}^{\mathcal{M},\text{ }%
\theta }\right) _{i\in \mathcal{I}}$. We say $m_{i}\in M_{i}$ is
rationalizable in $\mathcal{M}$ at state $\theta $ if and only if $m_{i}\in
S_{i}^{\mathcal{M},\text{ }\theta }$.

We say that $S\in \mathcal{S}$ satisfies the best reply property in $%
\mathcal{M}$ at $\theta $ if and only if $S\subset b^{\mathcal{M},\text{ }%
\theta }\left( S\right) $. It is straightforward to show that $S\subset S^{%
\mathcal{M},\text{ }\theta }$ if $S$ satisfies the best reply property.

\begin{define}[\protect\cite{rjain}]
\label{def:rationalizable}An SCC $F:\Theta \longrightarrow 2^{Z}\diagdown
\left\{ \varnothing \right\} $ is rationalizably implementable if there
exists a mechanism $\mathcal{M}=\left\langle M\text{, \ }g:M\longrightarrow
Y\right\rangle $ such that%
\begin{equation*}
\dbigcup\limits_{m\in S^{\mathcal{M},\text{ }\theta }}\text{SUPP}\left[
g\left( m\right) \right] =F\left( \theta \right) \text{, }\forall \theta \in
\Theta \text{.}
\end{equation*}
\end{define}

\begin{theo}
\label{theorem:rationalizable}An SCC $F:\Theta \longrightarrow
2^{Z}\diagdown \left\{ \varnothing \right\} $ is mixed-Nash-A-implementatble
if $F$ is rationalizably-implementable.
\end{theo}

The detailed proof of Theorem \ref{theorem:rationalizable} is relegated to 
\cite{sx2022b}.

\section{Connected to \protect\cite{jmrr} and \protect\cite{tomas}}

\label{sec:connection}

In this section, we illustrate \cite{jmrr} and \cite{tomas}. We show that
our full characterization share the same conceptual ideas as those in \cite%
{jmrr} and \cite{tomas}, and furthermore, we show why their full
characterization is complicated, and why ours is simple.

\subsection{A common conceptual idea}

\label{sec:connection:concept}

\cite{em} proves that Maskin monotonicity almost fully characterizes Nash
implementation. As being showed in \cite{jmrr} and illustrated in \cite%
{tomas}, in order to fully characterize Nash implementation, we need to take
two additional steps before defining Maskin monotonicity. All of these
papers use the canonical mechanism in \cite{em} to achieve Nash
implementation, and the two additional steps corresponds to eliminating bad
equilibria in Case (3) and Case (2) of the canonical mechanism. Roughly, we
have the following two additional steps.%
\begin{gather}
\text{Step (I): select }\widehat{Z}\in 2^{Z}\diagdown \left\{ \varnothing
\right\} \text{ such that }\cup _{\theta \in \Theta }F\left( \theta \right)
\subset \widehat{Z}\text{,}  \notag \\
\text{and }\widehat{Z}\text{ satisfies a unanimity condition:}  \notag \\
\forall \left( \theta ^{\ast },y\right) \in \Theta \times \widehat{Z}\text{, 
}  \notag \\
\left[ 
\begin{array}{c}
y\in \arg \max_{z\in \widehat{Z}}U_{i}^{\theta ^{\ast }}\left( z\right) 
\text{,} \\ 
\forall i\in \mathcal{I}%
\end{array}%
\right] \Longrightarrow \text{"}y\text{ is a good outcome at }\theta ^{\ast }%
\text{."}  \label{tgr1}
\end{gather}%
where $\widehat{Z}$ is the set of outcomes that agents can choose when Case
(3) is triggered in the canonical mechanism. In particular, "$y$ is a good
outcome at $\theta ^{\ast }$" in (\ref{tgr1}) may have different
formalization in different environments and under different implementation
notions, which will be illustrated in Sections \ref{sec:connection:M-R}, \ref%
{sec:illustration:SCF} and \ref{sec:connection:A}.

To see the necessity of Step (I), suppose an SCC $F$ is Nash implemented by $%
\mathcal{M}=\left\langle M\text{, \ }g:M\longrightarrow Y\right\rangle $,
and define $\widehat{Z}=g\left( M\right) $. If $y=g\left( m\right) $
satisfies the left-hand side of (\ref{tgr1}) for some $\theta ^{\ast }\in
\Theta $, then, $m$ must be a Nash equilibrium at $\theta ^{\ast }$, and
hence, $y=g\left( m\right) $ must be a good outcome $\theta ^{\ast }$, i.e.,
the right-hand side of (\ref{tgr1}) holds.

To see the sufficiency of Step (I), suppose the true state is $\theta ^{\ast
}$. Consider any Nash equilibrium in Case (3) of the canonical mechanism,
which induces $y\in \widehat{Z}$. Then, $y$ must be a top outcome in $%
\widehat{Z}$ for all agents, i.e., the left-hand side of (\ref{tgr1}) holds.
Thus, by (\ref{tgr1}), $y$ is a good outcome at $\theta ^{\ast }$.

\begin{gather}
\text{Step (II): for each }\left( \theta ,i\right) \in \Theta \times 
\mathcal{I}\text{ and each }a\in F\left( \theta \right) \text{, select }%
\widehat{\mathcal{L}}_{i}\left( a,\theta \right) \in 2^{\left[ \widehat{Z}%
\cap \mathcal{L}_{i}\left( a,\theta \right) \right] }\diagdown \left\{
\varnothing \right\} \text{ }  \notag \\
\text{such that }a\in \widehat{\mathcal{L}}_{i}\left( a,\theta \right) \text{%
,}  \notag \\
\text{and }\widehat{\mathcal{L}}_{i}\left( a,\theta \right) \text{ satisfies
a weak no-veto condition:}  \notag \\
\forall \left( \theta ^{\ast },y\right) \in \Theta \times \widehat{\mathcal{L%
}}_{i}\left( a,\theta \right) \text{, }  \notag \\
\left[ 
\begin{array}{c}
y\in \arg \max_{z\in \widehat{Z}}U_{j}^{\theta ^{\ast }}\left( z\right) 
\text{, }\forall j\in \mathcal{I}\diagdown \left\{ i\right\} \text{,} \\ 
y\in \arg \max_{z\in \widehat{\mathcal{L}}_{i}\left( a,\theta \right)
}U_{i}^{\theta ^{\ast }}\left( z\right) \text{,}%
\end{array}%
\right] \Longrightarrow \text{"}y\text{ is a good outcome at }\theta ^{\ast }%
\text{." }  \label{tgr2}
\end{gather}%
where $\widehat{\mathcal{L}}_{i}\left( a,\theta \right) $ is the set of
outcomes that agent $i$ can choose when agent $i$ is the whistle-blower and
agents $-i$ report $\left( a,\theta \right) $, i.e., Case (2) is triggered
in the canonical mechanism.

To see the necessity of Step (II), suppose an SCC $F$ is Nash implemented by 
$\mathcal{M}=\left\langle M\text{, \ }g:M\longrightarrow Y\right\rangle $,
and define%
\begin{equation*}
\widehat{\mathcal{L}}_{i}\left( a,\theta \right) =\left\{ g\left(
m_{i},\lambda _{-i}\right) :%
\begin{tabular}{l}
$m_{i}\in M_{i}\text{,}$ \\ 
$g\left( \lambda _{i},\lambda _{-i}\right) =a\text{,}$ \\ 
$\text{and }\left( \lambda _{i},\lambda _{-i}\right) \text{ is a Nash
equilibrium}$%
\end{tabular}%
\right\} \text{.}
\end{equation*}%
If $y=g\left( m_{i},\lambda _{-i}\right) $ satisfies the left-hand side of (%
\ref{tgr2}) for some $\theta ^{\ast }\in \Theta $, $\left( m_{i},\lambda
_{-i}\right) $ must be a Nash equilibrium at $\theta ^{\ast }$, and $%
y=g\left( m_{i},\lambda _{-i}\right) $ must be a good outcome at $\theta
^{\ast }$, i.e., the right-hand side of (\ref{tgr2}) holds.

To see the sufficiency of Step (II), suppose the true state is $\theta
^{\ast }$. Consider any Nash equilibrium in Case (2) of the canonical
mechanism such that $i$ is the whistle-blower and it induces $y\in \widehat{%
\mathcal{L}}_{i}\left( a,\theta \right) $. Then, $y$ must be a top outcome
in $\widehat{Z}$ for agents $-i$ (because they can deviate to Case (3)), and 
$y$ must be a top outcome in $\widehat{\mathcal{L}}_{i}\left( a,\theta
\right) $ for agents $i$ (because $i$ can deviate to any outcome in $%
\widehat{\mathcal{L}}_{i}\left( a,\theta \right) $), i.e., the left-hand
side of (\ref{tgr2}) holds. Thus, by (\ref{tgr2}), $y$ is a good outcome at $%
\theta ^{\ast }$.

Given Steps (I) and (II), we define a modified Maskin monotonicity as
follows.%
\begin{gather*}
\widehat{\mathcal{L}}\text{-Maskin-monotonicity holds iff} \\
\text{ }\left[ 
\begin{array}{c}
a\in F\left( \theta \right) \text{,} \\ 
\widehat{\mathcal{L}}_{i}\left( a,\theta \right) \subset \mathcal{L}%
_{i}\left( a,\theta \right) \text{, }\forall i\in \mathcal{I}%
\end{array}%
\right] \text{ }\Longrightarrow a\in F\left( \theta ^{\prime }\right) \text{%
, }\forall \left( \theta ,\theta ^{\prime },a\right) \in \Theta \times
\Theta \times Z\text{.}
\end{gather*}%
This leads to a full characterization of Nash implementation: an SCC $F$ is
Nash implementable if and only if $\widehat{\mathcal{L}}$%
-Maskin-monotonicity holds. In particular, in the canonical mechanism, Step
(I) eliminates bad equilibria in Case (3), and Step (II) eliminates bad
equilibria in Case (2), and $\widehat{\mathcal{L}}$-Maskin-monotonicity
eliminates bad equilibria in Case (1).

\subsection{The full characterization in \protect\cite{jmrr} and 
\protect\cite{tomas}}

\label{sec:connection:M-R}

In the environment of \cite{jmrr}, the two steps becomes 
\begin{gather}
\text{Step (I): select }\widehat{Z}\in 2^{Z}\diagdown \left\{ \varnothing
\right\} \text{ such that }\cup _{\theta \in \Theta }F\left( \theta \right)
\subset \widehat{Z}\text{ and}  \notag \\
\forall \left( \theta ^{\ast },y\right) \in \Theta \times \widehat{Z}\text{, 
}  \notag \\
\left[ 
\begin{array}{c}
y\in \arg \max_{z\in \widehat{Z}}U_{i}^{\theta ^{\ast }}\left( z\right) 
\text{,} \\ 
\forall i\in \mathcal{I}%
\end{array}%
\right] \Longrightarrow y\in F\left( \theta ^{\ast }\right) \text{;}
\label{hhg1}
\end{gather}

\begin{gather}
\text{Step (II): for each }\left( \theta ,i\right) \in \Theta \times 
\mathcal{I}\text{ and each }a\in F\left( \theta \right) \text{, select }%
\widehat{\mathcal{L}}_{i}\left( a,\theta \right) \in 2^{\left[ \widehat{Z}%
\cap \mathcal{L}_{i}^{Z}\left( a,\theta \right) \right] }\diagdown \left\{
\varnothing \right\}  \notag \\
\text{such that }a\in \widehat{\mathcal{L}}_{i}\left( a,\theta \right) \text{
and}  \notag \\
\forall \left( \theta ^{\ast },y\right) \in \Theta \times \widehat{\mathcal{L%
}}_{i}\left( a,\theta \right) \text{, }  \notag \\
\left[ 
\begin{array}{c}
y\in \arg \max_{z\in \widehat{Z}}U_{j}^{\theta ^{\ast }}\left( z\right) 
\text{, }\forall j\in \mathcal{I}\diagdown \left\{ i\right\} \text{,} \\ 
y\in \arg \max_{z\in \widehat{\mathcal{L}}_{i}\left( a,\theta \right)
}U_{i}^{\theta ^{\ast }}\left( z\right) \text{,}%
\end{array}%
\right] \Longrightarrow y\in F\left( \theta ^{\ast }\right) \text{ }
\label{hhg2}
\end{gather}%
This leads to the full characterization in \cite{jmrr}: an SCC $F$ is Nash
implementable if and only if there exists such $\left[ \widehat{Z}\text{, }%
\left( \widehat{\mathcal{L}}_{i}\left( a,\theta \right) \right) _{i\in 
\mathcal{I}\text{, }\theta \in \Theta \text{, }a\in F\left( \theta \right) }%
\right] $ and $\widehat{\mathcal{L}}$-Maskin-monotonicity hold. However, 
\cite{jmrr} is silent regarding how to find such $\left[ \widehat{Z}\text{, }%
\left( \widehat{\mathcal{L}}_{i}\left( a,\theta \right) \right) _{i\in 
\mathcal{I}\text{, }\theta \in \Theta \text{, }a\in F\left( \theta \right) }%
\right] $, while \cite{tomas} provides an algorithm to find the largest such
sets.\footnote{%
There may be multiple candidates of $\widehat{Z}$ which satisfies (\ref{hhg1}%
). The union of these candidates is the largest $\widehat{Z}$ satisfying (%
\ref{hhg1}), which is identified by \cite{tomas}. Similarly, \cite{tomas}
identifies the largest such $\widehat{\mathcal{L}}_{i}\left( a,\theta
\right) $ for each $i\in \mathcal{I}$, $\theta \in \Theta $ and $a\in
F\left( \theta \right) $.} In order to find $\widehat{Z}$, we need an
iterative process of elimination. At round 1, define $Z^{1}\equiv Z$. If $%
\widehat{Z}=Z^{1}$ does not satisfy (\ref{hhg1}), we eliminate any $z$ that
is a top outcome in $Z^{1}$ for all agents at some state $\theta $, but $%
z\notin F\left( \theta \right) $, and let $Z^{2}$ denote the set of outcomes
that survive round 1. We have found the appropriate $\widehat{Z}$ if $%
\widehat{Z}=Z^{2}$ satisfies (\ref{hhg1}). However, $\widehat{Z}=Z^{2}$ may
not satisfy (\ref{hhg1}), i.e., it may happen that some $z\in Z^{1}$ is not
a top outcome of some agent in $Z^{1}$ at some state $\theta $, but becomes
a top outcome in $Z^{2}$ for all agents at $\theta $, while $z\notin F\left(
\theta \right) $. In this case, we need another round of elimination. At
round 2, if $\widehat{Z}=Z^{2}$ does not satisfy (\ref{hhg1}), we eliminate
any $z$ that is a top outcome in $Z^{2}$ for all agents at some state $%
\theta $, but $z\notin F\left( \theta \right) $, and let $Z^{3}$ denote the
set of outcomes that survive round 2.... We continue this process until we
find $Z^{n}$ such that $\widehat{Z}=Z^{n}$ satisfies (\ref{hhg1}).

After finding $\widehat{Z}$, we need a similar iterative process to find
each $\widehat{\mathcal{L}}_{i}\left( a,\theta \right) $. At round 1, define 
$\mathcal{L}_{i}^{1}\left( a,\theta \right) \equiv \mathcal{L}_{i}\left(
a,\theta \right) $. If $\widehat{\mathcal{L}}_{i}\left( a,\theta \right) =%
\mathcal{L}_{i}^{1}\left( a,\theta \right) $ does not satisfy (\ref{hhg2}),
we eliminate any $z$ that is a top outcome in $\widehat{Z}$ for agents $-i$
at some state $\theta ^{\ast }$ and a top outcome in $\mathcal{L}%
_{i}^{1}\left( a,\theta \right) $ for agent $i$ at $\theta ^{\ast }$, but $%
z\notin F\left( \theta ^{\ast }\right) $. Let $\mathcal{L}_{i}^{2}\left(
a,\theta \right) $ denote the set of outcomes that survive round 1. At round
2, if $\widehat{\mathcal{L}}_{i}\left( a,\theta \right) =\mathcal{L}%
_{i}^{2}\left( a,\theta \right) $ does not satisfy (\ref{hhg2}), we
eliminate any $z$ that is a top outcome in $\widehat{Z}$ for agents $-i$ at
some state $\theta ^{\ast }$ and a top outcome in $\mathcal{L}_{i}^{2}\left(
a,\theta \right) $ for agent $i$ at $\theta ^{\ast }$, but $z\notin F\left(
\theta ^{\ast }\right) $. Let $\mathcal{L}_{i}^{3}\left( a,\theta \right) $
denote the set of outcomes that survive round 2.... We continue this process
until we find $\mathcal{L}_{i}^{n}\left( a,\theta \right) $ such that $%
\widehat{\mathcal{L}}_{i}\left( a,\theta \right) =\mathcal{L}_{i}^{n}\left(
a,\theta \right) $ satisfies (\ref{hhg2}).

Clearly, both the existential statement in \cite{jmrr} and the iterative
process in \cite{tomas} make their full characterization complicated.

\subsection{Our full characterization for SCFs}

\label{sec:illustration:SCF}

For simplicity, we focus on SCFs throughout this subsection. Given
stochastic mechanisms, the insight of this paper is that we can easily
select $\left[ \widehat{Z}\text{, }\left( \widehat{\mathcal{L}}_{i}\left(
f\left( \theta \right) ,\theta \right) \right) _{i\in \mathcal{I}\text{, }%
\theta \in \Theta }\right] $ in Step (I) and Step (II). Taking full
advantage of the convexity structure of lotteries, we assign the following
lottery, when Case (3) is triggered in the canonical mechanism.%
\begin{equation*}
g\left[ m\right] =\left( 1-\frac{1}{k_{j^{\ast }}^{2}}\right) \times
b_{j^{\ast }}+\frac{1}{k_{j^{\ast }}^{2}}\times \text{UNIF}\left( \widehat{Z}%
\right) \text{.}
\end{equation*}%
In particular, the winner of the integer game, i.e., $j^{\ast }$, can
increase the probability of her top outcome $b_{j^{\ast }}$ by increasing $%
k_{j^{\ast }}^{2}$. As a result, a Nash equilibrium in Case (3) at true
state $\theta ^{\ast }$ must require all agents be indifferent between any
two outcomes in $\widehat{Z}$ at $\theta ^{\ast }$. Given SCFs, this means%
\footnote{%
Given the canonical mechanism $\mathcal{M}=\left\langle M\text{, \ }%
g:M\longrightarrow Y\right\rangle $, all agents being indifferent between
any two outcomes in $\widehat{Z}$ at $\theta ^{\ast }$ implies any $m\in M$
is a Nash equilibrium at $\theta ^{\ast }$, i.e., $M\subset MNE^{\left( 
\mathcal{M},\text{ }\theta ^{\ast }\right) }$. Since $\cup _{m\in M}$SUPP$%
\left[ g\left( m\right) \right] =\widehat{Z}$, we conclude that $\widehat{Z}%
=\cup _{m\in M}$SUPP$\left[ g\left( m\right) \right] =\left\{ f\left( \theta
^{\ast }\right) \right\} $, i.e., $\left\vert \widehat{Z}\right\vert =1$.}%
\begin{gather*}
\text{Step (I): select }\widehat{Z}\in 2^{Z}\diagdown \left\{ \varnothing
\right\} \text{ such that }f\left( \Theta \right) \subset \widehat{Z}\text{
and} \\
\forall \theta ^{\ast }\in \Theta \text{, } \\
\left[ 
\begin{array}{c}
\widehat{Z}\subset \arg \max_{z\in \widehat{Z}}U_{i}^{\theta ^{\ast }}\left(
z\right) \text{,} \\ 
\forall i\in \mathcal{I}%
\end{array}%
\right] \Longrightarrow \left( 
\begin{array}{c}
\widehat{Z}\subset \left\{ f\left( \theta ^{\ast }\right) \right\} \text{,}
\\ 
\text{and hence, }\left\vert \widehat{Z}\right\vert =1\text{,}%
\end{array}%
\right) \text{.}
\end{gather*}%
Assumption \ref{assm:non-trivial} and $f\left( \Theta \right) \subset 
\widehat{Z}$ imply that $\left\vert \widehat{Z}\right\vert =1$ always fails.
As a result, Step (I)\ becomes%
\begin{gather*}
\text{Step (I): select }\widehat{Z}\in 2^{Z}\diagdown \left\{ \varnothing
\right\} \text{ such that }f\left( \Theta \right) \subset \widehat{Z}\text{
and} \\
\forall \theta ^{\ast }\in \Theta \text{, } \\
\left[ 
\begin{array}{c}
\widehat{Z}\subset \arg \max_{z\in \widehat{Z}}U_{i}^{\theta ^{\ast }}\left(
z\right) \text{,} \\ 
\forall i\in \mathcal{I}%
\end{array}%
\right] \text{ fails.}
\end{gather*}%
Since $\widehat{Z}\subset Z$, it is straightforward to show%
\begin{eqnarray*}
\left[ 
\begin{array}{c}
Z\text{ is not a }i\text{-max set} \\ 
\forall i\in \mathcal{I}%
\end{array}%
\right] &\Longleftrightarrow &\left( 
\begin{array}{c}
\left[ 
\begin{array}{c}
Z\subset \arg \max_{z\in \widehat{Z}}U_{i}^{\theta ^{\ast }}\left( z\right) 
\text{,} \\ 
\forall i\in \mathcal{I}%
\end{array}%
\right] \text{ fails} \\ 
\forall \theta ^{\ast }\in \Theta \text{, }%
\end{array}%
\right) \\
&\Longleftarrow &\left( 
\begin{array}{c}
\left[ 
\begin{array}{c}
\widehat{Z}\subset \arg \max_{z\in \widehat{Z}}U_{i}^{\theta ^{\ast }}\left(
z\right) \text{,} \\ 
\forall i\in \mathcal{I}%
\end{array}%
\right] \text{ fails} \\ 
\forall \theta ^{\ast }\in \Theta \text{, }%
\end{array}%
\right) \text{.}
\end{eqnarray*}%
Therefore, without an iterative process of elimination, we have already
found the largest such $\widehat{Z}$, i.e., $\widehat{Z}=Z$. In particular,
a necessary condition for Nash implementation is: $Z$ is not a $i$-max set
for any $i\in \mathcal{I}$.\footnote{%
This necessary condition is explicitly stated in (iii) of Theorem \ref%
{theorem:full:mix:SCC-E-F}, but omitted in Theorem \ref{theorem:full:mix},
because, with SCFs, it is implicitly encoded in $\widehat{\mathcal{L}}^{Y}$%
-monotonicity\ (see Lemma \ref{lem:no-veto:generalized}).}

Similarly, taking full advantage of the convexity structure of lotteries, we
assign the following lottery, when Case (2) is triggered in the canonical
mechanism.%
\begin{eqnarray*}
g\left[ m\right] &=&\left( 1-\frac{1}{k_{j}^{2}}\right) \times \phi
_{j}^{\theta }\left( \theta _{j}\right) \\
&&+\frac{1}{k_{j}^{2}}\times \left( 
\begin{tabular}{l}
$\varepsilon _{j}^{\theta }\times \left[ \left( 1-\frac{1}{k_{j}^{3}}\right)
\times \gamma _{j}\left( \widehat{\Gamma }_{j}\left( \theta \right) \right) +%
\frac{1}{k_{j}^{3}}\times \text{UNIF}\left( \widehat{\Gamma }_{j}\left(
\theta \right) \right) \right] $ \\ 
$+\left( 1-\varepsilon _{j}^{\theta }\right) \times y_{j}^{\theta }$%
\end{tabular}%
\right) \text{,}
\end{eqnarray*}%
In particular, the whistle-blower $j$ can increase the probability of her
top outcome $\gamma _{j}\left( \widehat{\Gamma }_{j}\left( \theta \right)
\right) $ in $\widehat{\Gamma }_{j}\left( \theta \right) $ by increasing $%
k_{j}^{3}$. As a result, a Nash equilibrium in Case (2) at true state $%
\theta ^{\ast }$ must require all outcomes in $\widehat{\Gamma }_{j}\left(
\theta \right) $ be top for agents $-j$ at $\theta ^{\ast }$ and agent $j$
be indifferent between any two outcomes in $\widehat{\Gamma }_{j}\left(
\theta \right) $ at $\theta ^{\ast }$. Given SCFs, we thus have

\begin{gather}
\text{Step (II): for each }\left( \theta ,i\right) \in \Theta \times 
\mathcal{I}\text{, select }\widehat{\mathcal{L}}_{i}^{Y}\left( f\left(
\theta \right) ,\theta \right) \in 2^{\left[ \mathcal{L}_{i}^{Y}\left(
f\left( \theta \right) ,\theta \right) \right] }\diagdown \left\{
\varnothing \right\}  \notag \\
\text{such that }f\left( \theta \right) \in \widehat{\mathcal{L}}%
_{i}^{Y}\left( f\left( \theta \right) ,\theta \right) \text{ and}  \notag \\
\forall \left( \theta ^{\ast },y\right) \in \Theta \times \widehat{\mathcal{L%
}}_{i}^{Y}\left( f\left( \theta \right) ,\theta \right) \text{, }  \notag \\
\left[ 
\begin{array}{c}
\widehat{\Gamma }_{i}\left( \theta \right) \subset \arg \max_{z\in
Z}u_{j}^{\theta ^{\ast }}\left( z\right) \text{, }\forall j\in \mathcal{I}%
\diagdown \left\{ i\right\} \text{,} \\ 
\widehat{\Gamma }_{i}\left( \theta \right) \subset \arg \max_{z\in \widehat{%
\Gamma }_{i}\left( \theta \right) }u_{i}^{\theta ^{\ast }}\left( z\right) 
\text{,}%
\end{array}%
\right] \Longrightarrow \left( 
\begin{array}{c}
\widehat{\Gamma }_{i}\left( \theta \right) \subset \left\{ f\left( \theta
^{\ast }\right) \right\} \text{,} \\ 
\text{and hence, }\left\vert \widehat{\Gamma }_{i}\left( \theta \right)
\right\vert =1\text{,}%
\end{array}%
\right) \text{,}  \notag \\
\text{or equivalently, }\left[ \widehat{\Gamma }_{i}\left( \theta \right) 
\text{ is an }i\text{-}\theta ^{\ast }\text{-max set}\right] \Longrightarrow
\left( 
\begin{array}{c}
\widehat{\Gamma }_{i}\left( \theta \right) \subset \left\{ f\left( \theta
^{\ast }\right) \right\} \text{,} \\ 
\text{and hence, }\left\vert \widehat{\Gamma }_{i}\left( \theta \right)
\right\vert =1\text{,}%
\end{array}%
\right) \text{,}  \label{bbt1} \\
\text{where }\widehat{\Gamma }_{i}\left( \theta \right) \equiv
\dbigcup\limits_{y\in \widehat{\mathcal{L}}_{i}^{Y}\left( f\left( \theta
\right) ,\theta \right) }\text{SUPP}\left[ y\right] \text{.}  \notag
\end{gather}%
Thus, it is straightforward to find the largest such $\widehat{\mathcal{L}}%
_{i}^{Y}\left( f\left( \theta \right) ,\theta \right) $ by considering three
scenarios:%
\begin{equation*}
\left( 
\begin{array}{c}
\text{scenario (I): }f\left( \theta \right) \in \arg \min_{z\in
Z}u_{i}^{\theta }\left( z\right) \text{ \ and }\mathcal{L}_{i}^{Z}\left(
f\left( \theta \right) ,\theta \right) \text{ is an }i\text{-max set,} \\ 
\text{scenario (II): }f\left( \theta \right) \in \arg \min_{z\in
Z}u_{i}^{\theta }\left( z\right) \text{ \ and }\mathcal{L}_{i}^{Z}\left(
f\left( \theta \right) ,\theta \right) \text{ is not an }i\text{-max set,}
\\ 
\text{scenario (III): }f\left( \theta \right) \notin \arg \min_{z\in
Z}u_{i}^{\theta }\left( z\right)%
\end{array}%
\right) \text{.}
\end{equation*}%
In scenario (I), we must have $\widehat{\mathcal{L}}_{i}^{Y}\left( f\left(
\theta \right) ,\theta \right) =\left\{ f\left( \theta \right) \right\} $ by
Lemma \ref{lem:mixed:deviation:SCF}. In scenarios (II) and (III), $\widehat{%
\mathcal{L}}_{i}^{Y}\left( f\left( \theta \right) ,\theta \right) =\mathcal{L%
}_{i}^{Y}\left( f\left( \theta \right) ,\theta \right) $ makes (\ref{bbt1})
hold \emph{vacuously}.\footnote{%
In scenario (III), $f\left( \theta \right) \notin \arg \min_{z\in
Z}u_{j}^{\theta }\left( z\right) $ implies $\widehat{\Gamma }_{j}\left(
\theta \right) =Z$, and $Z$ is not an $i$-$\theta ^{\ast }$-max set for any $%
i\in \mathcal{I}$, which would be implied by $\widehat{\mathcal{L}}^{Y}$%
-monotonicity imposed later (see Lemma \ref{lem:no-veto:generalized}).}%
---This leads to the definition of $\widehat{\mathcal{L}}_{i}^{Y}\left(
f\left( \theta \right) ,\theta \right) $ in (\ref{yjj8}).

\subsection{Our full characterization for mixed-Nash-A-implementation}

\label{sec:connection:A}

We consider SCCs and focus on mixed-Nash-A-implementation throughout this
subsection. We first illustrate why $\widehat{\mathcal{L}}^{Y}$%
-Maskin-monotonicity is defined on UNIF$\left[ F\left( \theta \right) \right]
$. Suppose that $F$ is mixed-Nash-A-implemented by $\mathcal{M}=\left\langle
M\text{, \ }g:M\longrightarrow Y\right\rangle $, and consider any $\lambda
\in MNE^{\left( \mathcal{M},\text{ }\theta \right) }$. By
mixed-Nash-A-implementation, $g\left( \lambda \right) $ could be any
lotteries in $\triangle \left[ F\left( \theta \right) \right] $, and as a
result, we need to define $\widehat{\mathcal{L}}$-Maskin-monotonicity as:%
\begin{equation*}
\left[ 
\begin{array}{c}
\widehat{\mathcal{L}}_{i}^{Y}\left( \eta ,\theta \right) \subset \mathcal{L}%
_{i}^{Y}\left( \eta ,\theta ^{\prime }\right) \text{, } \\ 
\forall \eta \in \triangle \left[ F\left( \theta \right) \right] \text{, }%
\forall i\in \mathcal{I}%
\end{array}%
\right] \text{ }\Longrightarrow F\left( \theta \right) \subset F\left(
\theta ^{\prime }\right) \text{, }\forall \left( \theta ,\theta ^{\prime
}\right) \in \Theta \times \Theta \text{.}
\end{equation*}%
However, the following lemma shows that it suffers no loss of generality to
define $\widehat{\mathcal{L}}$-Maskin-monotonicity on UNIF$\left[ F\left(
\theta \right) \right] $ only (i.e., Definition \ref{defin:A-B}). The proof
of Lemma \ref{lem:mixed-lottery:lower-contour} is relegated to Appendix \ref%
{sec:lem:mixed-lottery:lower-contour}.

\begin{lemma}
\label{lem:mixed-lottery:lower-contour}For any $E\in 2^{Z}\diagdown \left\{
\varnothing \right\} $ and any $\left( \gamma ,i,\theta ,\theta ^{\prime
}\right) \in \triangle ^{\circ }\left( E\right) \times \mathcal{I}\times
\Theta \times \Theta $, we have%
\begin{equation}
\left( 
\begin{array}{c}
\mathcal{L}_{i}^{Y}\left( \eta ,\theta \right) \subset \mathcal{L}%
_{i}^{Y}\left( \eta ,\theta ^{\prime }\right) \text{,} \\ 
\forall \eta \in \triangle \left[ E\right]%
\end{array}%
\right) \Longleftrightarrow \mathcal{L}_{i}^{Y}\left( \gamma ,\theta \right)
\subset \mathcal{L}_{i}^{Y}\left( \gamma ,\theta ^{\prime }\right) \text{.}
\label{kkg1a}
\end{equation}
\end{lemma}

Second, by a similar argument as in Section \ref{sec:illustration:SCF},
mixed-Nash-A-implementation implies

\begin{gather}
\text{Step (I): select }\widehat{Z}\in 2^{Z}\diagdown \left\{ \varnothing
\right\} \text{ such that }\cup _{\theta \in \Theta }F\left( \theta \right)
\subset \widehat{Z}\text{ and}  \notag \\
\forall \theta ^{\ast }\in \Theta \text{, }  \notag \\
\left[ 
\begin{array}{c}
\widehat{Z}\subset \arg \max_{z\in \widehat{Z}}U_{i}^{\theta ^{\ast }}\left(
z\right) \text{,} \\ 
\forall i\in \mathcal{I}%
\end{array}%
\right] \Longrightarrow \widehat{Z}\subset F\left( \theta ^{\ast }\right) 
\text{.}  \label{hhe1}
\end{gather}%
Without an iterative process of elimination, it is straightforward to find
the largest such $\widehat{Z}$ by consider two scenarios. If $Z$ is not an $%
i $-max set for any $i\in \mathcal{I}$, $\widehat{Z}=Z$ makes (\ref{hhe1})
hold \emph{vacuously}. Otherwise, we must have $\widehat{Z}=\cup _{\theta
\in \Theta }F\left( \theta \right) $.\footnote{%
Given the canonical mechanism $\mathcal{M}=\left\langle M\text{, \ }%
g:M\longrightarrow Y\right\rangle $, all agents being indifferent between
any two outcomes in $\widehat{Z}$ at $\theta ^{\ast }$ implies any $m\in M$
is a Nash equilibrium. As a result, we have $\widehat{Z}=\cup _{m\in M}$SUPP$%
\left[ g\left( m\right) \right] \subset F\left( \theta ^{\ast }\right) $,
which further implies $\cup _{\theta \in \Theta }F\left( \theta \right)
\subset \widehat{Z}\subset F\left( \theta ^{\ast }\right) \subset \cup
_{\theta \in \Theta }F\left( \theta \right) $, i.e., $\widehat{Z}=\cup
_{\theta \in \Theta }F\left( \theta \right) $.} Therefore, we must consider $%
Z^{\ast }$ defined in (\ref{hhe1}) for mixed-Nash-A-implementation.

Third, by a similar argument as in Section \ref{sec:illustration:SCF},
mixed-Nash-A-implementation implies

\begin{gather}
\text{Step (II): for each }\left( \theta ,i\right) \in \Theta \times 
\mathcal{I}\text{, select }\widehat{\mathcal{L}}_{i}^{Y}\left( \text{UNIF}%
\left[ F\left( \theta \right) \right] ,\theta \right) \in 2^{\left[
\triangle \left( Z^{\ast }\right) \cap \mathcal{L}_{i}^{Y}\left( \text{UNIF}%
\left[ F\left( \theta \right) \right] ,\theta \right) \right] }\diagdown
\left\{ \varnothing \right\}  \notag \\
\text{such that UNIF}\left[ F\left( \theta \right) \right] \in \widehat{%
\mathcal{L}}_{i}^{Y}\left( \text{UNIF}\left[ F\left( \theta \right) \right]
,\theta \right) \text{ and}  \notag \\
\forall \left( \theta ^{\ast },y\right) \in \Theta \times \widehat{\mathcal{L%
}}_{i}^{Y}\left( \text{UNIF}\left[ F\left( \theta \right) \right] ,\theta
\right) \text{, }  \notag \\
\left[ \widehat{\Gamma }_{i}\left( \theta \right) \text{ is an }i\text{-}%
Z^{\ast }\text{-}\theta ^{\ast }\text{-max set}\right] \Longrightarrow 
\widehat{\Gamma }_{i}\left( \theta \right) \subset F\left( \theta ^{\ast
}\right) \text{.}  \label{bbt2}
\end{gather}%
Thus, we can find the largest such $\widehat{\mathcal{L}}_{i}^{Y}\left(
a,\theta \right) $ by considering three scenarios:%
\begin{equation*}
\left( 
\begin{array}{c}
\text{scenario (I): }F\left( \theta \right) \subset \arg \min_{z\in Z^{\ast
}}u_{i}^{\theta }\left( z\right) \text{ \ and }Z^{\ast }\cap \mathcal{L}%
_{i}^{Z}\left( F\left( \theta \right) ,\theta \right) \text{ is an }i\text{-}%
Z^{\ast }\text{-max set,} \\ 
\text{scenario (II): }F\left( \theta \right) \subset \arg \min_{z\in Z^{\ast
}}u_{i}^{\theta }\left( z\right) \text{ \ and }Z^{\ast }\cap \mathcal{L}%
_{i}^{Z}\left( F\left( \theta \right) ,\theta \right) \text{ is not an }i%
\text{-}Z^{\ast }\text{-max set,} \\ 
\text{scenario (III): }F\left( \theta \right) \diagdown \arg \min_{z\in
Z^{\ast }}u_{i}^{\theta }\left( z\right) \neq \varnothing%
\end{array}%
\right) \text{.}
\end{equation*}%
In scenario (I), Lemma \ref{lem:mixed:deviation:SCC} implies that we must
have%
\begin{equation*}
\widehat{\mathcal{L}}_{i}^{Y}\left( \text{UNIF}\left[ F\left( \theta \right) %
\right] ,\theta \right) =\triangle \left[ Z^{\ast }\cap \mathcal{L}%
_{i}^{Z}\left( F\left( \theta \right) ,\theta \right) \cap \left(
\dbigcup\limits_{K\in \Xi _{i}\left( \theta \right) }\dbigcap\limits_{\theta
^{\prime }\in K}F\left( \theta ^{\prime }\right) \right) \right] \text{.}
\end{equation*}%
In scenario (II), $\widehat{\mathcal{L}}_{i}^{Y}\left( \text{UNIF}\left[
F\left( \theta \right) \right] ,\theta \right) =\mathcal{L}_{i}^{Y}\left( 
\text{UNIF}\left[ F\left( \theta \right) \right] ,\theta \right) $ makes (%
\ref{bbt2}) hold \emph{vacuously}. In scenario (III), $\widehat{\mathcal{L}}%
_{i}^{Y}\left( \text{UNIF}\left[ F\left( \theta \right) \right] ,\theta
\right) =\mathcal{L}_{i}^{Y}\left( \text{UNIF}\left[ F\left( \theta \right) %
\right] ,\theta \right) $ implies $\widehat{\Gamma }_{i}\left( \theta
\right) =Z^{\ast }$, and (\ref{bbt2}) holds by Lemma \ref%
{lem:no-veto:generalized:SCC}.---This leads to the definition of $\widehat{%
\mathcal{L}}_{i}^{Y\text{-}A\text{-}B}\left( \text{UNIF}\left[ F\left(
\theta \right) \right] ,\theta \right) $ in (\ref{ddtt}).

\section{Conclusion}

\label{sec:conclude}

We study Nash implementation by stochastic mechanisms, and provide a
surprisingly simple full characterization. Even though our full
characterization is of a form similar to Maskin monotonicity \emph{\`{a} la} 
\cite{em}, it has an interpretation parallel to \cite{jmrr} and \cite{tomas}%
. In this sense, we build a bridge between \cite{em} and \cite{jmrr} (as
well as \cite{tomas}).

Furthermore, our full characterization shed light on%
\begin{equation*}
\left( 
\begin{array}{c}
\text{"mixed-Nash-implementation VS pure-Nash-implementation,"} \\ 
\text{"ordinal-approach VS cardinal-approach,"} \\ 
\text{"Nash-implementation VS rationalizable-implementation" }%
\end{array}%
\right) \text{.}
\end{equation*}

\bigskip

\newpage

\appendix

\section{Proofs}

\subsection{Proof of Lemma \protect\ref{lem:mixed:deviation:SCF}}

\label{sec:lem:mixed:deviation:SCF}

Suppose that $f$ is mixed-Nash implemented by $\mathcal{M}=\left\langle M%
\text{, \ }g:M\longrightarrow Y\right\rangle $, and fix any $\left( i,\theta
\right) \in \mathcal{I}\times \Theta $ and any $\lambda \in MNE^{\left( 
\mathcal{M},\text{ }\theta \right) }$ such that%
\begin{equation*}
\left[ 
\begin{array}{c}
f\left( \theta \right) \in \arg \min_{z\in Z}u_{i}^{\theta }\left( z\right) 
\text{ and} \\ 
\mathcal{L}_{i}^{Z}\left( f\left( \theta \right) ,\theta \right) \text{ is
an }i\text{-max set }%
\end{array}%
\right] \text{,}
\end{equation*}%
and we aim to show $\dbigcup\limits_{m_{i}\in M_{i}}$SUPP$\left[ g\left(
m_{i},\lambda _{-i}\right) \right] =\left\{ f\left( \theta \right) \right\} $%
. First, $f\left( \theta \right) \in \arg \min_{z\in Z}u_{i}^{\theta }\left(
z\right) $ implies%
\begin{equation*}
\mathcal{L}_{i}^{Y}\left( f\left( \theta \right) ,\theta \right) =\triangle %
\left[ \mathcal{L}_{i}^{Z}\left( f\left( \theta \right) ,\theta \right) %
\right] \text{.}
\end{equation*}%
Given $\lambda \in MNE^{\left( \mathcal{M},\text{ }\theta \right) }$, we
have 
\begin{equation}
\left\{ f\left( \theta \right) \right\} \subset \dbigcup\limits_{m_{i}\in
M_{i}}\text{SUPP}\left[ g\left( m_{i},\lambda _{-i}\right) \right] \subset 
\mathcal{L}_{i}^{Z}\left( f\left( \theta \right) ,\theta \right) \text{.}
\label{tit1}
\end{equation}%
Furthermore, $\mathcal{L}_{i}^{Z}\left( f\left( \theta \right) ,\theta
\right) $ being an $i$-max set implies existence of $\theta ^{\prime }\in
\Theta $ such that%
\begin{eqnarray*}
\mathcal{L}_{i}^{Z}\left( f\left( \theta \right) ,\theta \right) &\subset
&\arg \max_{z\in \mathcal{L}_{i}^{Z}\left( f\left( \theta \right) ,\theta
\right) }u_{i}^{\theta ^{\prime }}\left( z\right) \text{,} \\
\mathcal{L}_{i}^{Z}\left( f\left( \theta \right) ,\theta \right) &\subset
&\arg \max_{z\in Z}u_{j}^{\theta ^{\prime }}\left( z\right) \text{, }\forall
j\in \mathcal{I}\diagdown \left\{ i\right\} \text{,}
\end{eqnarray*}%
which, together with (\ref{tit1}), further implies%
\begin{equation*}
\left( m_{i},\lambda _{-i}\right) \in MNE^{\left( \mathcal{M},\text{ }\theta
^{\prime }\right) }\text{, }\forall m_{i}\in M_{i}\text{.}
\end{equation*}%
Since $f$ is mixed-Nash-implemented by $\mathcal{M}$, we have%
\begin{equation*}
\left\{ f\left( \theta \right) \right\} \subset \dbigcup\limits_{m_{i}\in
M_{i}}\text{SUPP}\left[ g\left( m_{i},\lambda _{-i}\right) \right] \subset
\left\{ f\left( \theta ^{\prime }\right) \right\} \text{,}
\end{equation*}%
and as a result, $f\left( \theta \right) =f\left( \theta ^{\prime }\right) $
and $\dbigcup\limits_{m_{i}\in M_{i}}$SUPP$\left[ g\left( m_{i},\lambda
_{-i}\right) \right] =\left\{ f\left( \theta \right) \right\} $.$%
\blacksquare $

\subsection{Proof of Lemma \protect\ref{lem:no-veto:generalized}}

\label{sec:lem:no-veto:generalized}

Suppose that $\widehat{\mathcal{L}}^{Y}$-monotonicity holds. We first prove
that $Z$ is not an $i$-max set for any $i\in \mathcal{I}$. Suppose
otherwise, i.e., $Z$ is an $i$-max set for some $i\in \mathcal{I}$, or
equivalently, there exists $\theta ^{\prime }\in \Theta $ such that all
agents are indifferent between any two outcomes in $Z$, and hence,%
\begin{equation*}
\mathcal{L}_{j}^{Y}\left( z,\theta ^{\prime }\right) =Y\text{, }\forall
\left( j,z\right) \in \mathcal{I}\times Z\text{.}
\end{equation*}
As a result, for any $\theta \in \Theta $, we have%
\begin{equation*}
\widehat{\mathcal{L}}_{j}^{Y}\left( f\left( \theta \right) ,\theta \right)
\subset Y=\mathcal{L}_{j}^{Y}\left( f\left( \theta \right) ,\theta ^{\prime
}\right) \text{, }\forall j\in \mathcal{I}\text{,}
\end{equation*}%
which, together with $\widehat{\mathcal{L}}^{Y}$-monotonicity, implies $%
f\left( \theta \right) =f\left( \theta ^{\prime }\right) $ for any $\theta
\in \Theta $, or equivalently, $f\left( \Theta \right) =\left\{ f\left(
\theta ^{\prime }\right) \right\} $, contradicting Assumption \ref%
{assm:non-trivial}.

Second, fix any $\left( j,\theta ,\theta ^{\prime }\right) \in \mathcal{I}%
\times \Theta \times \Theta $ such that%
\begin{equation}
\left[ \widehat{\Gamma }_{j}\left( \theta \right) \text{ is a }j\text{-}%
\theta ^{\prime }\text{-max set}\right] \text{,}  \label{tit3}
\end{equation}%
and we aim to prove $\widehat{\Gamma }_{j}\left( \theta \right) =\left\{
f\left( \theta ^{\prime }\right) \right\} $. By the definition of $\widehat{%
\Gamma }_{j}\left( \theta \right) $ in (\ref{kkl2}) and the definition of $%
\widehat{\mathcal{L}}_{j}^{Y}\left( f\left( \theta \right) ,\theta \right) $%
\ in (\ref{yjj8}), we have

\begin{equation}
\widehat{\Gamma }_{j}\left( \theta \right) =\left\{ 
\begin{tabular}{ll}
$\left\{ f\left( \theta \right) \right\} \text{,}$ & if $f\left( \theta
\right) \in \arg \min_{z\in Z}u_{j}^{\theta }\left( z\right) $ and $\mathcal{%
L}_{j}^{Z}\left( f\left( \theta \right) ,\theta \right) $ is an $j$-max set,
\\ 
$\mathcal{L}_{j}^{Z}\left( f\left( \theta \right) ,\theta \right) $, & if $%
f\left( \theta \right) \in \arg \min_{z\in Z}u_{j}^{\theta }\left( z\right) $
and $\mathcal{L}_{j}^{Z}\left( f\left( \theta \right) ,\theta \right) $ is
not an $j$-max set, \\ 
$Z\text{,}$ & if $f\left( \theta \right) \notin \arg \min_{z\in
Z}u_{j}^{\theta }\left( z\right) $%
\end{tabular}%
\right. \text{.}  \label{tit2}
\end{equation}%
To see $\widehat{\Gamma }_{j}\left( \theta \right) =Z$ when $f\left( \theta
\right) \notin \arg \min_{z\in Z}u_{j}^{\theta }\left( z\right) $, pick any $%
z^{\prime }\in \arg \min_{z\in Z}u_{j}^{\theta }\left( z\right) $, and we
have $u_{j}^{\theta }\left( f\left( \theta \right) \right) >u_{j}^{\theta
}\left( z^{\prime }\right) $, which further implies%
\begin{equation*}
u_{j}^{\theta }\left( f\left( \theta \right) \right) >U_{j}^{\theta }\left[
\left( 1-\varepsilon \right) \times z^{\prime }+\varepsilon \times \text{UNIF%
}\left( Z\right) \right] \text{ for sufficiently small }\varepsilon >0\text{,%
}
\end{equation*}%
i.e., $\left[ \left( 1-\varepsilon \right) \times z^{\prime }+\varepsilon
\times \text{UNIF}\left( Z\right) \right] \in \mathcal{L}_{j}^{Y}\left(
f\left( \theta \right) ,\theta \right) $ and $\widehat{\Gamma }_{j}\left(
\theta \right) =Z$.

As proved above, $Z$ is not a $j$-max set. As a result, (\ref{tit3}) and (%
\ref{tit2}) imply that we must have $\widehat{\Gamma }_{j}\left( \theta
\right) =\left\{ f\left( \theta \right) \right\} $, which, together with (%
\ref{tit3}), further implies%
\begin{eqnarray}
\widehat{\mathcal{L}}_{i}^{Y}\left( f\left( \theta \right) ,\theta \right)
&\subset &Y=\mathcal{L}_{i}^{Y}\left( f\left( \theta \right) ,\theta
^{\prime }\right) \text{, }\forall i\in \mathcal{I}\diagdown \left\{
j\right\} \text{,}  \label{tit4} \\
\widehat{\mathcal{L}}_{j}^{Y}\left( f\left( \theta \right) ,\theta \right)
&=&\left\{ f\left( \theta \right) \right\} \subset \mathcal{L}_{j}^{Y}\left(
f\left( \theta \right) ,\theta ^{\prime }\right) \text{.}  \notag
\end{eqnarray}%
Finally, (\ref{tit4}) and $\widehat{\mathcal{L}}^{Y}$-monotonicity imply $%
f\left( \theta \right) =f\left( \theta ^{\prime }\right) $, i.e., $\widehat{%
\Gamma }_{j}\left( \theta \right) =\left\{ f\left( \theta \right) \right\}
=\left\{ f\left( \theta ^{\prime }\right) \right\} $.$\blacksquare $

\subsection{Proof of Lemma \protect\ref{lem:if}}

\label{sec:lem:if}

By the definition of $\widehat{\Gamma }_{j}\left( \theta \right) $ in (\ref%
{kkl2}) and the definition of $\widehat{\mathcal{L}}_{j}^{Y}\left( f\left(
\theta \right) ,\theta \right) $\ in (\ref{yjj8}), we have

\begin{equation*}
\widehat{\Gamma }_{j}\left( \theta \right) =\left\{ 
\begin{tabular}{ll}
$\left\{ f\left( \theta \right) \right\} \text{,}$ & if $f\left( \theta
\right) \in \arg \min_{z\in Z}u_{j}^{\theta }\left( z\right) $ and $\mathcal{%
L}_{j}^{Z}\left( f\left( \theta \right) ,\theta \right) $ is an $j$-max set,
\\ 
$\mathcal{L}_{j}^{Z}\left( f\left( \theta \right) ,\theta \right) $, & if $%
f\left( \theta \right) \in \arg \min_{z\in Z}u_{j}^{\theta }\left( z\right) $
and $\mathcal{L}_{j}^{Z}\left( f\left( \theta \right) ,\theta \right) $ is
not an $j$-max set, \\ 
$Z\text{,}$ & if $f\left( \theta \right) \notin \arg \min_{z\in
Z}u_{j}^{\theta }\left( z\right) $%
\end{tabular}%
\right. \text{.}
\end{equation*}%
Fix any $\left( \theta ,j\right) \in \Theta \times \mathcal{I}$, and we
consider three cases. First, suppose $f\left( \theta \right) \in \arg
\min_{z\in Z}u_{j}^{\theta }\left( z\right) $ and $\mathcal{L}_{j}^{Z}\left(
f\left( \theta \right) ,\theta \right) $ is an $j$-max set, i.e., $\widehat{%
\Gamma }_{j}\left( \theta \right) =\left\{ f\left( \theta \right) \right\} $%
. Thus, we can choose $\varepsilon _{j}^{\theta }=\frac{1}{2}$ and $%
y_{j}^{\theta }=f\left( \theta \right) $, and (\ref{yuy2}) holds.

Second, suppose $f\left( \theta \right) \in \arg \min_{z\in Z}u_{j}^{\theta
}\left( z\right) $ and $\mathcal{L}_{j}^{Z}\left( f\left( \theta \right)
,\theta \right) $ is not an $j$-max set, i.e., $\widehat{\Gamma }_{j}\left(
\theta \right) =\mathcal{L}_{j}^{Z}\left( f\left( \theta \right) ,\theta
\right) $ and%
\begin{equation*}
\widehat{\mathcal{L}}_{j}^{Y}\left( f\left( \theta \right) ,\theta \right) =%
\mathcal{L}_{j}^{Y}\left( f\left( \theta \right) ,\theta \right) =\triangle %
\left[ \mathcal{L}_{j}^{Z}\left( f\left( \theta \right) ,\theta \right) %
\right] =\triangle \left[ \widehat{\Gamma }_{j}\left( \theta \right) \right] 
\text{.}
\end{equation*}
Thus, we can choose $\varepsilon _{j}^{\theta }=\frac{1}{2}$ and $%
y_{j}^{\theta }=f\left( \theta \right) $, and (\ref{yuy2}) holds.

Third, suppose $f\left( \theta \right) \notin \arg \min_{z\in
Z}u_{j}^{\theta }\left( z\right) $, i.e., $\widehat{\Gamma }_{j}\left(
\theta \right) =Z$. By $f\left( \theta \right) \notin \arg \min_{z\in
Z}u_{j}^{\theta }\left( z\right) $, we can pick any $y_{j}^{\theta }\in \arg
\min_{z\in Z}u_{j}^{\theta }\left( z\right) \subset \mathcal{L}%
_{j}^{Y}\left( f\left( \theta \right) ,\theta \right) =\widehat{\mathcal{L}}%
_{j}^{Y}\left( f\left( \theta \right) ,\theta \right) $, and we have%
\begin{equation*}
U_{j}^{\theta }\left( y_{j}^{\theta }\right) <u_{j}^{\theta }\left( f\left(
\theta \right) \right) \text{.}
\end{equation*}%
Thus, there exists sufficiently small $\varepsilon _{j}^{\theta }>0$ such
that%
\begin{equation*}
U_{j}^{\theta }\left( \varepsilon _{j}^{\theta }\times y+\left(
1-\varepsilon _{j}^{\theta }\right) \times y_{j}^{\theta }\right)
<u_{j}^{\theta }\left( f\left( \theta \right) \right) \text{, }\forall y\in
Y=\triangle \left( Z\right) =\triangle \left( \widehat{\Gamma }_{j}\left(
\theta \right) \right) \text{,}
\end{equation*}%
and hence,%
\begin{equation*}
\left[ \varepsilon _{j}^{\theta }\times y+\left( 1-\varepsilon _{j}^{\theta
}\right) \times y_{j}^{\theta }\right] \in \mathcal{L}_{j}^{Y}\left( f\left(
\theta \right) ,\theta \right) =\widehat{\mathcal{L}}_{j}^{Y}\left( f\left(
\theta \right) ,\theta \right) \text{, }\forall y\in \triangle \left( 
\widehat{\Gamma }_{j}\left( \theta \right) \right) \text{,}
\end{equation*}%
i.e., (\ref{yuy2}) holds.$\blacksquare $

\subsection{Proof of Lemma \protect\ref{lem:mixed:canonical:pure}}

\label{sec:lem:mixed:canonical:pure}

For each $\left( \theta ,i\right) \in \Theta \times \mathcal{I}$, fix any%
\begin{equation*}
\widehat{b_{i}^{\theta }}\in \arg \max_{z\in Z}u_{i}^{\theta }\left(
z\right) \text{ and }\widehat{\gamma _{i}}\in \left( Z\right) ^{\left[
2^{Z}\diagdown \left\{ \varnothing \right\} \right] }\text{ such }\widehat{%
\gamma _{i}}\left( E\right) \in E\text{, }\forall E\in \left[ 2^{Z}\diagdown
\left\{ \varnothing \right\} \right] \text{,}
\end{equation*}%
i.e., $\widehat{b_{i}^{\theta }}$ is a top outcome for $i$ at $\theta $. We
need the following lemma to prove Lemma \ref{lem:mixed:canonical:pure}.

\begin{lemma}
\label{lem:mixed:canonical:pure:dominant}Consider the canonical mechanism $%
\mathcal{M}^{\ast }$ in Section \ref{sec:canonic}. For any $\left( \theta
,i\right) \in \Theta \times \mathcal{I}$, define%
\begin{equation*}
m_{i}^{n}\equiv \left( \theta ,k_{i}^{2}=n,k_{i}^{3}=n,\widehat{\gamma _{i}},%
\widehat{b_{i}^{\theta }}\right) \in M_{i}\text{, }\forall n\in \mathbb{N}%
\text{.}
\end{equation*}%
Then,%
\begin{equation*}
\lim_{n\rightarrow \infty }U_{i}^{\theta }\left[ g\left(
m_{i}^{n},m_{-i}\right) \right] \geq U_{i}^{\theta }\left[ g\left(
m_{i},m_{-i}\right) \right] \text{, }\forall \left( m_{i},m_{-i}\right) \in
M_{i}\times M_{-i}\text{.}
\end{equation*}
\end{lemma}

\noindent \textbf{Proof of Lemma \ref{lem:mixed:canonical:pure:dominant}:}
Fix any $\left( \theta ,i\right) \in \Theta \times \mathcal{I}$ and any $%
\left( m_{i},m_{-i}\right) \in M_{i}\times M_{-i}$. We consider two
scenarios: (A) there exists $\theta ^{\prime }\in \Theta $ such that 
\begin{equation*}
m_{j}=\left( \theta ^{\prime },k_{i}^{2}=1,\ast ,\ast ,\ast \right) \text{, }%
\forall j\in \mathcal{I}\diagdown \left\{ i\right\} \text{,}
\end{equation*}%
i.e., $\left( m_{i},m_{-i}\right) $ triggers either Case (1) or Case (2),
while $\left( m_{i}^{n},m_{-i}\right) $ triggers Case (2) if $n\geq 2$; and
(B) otherwise, i.e., $\left( m_{i}^{n},m_{-i}\right) $ triggers Case (3).

In Scenario (A), $g\left( m_{i},m_{-i}\right) \in \widehat{\mathcal{L}}%
_{i}^{Y}\left( f\left( \theta ^{\prime }\right) ,\theta ^{\prime }\right) $,
and $g\left( m_{i}^{n},m_{-i}\right) $ induces $\phi _{i}^{\theta ^{\prime
}}\left( \theta \right) $ with probability $\left( 1-\frac{1}{n}\right) $.
As a result,%
\begin{equation*}
\lim_{n\rightarrow \infty }U_{i}^{\theta }\left[ g\left(
m_{i}^{n},m_{-i}\right) \right] =U_{i}^{\theta }\left[ \phi _{i}^{\theta
^{\prime }}\left( \theta \right) \right] =\max_{y\in \widehat{\mathcal{L}}%
_{i}^{Y}\left( f\left( \theta ^{\prime }\right) ,\theta ^{\prime }\right)
}U_{i}^{\theta }\left[ y\right] \geq U_{i}^{\theta }\left[ g\left(
m_{i},m_{-i}\right) \right] \text{,}
\end{equation*}%
where the second equality follows from the definition of $\phi _{i}^{\theta
^{\prime }}\left( \theta \right) $ in (\ref{yuy1a}) and the inequality from $%
g\left( m_{i},m_{-i}\right) \in \widehat{\mathcal{L}}_{i}^{Y}\left( f\left(
\theta ^{\prime }\right) ,\theta ^{\prime }\right) $.

In Scenario (B), $\left( m_{i}^{n},m_{-i}\right) $ trigger Case (3), and
with sufficient large $n$, $\left( m_{i}^{n},m_{-i}\right) $ induces $%
\widehat{b_{i}^{\theta }}$ with probability $\left( 1-\frac{1}{n}\right) $.
As a result,%
\begin{equation*}
\lim_{n\rightarrow \infty }U_{i}^{\theta }\left[ g\left(
m_{i}^{n},m_{-i}\right) \right] =U_{i}^{\theta }\left[ \widehat{%
b_{i}^{\theta }}\right] =\max_{z\in Z}u_{i}^{\theta }\left[ z\right] \geq
U_{i}^{\theta }\left[ g\left( m_{i},m_{-i}\right) \right] \text{.}
\end{equation*}%
$\blacksquare $

\noindent \textbf{Proof of Lemma \ref{lem:mixed:canonical:pure}:} Fix any $%
\theta \in \Theta $, any $\lambda \in MNE^{\left( \mathcal{M}^{\ast },\text{ 
}\theta \right) }$ and any $\widehat{m}\in $SUPP$\left[ \lambda \right] $,
i.e.,%
\begin{equation}
\Pi _{i\in \mathcal{I}}\lambda _{i}\left( \widehat{m}_{i}\right) >0\text{.}
\label{tui2}
\end{equation}%
We aim to prove $\widehat{m}\in PNE^{\left( \mathcal{M}^{\ast },\text{ }%
\theta \right) }$. Suppose $\widehat{m}\notin PNE^{\left( \mathcal{M}^{\ast
},\text{ }\theta \right) }$, i.e., there exists $j\in \mathcal{I}$ and $%
m_{j}^{\prime }\in M_{i}$ such that%
\begin{equation*}
U_{j}^{\theta }\left[ g\left( m_{j}^{\prime },\widehat{m}_{-j}\right) \right]
>U_{j}^{\theta }\left[ g\left( \widehat{m}_{j},\widehat{m}_{-j}\right) %
\right] \text{,}
\end{equation*}%
which, together with Lemma \ref{lem:mixed:canonical:pure:dominant}, implies%
\begin{equation}
\lim_{n\rightarrow \infty }U_{j}^{\theta }\left[ g\left( m_{j}^{n},\widehat{m%
}_{-j}\right) \right] \geq U_{j}^{\theta }\left[ g\left( m_{j}^{\prime },%
\widehat{m}_{-j}\right) \right] >U_{j}^{\theta }\left[ g\left( \widehat{m}%
_{j},\widehat{m}_{-j}\right) \right] \text{.}  \label{tui}
\end{equation}%
We thus have%
\begin{eqnarray*}
&&\lim_{n\rightarrow \infty }U_{j}^{\theta }\left[ g\left( m_{j}^{n},\lambda
_{-j}\right) \right] -U_{j}^{\theta }\left[ g\left( \lambda _{j},\lambda
_{-j}\right) \right] \\
&=&\lim_{n\rightarrow \infty }\left( 
\begin{tabular}{l}
$\Sigma _{m\in M^{\ast }\diagdown \left\{ \widehat{m}\right\} }\left[ \Pi
_{i\in \mathcal{I}}\lambda _{i}\left( m_{i}\right) \times \left(
U_{j}^{\theta }\left[ g\left( m_{j}^{n},m_{-j}\right) \right] -U_{j}^{\theta
}\left[ g\left( m\right) \right] \right) \right] $ \\ 
$+\Pi _{i\in \mathcal{I}}\lambda _{i}\left( \widehat{m}_{i}\right) \times
\left( U_{j}^{\theta }\left[ g\left( m_{j}^{n},\widehat{m}_{-j}\right) %
\right] -U_{j}^{\theta }\left[ g\left( \widehat{m}_{j},\widehat{m}%
_{-j}\right) \right] \right) $%
\end{tabular}%
\right) \\
&\geq &0+\lim_{n\rightarrow \infty }\left[ 
\begin{tabular}{l}
$\Pi _{i\in \mathcal{I}}\lambda _{i}\left( \widehat{m}_{i}\right) \times
\left( U_{j}^{\theta }\left[ g\left( m_{j}^{n},\widehat{m}_{-j}\right) %
\right] -U_{j}^{\theta }\left[ g\left( \widehat{m}_{j},\widehat{m}%
_{-j}\right) \right] \right) $%
\end{tabular}%
\right] \\
&>&0\text{,}
\end{eqnarray*}%
where the first inequality follows from Lemma \ref%
{lem:mixed:canonical:pure:dominant}, and the second inequality from (\ref%
{tui2}) and (\ref{tui}). As a result, there exists $n\in \mathbb{N}$ such
that%
\begin{equation*}
U_{j}^{\theta }\left[ g\left( m_{j}^{n},\lambda _{-j}\right) \right]
>U_{j}^{\theta }\left[ g\left( \lambda _{j},\lambda _{-j}\right) \right] 
\text{,}
\end{equation*}%
contradicting $\lambda \in MNE^{\left( \mathcal{M}^{\ast },\text{ }\theta
\right) }$.$\blacksquare $

\subsection{Proof of Lemma \protect\ref{lem:ordinal}}

\label{sec:lem:ordinal}

For each $\left( i,z,\theta \right) \in \mathcal{I}\times Z\times \Theta
^{\ast }$, define%
\begin{eqnarray*}
SU_{i}^{\left( z,\theta \right) } &\equiv &\left\{ z^{\prime }\in
Z:z^{\prime }\succ _{i}^{\theta }z\right\} \text{,} \\
SL_{i}^{\left( z,\theta \right) } &\equiv &\left\{ z^{\prime }\in Z:z\succ
_{i}^{\theta }z^{\prime }\right\} \text{,} \\
ID_{i}^{\left( z,\theta \right) } &\equiv &\left\{ z^{\prime }\in Z:z\sim
_{i}^{\theta }z^{\prime }\right\} \text{.}
\end{eqnarray*}%
We need the following lemma to proceed.

\begin{lemma}
\label{lem:ordinal:bound}For any $\theta \in \Theta ^{\ast }$, any $%
u^{\theta }\in \Omega ^{\left[ \succeq ^{\theta },\text{ }%
\mathbb{R}
\right] }$, any $\left( i,z\right) \in \mathcal{I}\times Z$, consider%
\begin{equation*}
\rho _{\left( i,z\right) }\equiv \frac{u_{i}^{\theta }\left( z\right)
-\max_{z^{\prime }\in SL_{i}^{\left( z,\theta \right) }}u_{i}^{\theta
}\left( z^{\prime }\right) }{\max_{z^{\prime }\in SU_{i}^{\left( z,\theta
\right) }}u_{i}^{\theta }\left( z^{\prime }\right) -\max_{z^{\prime }\in
SL_{i}^{\left( z,\theta \right) }}u_{i}^{\theta }\left( z^{\prime }\right) }%
\text{ and }\rho ^{\left( i,z\right) }\equiv \frac{u_{i}^{\theta }\left(
z\right) -\min_{z^{\prime }\in SL_{i}^{\left( z,\theta \right)
}}u_{i}^{\theta }\left( z^{\prime }\right) }{\min_{z^{\prime }\in
SU_{i}^{\left( z,\theta \right) }}u_{i}^{\theta }\left( z^{\prime }\right)
-\min_{z^{\prime }\in SL_{i}^{\left( z,\theta \right) }}u_{i}^{\theta
}\left( z^{\prime }\right) }\text{.}
\end{equation*}%
Then, for any $z\in Z\diagdown \left[ \arg \max_{z^{\prime }\in
Z}u_{i}^{\theta }\left( z^{\prime }\right) \cup \arg \min_{z^{\prime }\in
Z}u_{i}^{\theta }\left( z^{\prime }\right) \right] $, we have 
\begin{equation}
\left\{ y\in \triangle \left[ Z\right] \diagdown \triangle \left[
ID_{i}^{\left( z,\theta \right) }\right] :\frac{\dsum\limits_{z^{\prime }\in
SU_{i}^{\left( z,\theta \right) }}y_{z^{\prime }}}{\dsum\limits_{z^{\prime
}\in SU_{i}^{\left( z,\theta \right) }\cup SL_{i}^{\left( z,\theta \right)
}}y_{z^{\prime }}}\leq \rho ^{\left( i,z\right) }\right\} \supset \mathcal{L}%
_{i}^{Y}\left( z,u^{\theta }\right) \diagdown \triangle \left[
ID_{i}^{\left( z,\theta \right) }\right] \text{,}  \label{tfp1}
\end{equation}%
\begin{equation}
\left\{ y\in \triangle \left[ Z\right] \diagdown \triangle \left[
ID_{i}^{\left( z,\theta \right) }\right] :\frac{\dsum\limits_{z^{\prime }\in
SU_{i}^{\left( z,\theta \right) }}y_{z^{\prime }}}{\dsum\limits_{z^{\prime
}\in SU_{i}^{\left( z,\theta \right) }\cup SL_{i}^{\left( z,\theta \right)
}}y_{z^{\prime }}}\leq \rho _{\left( i,z\right) }\right\} \subset \mathcal{L}%
_{i}^{Y}\left( z,u^{\theta }\right) \diagdown \triangle \left[
ID_{i}^{\left( z,\theta \right) }\right] \text{.}  \label{tfp2}
\end{equation}
\end{lemma}

\noindent \textbf{Proof of Lemma \ref{lem:ordinal:bound}:} Fix any $z\in
Z\diagdown \left[ \arg \max_{z^{\prime }\in Z}u_{i}^{\theta }\left(
z^{\prime }\right) \cup \arg \min_{z^{\prime }\in Z}u_{i}^{\theta }\left(
z^{\prime }\right) \right] $. For any $y\in \triangle \left[ Z\right] $, we
have%
\begin{gather}
\left( \dsum\limits_{z^{\prime }\in SU_{i}^{\left( z,\theta \right)
}}y_{z^{\prime }}\right) \times \min_{z^{\prime }\in SU_{i}^{\left( z,\theta
\right) }}u_{i}^{\theta }\left( z^{\prime }\right) +\left(
\dsum\limits_{z^{\prime }\in SL_{i}^{\left( z,\theta \right) }}y_{z^{\prime
}}\right) \times \min_{z^{\prime }\in SL_{i}^{\left( z,\theta \right)
}}u_{i}^{\theta }\left( z^{\prime }\right)  \label{thh1} \\
\leq \dsum\limits_{z^{\prime }\in SU_{i}^{\left( z,\theta \right) }}\left[
y_{z^{\prime }}\times u_{i}^{\theta }\left( z^{\prime }\right) \right]
+\dsum\limits_{z^{\prime }\in SL_{i}^{\left( z,\theta \right) }}\left[
y_{z^{\prime }}\times u_{i}^{\theta }\left( z^{\prime }\right) \right] \text{%
.}  \notag
\end{gather}%
Pick any $y\in \mathcal{L}_{i}^{Y}\left( z,u^{\theta }\right) \diagdown
\triangle \left[ ID_{i}^{\left( z,\theta \right) }\right] $, we have%
\begin{equation}
\dsum\limits_{z^{\prime }\in SU_{i}^{\left( z,\theta \right) }}\left[
y_{z^{\prime }}\times u_{i}^{\theta }\left( z^{\prime }\right) \right]
+\dsum\limits_{z^{\prime }\in SL_{i}^{\left( z,\theta \right) }}\left[
y_{z^{\prime }}\times u_{i}^{\theta }\left( z^{\prime }\right) \right] \leq
\left( \dsum\limits_{z^{\prime }\in SU_{i}^{\left( z,\theta \right) }\cup
SL_{i}^{\left( z,\theta \right) }}y_{z^{\prime }}\right) \times
u_{i}^{\theta }\left( z\right) \text{.}  \label{thh3}
\end{equation}%
(\ref{thh1}) and (\ref{thh3}) imply%
\begin{gather*}
\left( \dsum\limits_{z^{\prime }\in SU_{i}^{\left( z,\theta \right)
}}y_{z^{\prime }}\right) \times \min_{z^{\prime }\in SU_{i}^{\left( z,\theta
\right) }}u_{i}^{\theta }\left( z^{\prime }\right) +\left(
\dsum\limits_{z^{\prime }\in SL_{i}^{\left( z,\theta \right) }}y_{z^{\prime
}}\right) \times \min_{z^{\prime }\in SL_{i}^{\left( z,\theta \right)
}}u_{i}^{\theta }\left( z^{\prime }\right) \leq \left(
\dsum\limits_{z^{\prime }\in SU_{i}^{\left( z,\theta \right) }\cup
SL_{i}^{\left( z,\theta \right) }}y_{z^{\prime }}\right) \times
u_{i}^{\theta }\left( z\right) \text{,} \\
\text{or equivalently, }\frac{\dsum\limits_{z\in SU_{i}^{\left( z,\theta
\right) }}y_{z}}{\dsum\limits_{z\in SU_{i}^{\left( z,\theta \right) }\cup
SL_{i}^{\left( z,\theta \right) }}y_{z}}\leq \frac{u_{i}^{\theta }\left(
z\right) -\min_{z^{\prime }\in SL_{i}^{\left( z,\theta \right)
}}u_{i}^{\theta }\left( z^{\prime }\right) }{\min_{z^{\prime }\in
SU_{i}^{\left( z,\theta \right) }}u_{i}^{\theta }\left( z^{\prime }\right)
-\min_{z^{\prime }\in SL_{i}^{\left( z,\theta \right) }}u_{i}^{\theta
}\left( z^{\prime }\right) }=\rho ^{\left( i,z\right) }\text{,}
\end{gather*}%
i.e., (\ref{tfp1}) holds.

Furthermore, for any $y\in \triangle \left[ Z\right] $, we have%
\begin{gather}
\dsum\limits_{z^{\prime }\in SU_{i}^{\left( z,\theta \right) }}\left[
y_{z^{\prime }}\times u_{i}^{\theta }\left( z^{\prime }\right) \right]
+\dsum\limits_{z^{\prime }\in SL_{i}^{\left( z,\theta \right) }}\left[
y_{z^{\prime }}\times u_{i}^{\theta }\left( z^{\prime }\right) \right] \text{%
,}  \label{tfp3} \\
\leq \dsum\limits_{z^{\prime }\in SU_{i}^{\left( z,\theta \right)
}}y_{z^{\prime }}\times \max_{z^{\prime }\in SU_{i}^{\left( z,\theta \right)
}}u_{i}^{\theta }\left( z^{\prime }\right) +\dsum\limits_{z^{\prime }\in
SL_{i}^{\left( z,\theta \right) }}y_{z^{\prime }}\times \max_{z^{\prime }\in
SL_{i}^{\left( z,\theta \right) }}u_{i}^{\theta }\left( z^{\prime }\right) 
\text{.}  \notag
\end{gather}%
Pick any $y\in \left\{ \overline{y}\in \triangle \left[ Z\right] \diagdown
\triangle \left[ ID_{i}^{\left( z,\theta \right) }\right] :\frac{%
\dsum\limits_{z^{\prime }\in SU_{i}^{\left( z,\theta \right) }}\overline{y}%
_{z^{\prime }}}{\dsum\limits_{z^{\prime }\in SU_{i}^{\left( z,\theta \right)
}\cup SL_{i}^{\left( z,\theta \right) }}\overline{y}_{z^{\prime }}}\leq \rho
_{\left( i,z\right) }\right\} $, we have%
\begin{gather}
\frac{\dsum\limits_{z^{\prime }\in SU_{i}^{\left( z,\theta \right)
}}y_{z^{\prime }}}{\dsum\limits_{z^{\prime }\in SU_{i}^{\left( z,\theta
\right) }\cup SL_{i}^{\left( z,\theta \right) }}y_{z^{\prime }}}\leq \rho
_{\left( i,z\right) }=\frac{u_{i}^{\theta }\left( z\right) -\max_{z^{\prime
}\in SL_{i}^{\left( z,\theta \right) }}u_{i}^{\theta }\left( z^{\prime
}\right) }{\max_{z^{\prime }\in SU_{i}^{\left( z,\theta \right)
}}u_{i}^{\theta }\left( z^{\prime }\right) -\max_{z^{\prime }\in
SL_{i}^{\left( z,\theta \right) }}u_{i}^{\theta }\left( z^{\prime }\right) }%
\text{,}  \notag \\
\text{or equivalently,}  \notag \\
\left( \dsum\limits_{z\in SU_{i}^{\left( z,\theta \right) }}y_{z}\right)
\times \max_{z^{\prime }\in SU_{i}^{\left( z,\theta \right) }}u_{i}^{\theta
}\left( z^{\prime }\right) +\left( \dsum\limits_{z^{\prime }\in
SL_{i}^{\left( z,\theta \right) }}y_{z^{\prime }}\right) \times
\max_{z^{\prime }\in SL_{i}^{\left( z,\theta \right) }}u_{i}^{\theta }\left(
z^{\prime }\right) \leq \left( \dsum\limits_{z^{\prime }\in SU_{i}^{\left(
z,\theta \right) }\cup SL_{i}^{\left( z,\theta \right) }}y_{z^{\prime
}}\right) \times u_{i}^{\theta }\left( z\right) \text{,}  \label{tfp5}
\end{gather}%
and (\ref{tfp3}) and (\ref{tfp5}) imply%
\begin{equation*}
\dsum\limits_{z^{\prime }\in SU_{i}^{\left( z,\theta \right) }}\left[
y_{z^{\prime }}\times u_{i}^{\theta }\left( z^{\prime }\right) \right]
+\dsum\limits_{z^{\prime }\in SL_{i}^{\left( z,\theta \right) }}\left[
y_{z^{\prime }}\times u_{i}^{\theta }\left( z^{\prime }\right) \right] \leq
\left( \dsum\limits_{z^{\prime }\in SU_{i}^{\left( z,\theta \right) }\cup
SL_{i}^{\left( z,\theta \right) }}y_{z^{\prime }}\right) \times
u_{i}^{\theta }\left( z\right) \text{,}
\end{equation*}%
and as a result, , $y\in \mathcal{L}_{i}^{Y}\left( z,u^{\theta }\right)
\diagdown \triangle \left[ ID_{i}^{\left( z,\theta \right) }\right] $ i.e., (%
\ref{tfp2}) holds.$\blacksquare $

\noindent \textbf{Proof of Lemma \ref{lem:ordinal}:} Fix any $\theta \in
\Theta ^{\ast }$ and pick any $\overline{u^{\theta }}\in \Omega ^{\left[
\succeq ^{\theta },\text{ }%
\mathbb{R}
\right] }$. For each $i\in \mathcal{I}$, define $r_{i}^{\theta
}:Z\longrightarrow 
\mathbb{R}
$ as follows.%
\begin{equation*}
r_{i}^{\theta }\left( z\right) \equiv \left\vert \left\{ \overline{u^{\theta
}}_{i}\left( z^{\prime }\right) :z^{\prime }\in SL_{i}^{z}\right\}
\right\vert +1
\end{equation*}%
That is,%
\begin{equation*}
\left( 
\begin{array}{c}
r_{i}^{\theta }\left( z\right) =1\text{ if }\overline{u^{\theta }}_{i}\left(
z\right) \text{ achieves the lowest value in }\overline{u^{\theta }}%
_{i}\left( Z\right) \text{;} \\ 
r_{i}^{\theta }\left( z\right) =2\text{ if }\overline{u^{\theta }}_{i}\left(
z\right) \text{ achieves the second-lowest value in }\overline{u^{\theta }}%
_{i}\left( Z\right) \text{;} \\ 
...%
\end{array}%
\right)
\end{equation*}%
Clearly, $\left( r_{i}^{\theta }\right) _{i\in \mathcal{I}}\in \Omega ^{%
\left[ \succeq ^{\theta },\text{ }\mathbb{Q}\right] }$.

We now define $\widehat{u^{\theta }}\in \Omega ^{\left[ \succeq ^{\theta },%
\text{ }\mathbb{Q}\right] }$ such that (\ref{ggi6})\ holds.%
\begin{equation*}
\widehat{u^{\theta }}_{i}\left( z\right) \equiv r_{i}^{\theta }\left(
z\right) \times 10^{r_{i}^{\theta }\left( z\right) \times n}\text{,}
\end{equation*}%
where $n$ is a positive integer which will be determined later.

We first consier $z\in Z\diagdown \left[ \arg \max_{z^{\prime }\in
Z}u_{i}^{\theta }\left( z^{\prime }\right) \cup \arg \min_{z^{\prime }\in
Z}u_{i}^{\theta }\left( z^{\prime }\right) \right] $. As $n\rightarrow
\infty $, we have%
\begin{equation*}
\frac{\widehat{u^{\theta }}_{i}\left( z\right) -\min_{z^{\prime }\in
SL_{i}^{\left( z,\theta \right) }}\widehat{u^{\theta }}_{i}\left( z^{\prime
}\right) }{\min_{z^{\prime }\in SU_{i}^{\left( z,\theta \right) }}\widehat{%
u^{\theta }}_{i}\left( z^{\prime }\right) -\min_{z^{\prime }\in
SL_{i}^{\left( z,\theta \right) }}\widehat{u^{\theta }}_{i}\left( z^{\prime
}\right) }\rightarrow 0\text{, }\forall z\in Z\diagdown \left( \arg
\max_{z^{\prime }\in Z}u_{i}^{\theta }\left( z^{\prime }\right) \cup \arg
\min_{z^{\prime }\in Z}u_{i}^{\theta }\left( z^{\prime }\right) \right) 
\text{.}
\end{equation*}%
We thus fix any positive $n$ such that%
\begin{gather*}
\frac{\widehat{u^{\theta }}_{i}\left( z\right) -\min_{z^{\prime }\in
SL_{i}^{\left( z,\theta \right) }}\widehat{u^{\theta }}_{i}\left( z^{\prime
}\right) }{\min_{z^{\prime }\in SU_{i}^{\left( z,\theta \right) }}\widehat{%
u^{\theta }}_{i}\left( z^{\prime }\right) -\min_{z^{\prime }\in
SL_{i}^{\left( z,\theta \right) }}\widehat{u^{\theta }}_{i}\left( z^{\prime
}\right) }<\frac{\overline{u^{\theta }}_{i}\left( z\right) -\max_{z^{\prime
}\in SL_{i}^{\left( z,\theta \right) }}\overline{u^{\theta }}_{i}\left(
z^{\prime }\right) }{\max_{z^{\prime }\in SU_{i}^{\left( z,\theta \right) }}%
\overline{u^{\theta }}_{i}\left( z^{\prime }\right) -\max_{z^{\prime }\in
SL_{i}^{\left( z,\theta \right) }}\overline{u^{\theta }}_{i}\left( z^{\prime
}\right) }\text{,} \\
\forall z\in Z\diagdown \left[ \arg \max_{z^{\prime }\in Z}u_{i}^{\theta
}\left( z^{\prime }\right) \cup \arg \min_{z^{\prime }\in Z}u_{i}^{\theta
}\left( z^{\prime }\right) \right] \text{.}
\end{gather*}%
which, together with Lemma \ref{lem:ordinal:bound}, implies%
\begin{equation*}
\mathcal{L}_{i}^{Y}\left( z,\widehat{u^{\theta }}\right) \diagdown \triangle %
\left[ ID_{i}^{z}\right] \subset \mathcal{L}_{i}^{Y}\left( z,\overline{%
u^{\theta }}\right) \diagdown \triangle \left[ ID_{i}^{z}\right] \text{.}
\end{equation*}%
and as a result,%
\begin{equation}
\mathcal{L}_{i}^{Y}\left( z,\widehat{u^{\theta }}\right) =\left[ \mathcal{L}%
_{i}^{Y}\left( z,\widehat{u^{\theta }}\right) \diagdown \triangle \left[
ID_{i}^{z}\right] \right] \cup \triangle \left( ID_{i}^{z}\right) \subset %
\left[ \mathcal{L}_{i}^{Y}\left( z,\overline{u^{\theta }}\right) \diagdown
\triangle \left[ ID_{i}^{z}\right] \right] \cup \triangle \left(
ID_{i}^{z}\right) =\mathcal{L}_{i}^{Y}\left( z,\overline{u^{\theta }}\right) 
\text{.}  \label{tfp6}
\end{equation}%
Second, consider $z\in \left[ \arg \max_{z^{\prime }\in Z}u_{i}^{\theta
}\left( z^{\prime }\right) \cup \arg \min_{z^{\prime }\in Z}u_{i}^{\theta
}\left( z^{\prime }\right) \right] $, and we have%
\begin{eqnarray}
\mathcal{L}_{i}^{Y}\left( z,\widehat{u^{\theta }}\right) &=&Y=\mathcal{L}%
_{i}^{Y}\left( z,\overline{u^{\theta }}\right) \text{, }\forall z\in \arg
\max_{z^{\prime }\in Z}u_{i}^{\theta }\left( z^{\prime }\right) \text{,}
\label{tfp7} \\
\mathcal{L}_{i}^{Y}\left( z,\widehat{u^{\theta }}\right) &=&\triangle \left[
ID_{i}^{\left( z,\theta \right) }\right] =\mathcal{L}_{i}^{Y}\left( z,%
\overline{u^{\theta }}\right) \text{, }\forall z\in \arg \min_{z^{\prime
}\in Z}u_{i}^{\theta }\left( z^{\prime }\right) \text{.}  \label{tfp8}
\end{eqnarray}%
(\ref{tfp6}), (\ref{tfp7}) and (\ref{tfp8}) prove (\ref{ggi6}) for $\widehat{%
u^{\theta }}$.

We now define $\widetilde{u^{\theta }}\in \Omega ^{\left[ \succeq ^{\theta },%
\text{ }\mathbb{Q}\right] }$ such that (\ref{ggi6})\ holds:%
\begin{equation*}
\widetilde{u^{\theta }}_{i}\left( z\right) \equiv -\frac{1}{\widehat{%
u^{\theta }}_{i}\left( z\right) }\text{,}
\end{equation*}%
where $n$ is a positive integer which will be determined later. As $%
n\rightarrow \infty $, we have%
\begin{equation*}
\frac{\widetilde{u^{\theta }}_{i}\left( z\right) -\max_{z^{\prime }\in
SL_{i}^{\left( z,\theta \right) }}\widetilde{u^{\theta }}_{i}\left(
z^{\prime }\right) }{\max_{z^{\prime }\in SU_{i}^{\left( z,\theta \right) }}%
\widetilde{u^{\theta }}_{i}\left( z^{\prime }\right) -\max_{z^{\prime }\in
SL_{i}^{\left( z,\theta \right) }}\widetilde{u^{\theta }}_{i}\left(
z^{\prime }\right) }\rightarrow 1\text{, }\forall z\in Z\diagdown \left[
\arg \max_{z^{\prime }\in Z}u_{i}^{\theta }\left( z^{\prime }\right) \cup
\arg \min_{z^{\prime }\in Z}u_{i}^{\theta }\left( z^{\prime }\right) \right] 
\text{.}
\end{equation*}%
We thus fix any positive $n$ such that%
\begin{gather*}
\frac{\overline{u^{\theta }}_{i}\left( z\right) -\min_{z^{\prime }\in
SL_{i}^{\left( z,\theta \right) }}\overline{u^{\theta }}_{i}\left( z^{\prime
}\right) }{\min_{z^{\prime }\in SU_{i}^{\left( z,\theta \right) }}\overline{%
u^{\theta }}_{i}\left( z^{\prime }\right) -\min_{z^{\prime }\in
SL_{i}^{\left( z,\theta \right) }}\overline{u^{\theta }}_{i}\left( z^{\prime
}\right) }<\frac{\widetilde{u^{\theta }}_{i}\left( z\right) -\max_{z^{\prime
}\in SL_{i}^{\left( z,\theta \right) }}\widetilde{u^{\theta }}_{i}\left(
z^{\prime }\right) }{\max_{z^{\prime }\in SU_{i}^{\left( z,\theta \right) }}%
\widetilde{u^{\theta }}_{i}\left( z^{\prime }\right) -\max_{z^{\prime }\in
SL_{i}^{\left( z,\theta \right) }}\widetilde{u^{\theta }}_{i}\left(
z^{\prime }\right) }\text{,} \\
\forall z\in Z\diagdown \left[ \arg \max_{z^{\prime }\in Z}u_{i}^{\theta
}\left( z^{\prime }\right) \cup \arg \min_{z^{\prime }\in Z}u_{i}^{\theta
}\left( z^{\prime }\right) \right] \text{.}
\end{gather*}%
As a above, Lemma \ref{lem:ordinal:bound}, implies%
\begin{equation*}
\mathcal{L}_{i}^{Y}\left( z,\overline{u^{\theta }}\right) \subset \mathcal{L}%
_{i}^{Y}\left( z,\widetilde{u^{\theta }}\right) \text{, }\forall z\in
Z\diagdown \left[ \arg \max_{z^{\prime }\in Z}u_{i}^{\theta }\left(
z^{\prime }\right) \cup \arg \min_{z^{\prime }\in Z}u_{i}^{\theta }\left(
z^{\prime }\right) \right] \text{,}
\end{equation*}%
and%
\begin{eqnarray*}
\mathcal{L}_{i}^{Y}\left( z,\overline{u^{\theta }}\right) &=&Y=\mathcal{L}%
_{i}^{Y}\left( z,\widetilde{u^{\theta }}\right) \text{, }\forall z\in \arg
\max_{z^{\prime }\in Z}\overline{u^{\theta }}_{i}\left( z^{\prime }\right) 
\text{,} \\
\mathcal{L}_{i}^{Y}\left( z,\overline{u^{\theta }}\right) &=&\triangle \left[
ID_{i}^{\left( z,\theta \right) }\right] =\mathcal{L}_{i}^{Y}\left( z,%
\widetilde{u^{\theta }}\right) \text{, }\forall z\in \arg \min_{z^{\prime
}\in Z}\overline{u^{\theta }}_{i}\left( z^{\prime }\right) \text{.}
\end{eqnarray*}%
i.e., (\ref{ggi6}) holds for $\widetilde{u^{\theta }}$.$\blacksquare $

\subsection{Proof of Lemma \protect\ref{lem:Z-A}}

\label{sec:lem:Z-A}

Suppose that an SCC $F$ is mixed-Nash-A-implemented by $\mathcal{M}%
=\left\langle M\text{, \ }g:M\longrightarrow Y\right\rangle $. We consider
two scenarios: (I) $Z$ is not an $i$-max set for any $i\in \mathcal{I}$ and
(II) $Z$ is an $i$-max set for some $i\in \mathcal{I}$. In scenario (I), we
have $Z^{\ast }=Z$ by (\ref{tth2}), and hence $g\left( M\right) \subset
\triangle \left( Z\right) =\triangle \left( Z^{\ast }\right) $. In scenario
(II), we have $Z^{\ast }=\cup _{\theta \in \Theta }F\left( \theta \right) $
by (\ref{tth2}). $Z$ being an $i$-max set implies existence of some state $%
\theta ^{\prime }\in \Theta $ such that all agents are indifferent between
any two elements in $Z$ at $\theta ^{\prime }$. Therefore, every $m\in M$ is
a Nash equilibrium at $\theta ^{\prime }$, and hence%
\begin{equation*}
\dbigcup\limits_{m\in M}\text{SUPP}\left[ g\left( m\right) \right] \subset
F\left( \theta ^{\prime }\right) \subset \cup _{\theta \in \Theta }F\left(
\theta \right) =Z^{\ast }\text{,}
\end{equation*}%
i.e., $g\left( M\right) \subset \triangle \left( Z^{\ast }\right) $.$%
\blacksquare $

\subsection{Proof of Lemma \protect\ref{lem:mixed:deviation:SCC}}

\label{sec:lem:mixed:deviation:SCC}

Suppose that an SCC $F$ is mixed-Nash-A-implemented by $\mathcal{M}%
=\left\langle M\text{, \ }g:M\longrightarrow Y\right\rangle $. Fix any $%
\left( i,\theta \right) \in \mathcal{I}\times \Theta $ and any $\lambda \in
MNE^{\left( \mathcal{M},\text{ }\theta \right) }$ such that%
\begin{gather}
F\left( \theta \right) \subset \arg \min_{z\in Z^{\ast }}u_{i}^{\theta
}\left( z\right) \text{,}  \label{ddf1} \\
\Xi _{i}\left( \theta \right) \neq \varnothing \text{,}  \notag \\
\text{and }Z^{\ast }\cap \mathcal{L}_{i}^{Z}\left( F\left( \theta \right)
,\theta \right) \text{ is an }i\text{-}Z^{\ast }\text{-max set.}
\label{ddfa1}
\end{gather}

First, with $Z^{\ast }\cap \mathcal{L}_{i}^{Z}\left( F\left( \theta \right)
,\theta \right) $ being an $i$-$Z^{\ast }$-max set, there exists $\widetilde{%
\theta }\in \Theta $ such that $Z^{\ast }\cap \mathcal{L}_{i}^{Z}\left(
F\left( \theta \right) ,\theta \right) $ is an $i$-$Z^{\ast }$-$\widetilde{%
\theta }$-max set, i.e.,%
\begin{eqnarray}
F\left( \theta \right) &\subset &Z^{\ast }\cap \mathcal{L}_{i}^{Z}\left(
F\left( \theta \right) ,\theta \right) \subset \arg \max_{z\in Z^{\ast
}}u_{j}^{\widetilde{\theta }}\left( z\right) \text{, }\forall j\in \mathcal{I%
}\diagdown \left\{ i\right\} \text{,}  \label{gge1} \\
F\left( \theta \right) &\subset &Z^{\ast }\cap \mathcal{L}_{i}^{Z}\left(
F\left( \theta \right) ,\theta \right) \subset \arg \max_{z\in Z^{\ast }\cap 
\mathcal{L}_{i}^{Z}\left( F\left( \theta \right) ,\theta \right) }u_{i}^{%
\widetilde{\theta }}\left( z\right) \text{.}  \label{gge2}
\end{eqnarray}%
(\ref{ddf1}) and Lemma \ref{lem:Z-A} imply%
\begin{equation*}
g\left( m_{i},\lambda _{-i}\right) \in \triangle \left[ Z^{\ast }\cap 
\mathcal{L}_{i}^{Z}\left( F\left( \theta \right) ,\theta \right) \right] 
\text{, }\forall m_{i}\in M_{i}\text{, }\forall \left( \lambda _{i},\lambda
_{-i}\right) \in MNE^{\left( \mathcal{M},\text{ }\theta \right) }\text{,}
\end{equation*}%
which, together with (\ref{gge1}) and (\ref{gge2}), immediately implies $%
MNE^{\left( \mathcal{M},\text{ }\theta \right) }\subset MNE^{\left( \mathcal{%
M},\text{ }\widetilde{\theta }\right) }$, and hence, $F\left( \theta \right)
\subset F\left( \widetilde{\theta }\right) $. Therefore, we have%
\begin{equation}
\left( 
\begin{array}{c}
\widetilde{\theta }\in \Theta _{i}^{\theta }\neq \varnothing \text{,} \\ 
\text{and }Z^{\ast }\cap \mathcal{L}_{i}^{Z}\left( F\left( \theta \right)
,\theta \right) \text{ is an }i\text{-}Z^{\ast }\text{-}\widetilde{\theta }%
\text{-max set}%
\end{array}%
\right) \text{.}  \label{hhty2}
\end{equation}

Second, we define an$\text{ algorithm.}$%
\begin{equation}
\left( 
\begin{array}{c}
\text{Step }1\text{: let }K^{1}\equiv \left\{ \theta ^{\prime }\in \Theta
_{i}^{\theta }:Z^{\ast }\cap \mathcal{L}_{i}^{Z}\left( F\left( \theta
\right) ,\theta \right) \text{ is an }i\text{-}Z^{\ast }\text{-}\theta
^{\prime }\text{-max set}\right\} \text{,} \\ 
\text{Step }2\text{: let }K^{2}\equiv \left\{ \theta ^{\prime }\in \Theta
_{i}^{\theta }:Z^{\ast }\cap \mathcal{L}_{i}^{Z}\left( F\left( \theta
\right) ,\theta \right) \cap \left[ \dbigcap\limits_{\theta ^{\prime }\in
K^{1}}F\left( \theta ^{\prime }\right) \right] \text{ is an }i\text{-}%
Z^{\ast }\text{-}\theta ^{\prime }\text{-max set}\right\} \text{,} \\ 
... \\ 
\text{Step }n\text{: let }K^{n}\equiv \left\{ \theta ^{\prime }\in \Theta
_{i}^{\theta }:Z^{\ast }\cap \mathcal{L}_{i}^{Z}\left( F\left( \theta
\right) ,\theta \right) \cap \left[ \dbigcap\limits_{\theta ^{\prime }\in
K^{n-1}}F\left( \theta ^{\prime }\right) \right] \text{ is an }i\text{-}%
Z^{\ast }\text{-}\theta ^{\prime }\text{-max set}\right\} \text{,} \\ 
...%
\end{array}%
\right)  \label{tpt3}
\end{equation}%
(\ref{hhty2}) implies $\widetilde{\theta }\in K^{1}\neq \varnothing $.\
Furthermore, we have $K^{1}\subset K^{2}$, and inductively, it is easy to
show%
\begin{eqnarray}
\forall n &\geq &1\text{,}  \notag \\
K^{n} &\subset &K^{n+1}\text{ and}  \notag \\
Z^{\ast }\cap \mathcal{L}_{i}^{Z}\left( F\left( \theta \right) ,\theta
\right) \cap \left[ \dbigcap\limits_{\theta ^{\prime }\in K^{n}}F\left(
\theta ^{\prime }\right) \right] &\supset &Z^{\ast }\cap \mathcal{L}%
_{i}^{Z}\left( F\left( \theta \right) ,\theta \right) \cap \left[
\dbigcap\limits_{\theta ^{\prime }\in K^{n+1}}F\left( \theta ^{\prime
}\right) \right] \text{.}  \label{tpt2}
\end{eqnarray}%
$\text{Since }Z$ is finite, (\ref{tpt2}) implies that there exists $n\geq 1$
such that%
\begin{equation}
Z^{\ast }\cap \mathcal{L}_{i}^{Z}\left( F\left( \theta \right) ,\theta
\right) \cap \left[ \dbigcap\limits_{\theta ^{\prime }\in K^{n}}F\left(
\theta ^{\prime }\right) \right] =Z^{\ast }\cap \mathcal{L}_{i}^{Z}\left(
F\left( \theta \right) ,\theta \right) \cap \left[ \dbigcap\limits_{\theta
^{\prime }\in K^{n+1}}F\left( \theta ^{\prime }\right) \right] \text{.}
\label{tpt1}
\end{equation}%
and as a result,%
\begin{eqnarray*}
K^{n+1} &\equiv &\left\{ \theta ^{\prime }\in \Theta _{i}^{\theta }:Z^{\ast
}\cap \mathcal{L}_{i}^{Z}\left( F\left( \theta \right) ,\theta \right) \cap %
\left[ \dbigcap\limits_{\theta ^{\prime }\in K^{n}}F\left( \theta ^{\prime
}\right) \right] \text{ is an }i\text{-}Z^{\ast }\text{-}\theta ^{\prime }%
\text{-max set}\right\} \\
&=&\left\{ \theta ^{\prime }\in \Theta _{i}^{\theta }:Z^{\ast }\cap \mathcal{%
L}_{i}^{Z}\left( F\left( \theta \right) ,\theta \right) \cap \left[
\dbigcap\limits_{\theta ^{\prime }\in K^{n+1}}F\left( \theta ^{\prime
}\right) \right] \text{ is an }i\text{-}Z^{\ast }\text{-}\theta ^{\prime }%
\text{-max set}\right\} \text{,}
\end{eqnarray*}%
where the second inequality follows from (\ref{tpt1}). Therefore, $%
\varnothing \neq K^{1}\subset K^{n+1}$ and%
\begin{equation}
K^{n+1}\in \Xi _{i}\left( \theta \right) \neq \varnothing \text{.}
\label{tptt1a}
\end{equation}

Third, we aim to prove%
\begin{equation}
\dbigcup\limits_{m_{i}\in M_{i}}\text{SUPP}\left[ g\left( m_{i},\lambda
_{-i}\right) \right] \subset \left[ Z^{\ast }\cap \mathcal{L}_{i}^{Z}\left(
F\left( \theta \right) ,\theta \right) \cap \left( \dbigcup\limits_{K\in \Xi
_{i}\left( \theta \right) }\dbigcap\limits_{\theta ^{\prime }\in K}F\left(
\theta ^{\prime }\right) \right) \right] \text{.}  \label{tpt7}
\end{equation}%
(\ref{tptt1a}) and (\ref{tpt7}) implies that it suffices to prove%
\begin{equation}
\dbigcup\limits_{m_{i}\in M_{i}}\text{SUPP}\left[ g\left( m_{i},\lambda
_{-i}\right) \right] \subset \left[ Z^{\ast }\cap \mathcal{L}_{i}^{Z}\left(
F\left( \theta \right) ,\theta \right) \cap \left( \dbigcap\limits_{\theta
^{\prime }\in K^{n+1}}F\left( \theta ^{\prime }\right) \right) \right] \text{%
.}  \label{tptt1}
\end{equation}%
We prove (\ref{tptt1}) inductively. First, 
\begin{equation*}
K^{1}\equiv \left\{ \theta ^{\prime }\in \Theta _{i}^{\theta }:Z^{\ast }\cap 
\mathcal{L}_{i}^{Z}\left( F\left( \theta \right) ,\theta \right) \text{ is
an }i\text{-}Z^{\ast }\text{-}\theta ^{\prime }\text{-max set}\right\} \text{%
,}
\end{equation*}%
and as a result,%
\begin{equation*}
g\left( m_{i},\lambda _{-i}\right) \in MNE^{\left( \mathcal{M},\text{ }%
\theta ^{\prime }\right) }\text{, }\forall m_{i}\in M_{i}\text{, }\forall
\theta ^{\prime }\in K^{1}\text{,}
\end{equation*}%
and hence,%
\begin{equation*}
g\left( m_{i},\lambda _{-i}\right) \in \triangle \left[ F\left( \theta
^{\prime }\right) \right] \text{, }\forall m_{i}\in M_{i}\text{, }\forall
\theta ^{\prime }\in K^{1}\text{,}
\end{equation*}%
which, together with Lemma \ref{lem:Z-A}, implies%
\begin{equation*}
\dbigcup\limits_{m_{i}\in M_{i}}\text{SUPP}\left[ g\left( m_{i},\lambda
_{-i}\right) \right] \subset \left[ Z^{\ast }\cap \mathcal{L}_{i}^{Z}\left(
F\left( \theta \right) ,\theta \right) \cap \left( \dbigcap\limits_{\theta
^{\prime }\in K^{1}}F\left( \theta ^{\prime }\right) \right) \right] \text{.}
\end{equation*}%
This completes the first step of the induction.

Suppose that for some $t<k+1$, we have proved%
\begin{equation*}
\dbigcup\limits_{m_{i}\in M_{i}}\text{SUPP}\left[ g\left( m_{i},\lambda
_{-i}\right) \right] \subset \left[ Z^{\ast }\cap \mathcal{L}_{i}^{Z}\left(
F\left( \theta \right) ,\theta \right) \cap \left( \dbigcap\limits_{\theta
^{\prime }\in K^{t}}F\left( \theta ^{\prime }\right) \right) \right] \text{.}
\end{equation*}%
Consider 
\begin{equation*}
K^{t+1}\equiv \left\{ \theta ^{\prime }\in \Theta _{i}^{\theta }:Z^{\ast
}\cap \mathcal{L}_{i}^{Z}\left( F\left( \theta \right) ,\theta \right) \cap %
\left[ \dbigcap\limits_{\theta ^{\prime }\in K^{t}}F\left( \theta ^{\prime
}\right) \right] \text{ is an }i\text{-}Z^{\ast }\text{-}\theta ^{\prime }%
\text{-max set}\right\} \text{,}
\end{equation*}%
and as a result,%
\begin{equation*}
g\left( m_{i},\lambda _{-i}\right) \in MNE^{\left( \mathcal{M},\text{ }%
\theta ^{\prime }\right) }\text{, }\forall m_{i}\in M_{i}\text{, }\forall
\theta ^{\prime }\in K^{t+1}\text{,}
\end{equation*}%
and hence,%
\begin{equation*}
g\left( m_{i},\lambda _{-i}\right) \in \triangle \left[ F\left( \theta
^{\prime }\right) \right] \text{, }\forall m_{i}\in M_{i}\text{, }\forall
\theta ^{\prime }\in K^{t+1}\text{,}
\end{equation*}%
which, together with Lemma \ref{lem:Z-A}, implies%
\begin{equation*}
\dbigcup\limits_{m_{i}\in M_{i}}\text{SUPP}\left[ g\left( m_{i},\lambda
_{-i}\right) \right] \subset \left[ Z^{\ast }\cap \mathcal{L}_{i}^{Z}\left(
F\left( \theta \right) ,\theta \right) \cap \left( \dbigcap\limits_{\theta
^{\prime }\in K^{t+1}}F\left( \theta ^{\prime }\right) \right) \right] \text{%
.}
\end{equation*}%
This completes the induction process, and proves (\ref{tptt1}).$\blacksquare 
$

\subsection{Two lemmas}

\begin{lemma}
\label{lem:mixed-lottery}For any $E\in 2^{Z}\diagdown \left\{ \varnothing
\right\} $ and any $\left( \gamma ,\eta \right) \in \triangle ^{\circ }\left[
E\right] \times \triangle \left[ E\right] $ and there exist $\beta \in
\left( 0,1\right) $ and $\mu \in \triangle \left[ E\right] $ such that%
\begin{equation}
\gamma =\beta \times \eta +\left( 1-\beta \right) \times \mu \text{.}
\label{kkg1}
\end{equation}
\end{lemma}

\noindent \textbf{Proof of Lemma \ref{lem:mixed-lottery}:} With $\alpha \in
\left( 0,1\right) $, consider $\left( -\alpha \right) \times \eta +\left(
1+\alpha \right) \times \gamma $. Since $\gamma $ is in the interior of $%
\triangle \left[ E\right] $, we have 
\begin{equation*}
\mu =\left( -\alpha ^{\ast }\right) \times \eta +\left( 1+\alpha ^{\ast
}\right) \times \gamma \in \triangle ^{\circ }\left[ E\right] \text{,}
\end{equation*}%
for some sufficiently small $\alpha ^{\ast }\in \left( 0,1\right) $. As a
result, we have%
\begin{equation*}
\gamma =\frac{\alpha ^{\ast }}{\left( 1+\alpha ^{\ast }\right) }\times \eta +%
\frac{1}{\left( 1+\alpha ^{\ast }\right) }\mu \text{,}
\end{equation*}%
i.e., $\beta =\frac{\alpha ^{\ast }}{\left( 1+\alpha ^{\ast }\right) }\in
\left( 0,1\right) $ and (\ref{kkg1}) holds.$\blacksquare $

\begin{lemma}
\label{lem:mixed:deviation:gamma}If $\widehat{\mathcal{L}}^{Y\text{-}A\text{-%
}B}$-monotonicity holds, we have%
\begin{equation*}
\left( 
\begin{array}{c}
F\left( \theta \right) \subset \arg \min_{z\in Z^{\ast }}u_{i}^{\theta
}\left( z\right) \text{,} \\ 
\text{and }Z^{\ast }\cap \mathcal{L}_{i}^{Z}\left( F\left( \theta \right)
,\theta \right) \text{ is an }i\text{-}Z^{\ast }\text{-max set}%
\end{array}%
\right) \Longrightarrow \Xi _{i}\left( \theta \right) \neq \varnothing \text{%
.}
\end{equation*}
\end{lemma}

\noindent \textbf{Proof of Lemma \ref{lem:mixed:deviation:gamma}:} Suppose $%
\widehat{\mathcal{L}}^{Y\text{-}A\text{-}B}$-monotonicity and%
\begin{equation*}
\left( 
\begin{array}{c}
F\left( \theta \right) \subset \arg \min_{z\in Z^{\ast }}u_{i}^{\theta
}\left( z\right) \text{,} \\ 
\text{and }Z^{\ast }\cap \mathcal{L}_{i}^{Z}\left( F\left( \theta \right)
,\theta \right) \text{ is an }i\text{-}Z^{\ast }\text{-max set}%
\end{array}%
\right)
\end{equation*}%
hold, and we aim to show $\Xi _{i}\left( \theta \right) \neq \varnothing $.

First, we prove $\Theta _{i}^{\theta }\neq \varnothing $. With $Z^{\ast
}\cap \mathcal{L}_{i}^{Z}\left( F\left( \theta \right) ,\theta \right) $
being an $i$-$Z^{\ast }$-max set, there exists $\widetilde{\theta }\in
\Theta $ such that $Z^{\ast }\cap \mathcal{L}_{i}^{Z}\left( F\left( \theta
\right) ,\theta \right) $ is an $i$-$Z^{\ast }$-$\widetilde{\theta }$-max
set, i.e.,%
\begin{eqnarray*}
Z^{\ast }\cap \mathcal{L}_{i}^{Z}\left( F\left( \theta \right) ,\theta
\right) &\subset &\arg \max_{z\in Z^{\ast }}u_{j}^{\widetilde{\theta }%
}\left( z\right) \text{, }\forall j\in \mathcal{I}\diagdown \left\{
i\right\} \text{,} \\
Z^{\ast }\cap \mathcal{L}_{i}^{Z}\left( F\left( \theta \right) ,\theta
\right) &\subset &\arg \max_{z\in Z^{\ast }\cap \mathcal{L}_{i}^{Z}\left(
F\left( \theta \right) ,\theta \right) }u_{i}^{\widetilde{\theta }}\left(
z\right) \text{.}
\end{eqnarray*}%
As a result, $F\left( \theta \right) \subset \arg \min_{z\in Z^{\ast
}}u_{i}^{\theta }\left( z\right) $ implies%
\begin{eqnarray*}
\widehat{\mathcal{L}}_{j}^{Y\text{-}A\text{-}B}\left( \text{UNIF}\left[
F\left( \theta \right) \right] ,\theta \right) &\subset &\triangle \left(
Z^{\ast }\right) \subset \mathcal{L}_{j}^{Y}\left( F\left( \theta \right) ,%
\widetilde{\theta }\right) \text{, }\forall j\in \mathcal{I}\diagdown
\left\{ i\right\} \text{,} \\
\widehat{\mathcal{L}}_{i}^{Y\text{-}A\text{-}B}\left( \text{UNIF}\left[
F\left( \theta \right) \right] ,\theta \right) &\subset &\triangle \left[
Z^{\ast }\cap \mathcal{L}_{i}^{Z}\left( F\left( \theta \right) ,\theta
\right) \right] \subset \mathcal{L}_{i}^{Y}\left( F\left( \theta \right) ,%
\widetilde{\theta }\right) \text{,}
\end{eqnarray*}%
which, together with $\widehat{\mathcal{L}}^{Y\text{-}A\text{-}B}$%
-monotonicity, implies $F\left( \theta \right) \subset F\left( \widetilde{%
\theta }\right) $. Therefore, we have%
\begin{equation}
\left( 
\begin{array}{c}
\widetilde{\theta }\in \Theta _{i}^{\theta }\neq \varnothing \text{,} \\ 
\text{and }Z^{\ast }\cap \mathcal{L}_{i}^{Z}\left( F\left( \theta \right)
,\theta \right) \text{ is an }i\text{-}Z^{\ast }\text{-}\widetilde{\theta }%
\text{-max set}%
\end{array}%
\right) \text{.}  \label{hhty1}
\end{equation}

Second, we prove $\Xi _{i}\left( \theta \right) \neq \varnothing $. Consider
the following $\text{algorithm.}$%
\begin{equation*}
\left( 
\begin{array}{c}
\text{Step }1\text{: let }K^{1}\equiv \left\{ \theta ^{\prime }\in \Theta
_{i}^{\theta }:Z^{\ast }\cap \mathcal{L}_{i}^{Z}\left( F\left( \theta
\right) ,\theta \right) \text{ is an }i\text{-}Z^{\ast }\text{-}\theta
^{\prime }\text{-max set}\right\} \text{,} \\ 
\text{Step }2\text{: let }K^{2}\equiv \left\{ \theta ^{\prime }\in \Theta
_{i}^{\theta }:Z^{\ast }\cap \mathcal{L}_{i}^{Z}\left( F\left( \theta
\right) ,\theta \right) \cap \left[ \dbigcap\limits_{\theta ^{\prime }\in
K^{1}}F\left( \theta ^{\prime }\right) \right] \text{ is an }i\text{-}%
Z^{\ast }\text{-}\theta ^{\prime }\text{-max set}\right\} \text{,} \\ 
... \\ 
\text{Step }n\text{: let }K^{n}\equiv \left\{ \theta ^{\prime }\in \Theta
_{i}^{\theta }:Z^{\ast }\cap \mathcal{L}_{i}^{Z}\left( F\left( \theta
\right) ,\theta \right) \cap \left[ \dbigcap\limits_{\theta ^{\prime }\in
K^{n-1}}F\left( \theta ^{\prime }\right) \right] \text{ is an }i\text{-}%
Z^{\ast }\text{-}\theta ^{\prime }\text{-max set}\right\} \text{,} \\ 
...%
\end{array}%
\right)
\end{equation*}%
(\ref{hhty1}) implies $\widetilde{\theta }\in K^{1}\neq \varnothing $.\
Furthermore, we have $K^{1}\subset K^{2}$, and inductively, it is easy to
show%
\begin{eqnarray}
\forall n &\geq &1\text{,}  \notag \\
K^{n} &\subset &K^{n+1}\text{ and}  \notag \\
Z^{\ast }\cap \mathcal{L}_{i}^{Z}\left( F\left( \theta \right) ,\theta
\right) \cap \left[ \dbigcap\limits_{\theta ^{\prime }\in K^{n}}F\left(
\theta ^{\prime }\right) \right] &\supset &Z^{\ast }\cap \mathcal{L}%
_{i}^{Z}\left( F\left( \theta \right) ,\theta \right) \cap \left[
\dbigcap\limits_{\theta ^{\prime }\in K^{n+1}}F\left( \theta ^{\prime
}\right) \right] \text{.}  \label{ddtp1}
\end{eqnarray}%
$\text{Since }Z$ is finite, (\ref{ddtp1}) implies that there exists $n\geq 1$
such that%
\begin{equation}
Z^{\ast }\cap \mathcal{L}_{i}^{Z}\left( F\left( \theta \right) ,\theta
\right) \cap \left[ \dbigcap\limits_{\theta ^{\prime }\in K^{n}}F\left(
\theta ^{\prime }\right) \right] =Z^{\ast }\cap \mathcal{L}_{i}^{Z}\left(
F\left( \theta \right) ,\theta \right) \cap \left[ \dbigcap\limits_{\theta
^{\prime }\in K^{n+1}}F\left( \theta ^{\prime }\right) \right] \text{.}
\label{ddtp2}
\end{equation}%
and as a result,%
\begin{eqnarray*}
K^{n+1} &\equiv &\left\{ \theta ^{\prime }\in \Theta _{i}^{\theta }:Z^{\ast
}\cap \mathcal{L}_{i}^{Z}\left( F\left( \theta \right) ,\theta \right) \cap %
\left[ \dbigcap\limits_{\theta ^{\prime }\in K^{n}}F\left( \theta ^{\prime
}\right) \right] \text{ is an }i\text{-}Z^{\ast }\text{-}\theta ^{\prime }%
\text{-max set}\right\} \\
&=&\left\{ \theta ^{\prime }\in \Theta _{i}^{\theta }:Z^{\ast }\cap \mathcal{%
L}_{i}^{Z}\left( F\left( \theta \right) ,\theta \right) \cap \left[
\dbigcap\limits_{\theta ^{\prime }\in K^{n+1}}F\left( \theta ^{\prime
}\right) \right] \text{ is an }i\text{-}Z^{\ast }\text{-}\theta ^{\prime }%
\text{-max set}\right\} \text{,}
\end{eqnarray*}%
where the second inequality follows from (\ref{ddtp2}). Therefore, we have $%
K^{n+1}\in \Xi _{i}\left( \theta \right) \neq \varnothing $.$\blacksquare $

\subsection{Proof of Lemma \protect\ref{lem:no-veto:generalized:SCC}}

\label{sec:lem:no-veto:generalized:SCC}

Suppose that $\widehat{\mathcal{L}}^{Y\text{-}A\text{-}B}$-monotonicity
holds. We first we prove%
\begin{equation}
\left[ Z^{\ast }\text{ is a }j\text{-}Z^{\ast }\text{-}\theta ^{\prime }%
\text{-max set}\right] \Longrightarrow Z^{\ast }\subset F\left( \theta
^{\prime }\right) \text{, }\forall \left( j,\theta ^{\prime }\right) \in 
\mathcal{I}\times \Theta \text{.}  \label{ddt2}
\end{equation}%
Fix any $\left( j,\theta ^{\prime }\right) \in \mathcal{I}\times \Theta $
such that $Z^{\ast }$ is a $j$-$Z^{\ast }$-$\theta ^{\prime }$-max set, and
we aim to show $Z^{\ast }\subset F\left( \theta ^{\prime }\right) $.

$Z^{\ast }$ being an $j$-$Z^{\ast }$-$\theta ^{\prime }$-max set implies
that all agents are indifferent between any two deterministic outcomes in $%
Z^{\ast }$ at state $\theta ^{\prime }$. This leads to two implications: (i) 
$Z^{\ast }=\cup _{\theta \in \Theta }F\left( \theta \right) $, because if $%
Z^{\ast }\neq \cup _{\theta \in \Theta }F\left( \theta \right) $, (\ref{tth2}%
) implies $Z=Z^{\ast }$ and $Z$ is not an $i$-max set for any $i\in \mathcal{%
I}$, contradicting all agents being indifferent between any two
deterministic outcomes in $Z^{\ast }=Z$ at state $\theta ^{\prime }$; (ii)
we have%
\begin{equation}
\widehat{\mathcal{L}}_{i}^{Y\text{-}A\text{-}B}\left( \text{UNIF}\left[
F\left( \widetilde{\theta }\right) \right] ,\widetilde{\theta }\right)
\subset \triangle \left( Z^{\ast }\right) =\mathcal{L}_{i}^{Y}\left( \text{%
UNIF}\left[ F\left( \widetilde{\theta }\right) \right] ,\theta ^{\prime
}\right) \text{, }\forall \widetilde{\theta }\in \Theta \text{, }\forall
i\in \mathcal{I}\text{,}  \label{ddt1}
\end{equation}%
where the equality follows from all agents being indifferent between any two
deterministic outcomes in $Z^{\ast }$ at $\theta ^{\prime }$. Thus, $%
\widehat{\mathcal{L}}^{Y\text{-}A\text{-}B}$-monotonicity and (\ref{ddt1})
imply%
\begin{equation*}
\cup _{\widetilde{\theta }\in \Theta }F\left( \widetilde{\theta }\right)
\subset F\left( \theta ^{\prime }\right) \text{,}
\end{equation*}%
i.e., $Z^{\ast }=\cup _{\widetilde{\theta }\in \Theta }F\left( \widetilde{%
\theta }\right) \subset F\left( \theta ^{\prime }\right) $, and (\ref{ddt2})
holds.

Second, we prove 
\begin{equation*}
\left[ \widehat{\Gamma }_{j}^{A\text{-}B}\left( \theta \right) \text{ is a }j%
\text{-}Z^{\ast }\text{-}\theta ^{\prime }\text{-max set}\right]
\Longrightarrow \widehat{\Gamma }_{j}^{A\text{-}B}\left( \theta \right)
\subset F\left( \theta ^{\prime }\right) \text{, }\forall \left( j,\theta
,\theta ^{\prime }\right) \in \mathcal{I}\times \Theta \times \Theta \text{.}
\end{equation*}%
Fix any $\left( j,\theta ,\theta ^{\prime }\right) \in \mathcal{I}\times
\Theta \times \Theta $ such that $\widehat{\Gamma }_{j}^{A\text{-}B}\left(
\theta \right) $ is a $j$-$Z^{\ast }$-$\theta ^{\prime }$-max set, and we
aim to show $\widehat{\Gamma }_{j}^{A\text{-}B}\left( \theta \right) \subset
F\left( \theta ^{\prime }\right) $. We consider three scenarios:%
\begin{equation*}
\left[ 
\begin{array}{c}
\text{scenario 1: }\left( 
\begin{tabular}{l}
$F\left( \theta \right) \subset \arg \min_{z\in Z^{\ast }}u_{j}^{\theta
}\left( z\right) $, \\ 
$Z^{\ast }\cap \mathcal{L}_{j}^{Z}\left( F\left( \theta \right) ,\theta
\right) \text{ is a }j\text{-}Z^{\ast }\text{-max set}$%
\end{tabular}%
\right) \\ 
\text{scenario 2: }\left( 
\begin{tabular}{l}
$F\left( \theta \right) \subset \arg \min_{z\in Z^{\ast }}u_{j}^{\theta
}\left( z\right) $, \\ 
$Z^{\ast }\cap \mathcal{L}_{j}^{Z}\left( F\left( \theta \right) ,\theta
\right) \text{ is not an }j\text{-}Z^{\ast }\text{-max set}$%
\end{tabular}%
\right) \\ 
\text{scenario 3: }F\left( \theta \right) \diagdown \arg \min_{z\in Z^{\ast
}}u_{j}^{\theta }\left( z\right) \neq \varnothing \text{.}%
\end{array}%
\right]
\end{equation*}%
By Lemma \ref{lem:mixed:deviation:gamma}, we have $\Xi _{j}\left( \theta
\right) \neq \varnothing $\ if scenario 1 occurs. Given this, (\ref{ddtt})\
implies%
\begin{equation}
\widehat{\Gamma }_{j}^{A\text{-}B}\left( \theta \right) =\left\{ 
\begin{tabular}{ll}
$Z^{\ast }\cap \mathcal{L}_{j}^{Z}\left( F\left( \theta \right) ,\theta
\right) \cap \left( \dbigcup\limits_{K\in \Xi _{j}\left( \theta \right)
}\dbigcap\limits_{\widetilde{\theta }\in K}F\left( \widetilde{\theta }%
\right) \right) $, & if scenario 1 occurs, \\ 
$Z^{\ast }\cap \mathcal{L}_{j}^{Z}\left( F\left( \theta \right) ,\theta
\right) $, & if scenario 2 occurs, \\ 
$Z^{\ast }\text{,}$ & if scenario 3 occurs%
\end{tabular}%
\right. \text{.}  \label{ddtt5}
\end{equation}%
Since $\widehat{\Gamma }_{j}^{A\text{-}B}\left( \theta \right) $ is a $j$-$%
Z^{\ast }$-$\theta ^{\prime }$-max set, scenario 2 cannot happen.

Suppose scenario 3 occurs, i.e., $\widehat{\Gamma }_{j}^{A\text{-}B}\left(
\theta \right) =Z^{\ast }$ is an $j$-$Z^{\ast }$-$\theta ^{\prime }$-max
set. By (\ref{ddt2}), we have $\widehat{\Gamma }_{j}^{A\text{-}B}\left(
\theta \right) =Z^{\ast }\subset F\left( \theta ^{\prime }\right) $.

Suppose scenario 1 occurs. We thus have $\Xi _{j}\left( \theta \right) \neq
\varnothing $ and%
\begin{equation*}
\widehat{\Gamma }_{j}^{A\text{-}B}\left( \theta \right) =Z^{\ast }\cap 
\mathcal{L}_{j}^{Z}\left( F\left( \theta \right) ,\theta \right) \cap \left(
\dbigcup\limits_{K\in \Xi _{j}\left( \theta \right) }\dbigcap\limits_{%
\widetilde{\theta }\in K}F\left( \widetilde{\theta }\right) \right) \text{,}
\end{equation*}%
and $Z^{\ast }\cap \mathcal{L}_{j}^{Z}\left( F\left( \theta \right) ,\theta
\right) $ is an $j$-$Z^{\ast }$-max set. Pick any $K^{\ast }\in \Xi
_{j}\left( \theta \right) $. Recall the definition of $\Xi _{j}\left( \theta
\right) $ in (\ref{ddttaa}), we have%
\begin{equation}
\Theta _{j}^{\theta }\cap \left[ \Lambda ^{j\text{-}Z^{\ast }\text{-}\Theta
}\left( Z^{\ast }\cap \mathcal{L}_{j}^{Z}\left( F\left( \theta \right)
,\theta \right) \cap \left[ \dbigcap\limits_{\widetilde{\theta }\in K^{\ast
}}F\left( \widetilde{\theta }\right) \right] \right) \right] =K^{\ast }\text{%
.}  \label{jju1}
\end{equation}%
Since%
\begin{equation*}
Z^{\ast }\cap \mathcal{L}_{j}^{Z}\left( F\left( \theta \right) ,\theta
\right) \cap \left( \dbigcap\limits_{\widetilde{\theta }\in K^{\ast
}}F\left( \widetilde{\theta }\right) \right) \subset Z^{\ast }\cap \mathcal{L%
}_{j}^{Z}\left( F\left( \theta \right) ,\theta \right) \cap \left(
\dbigcup\limits_{K\in \Xi _{j}\left( \theta \right) }\dbigcap\limits_{%
\widetilde{\theta }\in K}F\left( \widetilde{\theta }\right) \right) =%
\widehat{\Gamma }_{j}^{A\text{-}B}\left( \theta \right) \text{,}
\end{equation*}%
$\widehat{\Gamma }_{j}^{A\text{-}B}\left( \theta \right) $ being an $j$-$%
Z^{\ast }$-$\theta ^{\prime }$-max set implies $Z^{\ast }\cap \mathcal{L}%
_{j}^{Z}\left( F\left( \theta \right) ,\theta \right) \cap \left(
\dbigcap\limits_{\theta ^{\prime }\in K^{\ast }}F\left( \theta ^{\prime
}\right) \right) $ is a $j$-$Z^{\ast }$-$\theta ^{\prime }$-max set. As a
result, we have%
\begin{equation}
\theta ^{\prime }\in \left[ \Lambda ^{j\text{-}Z^{\ast }\text{-}\Theta
}\left( Z^{\ast }\cap \mathcal{L}_{j}^{Z}\left( F\left( \theta \right)
,\theta \right) \cap \left[ \dbigcap\limits_{\widetilde{\theta }\in K^{\ast
}}F\left( \widetilde{\theta }\right) \right] \right) \right] \text{.}
\label{jju2}
\end{equation}%
We now show%
\begin{equation}
\theta ^{\prime }\in \Theta _{j}^{\theta }\text{.}  \label{jju3}
\end{equation}%
Given $F\left( \theta \right) \subset \widehat{\Gamma }_{j}^{A\text{-}%
B}\left( \theta \right) $, $\widehat{\Gamma }_{j}^{A\text{-}B}\left( \theta
\right) $\ being an $j$-$Z^{\ast }$-$\theta ^{\prime }$-max set implies two
things: (1) $F\left( \theta \right) $ is a $j$-$Z^{\ast }$-$\theta ^{\prime
} $-max set, and (2) 
\begin{eqnarray*}
\widehat{\mathcal{L}}_{i}^{Y\text{-}A\text{-}B}\left( \text{UNIF}\left[
F\left( \theta \right) \right] ,\theta \right) &\subset &\triangle \left(
Z^{\ast }\right) \subset \mathcal{L}_{i}^{Y}\left( F\left( \theta \right)
,\theta ^{\prime }\right) \text{, }\forall i\in \mathcal{I}\diagdown \left\{
j\right\} \text{,} \\
\widehat{\mathcal{L}}_{j}^{Y\text{-}A\text{-}B}\left( \text{UNIF}\left[
F\left( \theta \right) \right] ,\theta \right) &=&\triangle \left[ \widehat{%
\Gamma }_{j}^{A\text{-}B}\left( \theta \right) \right] \subset \mathcal{L}%
_{j}^{Y}\left( F\left( \theta \right) ,\theta ^{\prime }\right) \text{,}
\end{eqnarray*}%
which, together with $\widehat{\mathcal{L}}^{Y\text{-}A\text{-}B}$%
-monotonicity implies $F\left( \theta \right) \subset F\left( \theta
^{\prime }\right) $. Therefore, (\ref{jju3}) holds.

(\ref{jju1}), (\ref{jju2}) and (\ref{jju3}) show $\theta ^{\prime }\in
K^{\ast }$. We thus have%
\begin{eqnarray*}
\widehat{\Gamma }_{j}^{A\text{-}B}\left( \theta \right) &=&Z^{\ast }\cap 
\mathcal{L}_{j}^{Z}\left( F\left( \theta \right) ,\theta \right) \cap \left(
\dbigcup\limits_{K\in \Xi _{j}\left( \theta \right) }\dbigcap\limits_{%
\widetilde{\theta }\in K}F\left( \widetilde{\theta }\right) \right) \\
&\subset &Z^{\ast }\cap \mathcal{L}_{j}^{Z}\left( F\left( \theta \right)
,\theta \right) \cap \left( \dbigcap\limits_{\widetilde{\theta }\in K^{\ast
}}F\left( \widetilde{\theta }\right) \right) \\
&\subset &F\left( \theta ^{\prime }\right) \text{.}
\end{eqnarray*}%
where the first "$\subset $" follows from $K^{\ast }\in \Xi _{i}\left(
\theta \right) $ and the second "$\subset $" follows from $\theta ^{\prime
}\in K^{\ast }$.$\blacksquare $

\subsection{Proof of Theorem \protect\ref{theorem:full:mix:SCC-A}}

\label{sec:theorem:full:mix:SCC-A}

By their definitions, mixed-Nash-B-implementation implies
mixed-Nash-A-implementable, i.e., (ii)$\Longrightarrow $(i). We show (i)$%
\Longrightarrow $(iii) and (iii)$\Longrightarrow $(ii) in Appendix \ref%
{sec:theorem:full:mix:SCC-A:1->3}\ and \ref{sec:theorem:full:mix:SCC-A:3->2}%
, respectively.

\subsubsection{The proof of (i)$\Longrightarrow $(iii)}

\label{sec:theorem:full:mix:SCC-A:1->3}

Suppose that an SCC $F$ is mixed-Nash-A-implemented by $\mathcal{M}%
=\left\langle M\text{, \ }g:M\longrightarrow Y\right\rangle $. Fix any $%
\left( \theta ,\theta ^{\prime }\right) \in \Theta \times \Theta $ such that%
\begin{equation}
\left[ 
\begin{array}{c}
\widehat{\mathcal{L}}_{i}^{Y\text{-}A\text{-}B}\left( \text{UNIF}\left[
F\left( \theta \right) \right] ,\theta \right) \subset \mathcal{L}%
_{i}^{Y}\left( \text{UNIF}\left[ F\left( \theta \right) \right] ,\theta
^{\prime }\right) \text{, } \\ 
\forall i\in \mathcal{I}%
\end{array}%
\right] \text{ ,}  \label{tth7}
\end{equation}%
and we aim to show $F\left( \theta \right) \subset F\left( \theta ^{\prime
}\right) $, i.e., $\widehat{\mathcal{L}}^{Y\text{-}A\text{-}B}$%
-uniform-monotonicity holds. In particular, fix any $a\in F\left( \theta
\right) $, and we aim to show $a\in F\left( \theta ^{\prime }\right) $.
Since $a\in F\left( \theta \right) $, there exists some $\lambda \in
MNE^{\left( \mathcal{M},\text{ }\theta \right) }$, such that $a$ is induced
with positive probability by $\lambda $. Thus, SUPP$\left[ g\left( \lambda
\right) \right] \subset F\left( \theta \right) $. We now prove $\lambda \in
MNE^{\left( \mathcal{M},\text{ }\theta ^{\prime }\right) }$ by
contradiction, which further implies $a\in F\left( \theta ^{\prime }\right) $%
. Suppose $\lambda \notin MNE^{\left( \mathcal{M},\text{ }\theta ^{\prime
}\right) }$, i.e., there exist $i\in \mathcal{I}$, and $m_{i}^{\prime }\in
M_{i}$ such that%
\begin{eqnarray}
U_{i}^{\theta }\left[ g\left( m_{i}^{\prime },\lambda _{-i}\right) \right]
&\leq &U_{i}^{\theta }\left[ g\left( \lambda \right) \right] \text{,}
\label{cch1} \\
U_{i}^{\theta ^{\prime }}\left[ g\left( m_{i}^{\prime },\lambda _{-i}\right) %
\right] &>&U_{i}^{\theta ^{\prime }}\left[ g\left( \lambda \right) \right] 
\text{.}  \label{cch2}
\end{eqnarray}%
Lemma \ref{lem:mixed-lottery} and SUPP$\left[ g\left( \lambda \right) \right]
\subset F\left( \theta \right) $ imply existence of $\mu \in \triangle \left[
F\left( \theta \right) \right] $ and $\beta \in \left( 0,1\right) $ such that%
\begin{equation}
\text{UNIF}\left[ F\left( \theta \right) \right] =\beta \times g\left(
\lambda \right) +\left( 1-\beta \right) \times \mu \text{.}  \label{cch3}
\end{equation}%
Thus, (\ref{cch2}) and (\ref{cch3})\ imply%
\begin{equation}
U_{i}^{\theta ^{\prime }}\left[ \beta \times g\left( m_{i}^{\prime },\lambda
_{-i}\right) +\left( 1-\beta \right) \times \mu \right] >U_{i}^{\theta
^{\prime }}\left[ \beta \times g\left( \lambda \right) +\left( 1-\beta
\right) \times \mu \right] =U_{i}^{\theta ^{\prime }}\left( \text{UNIF}\left[
F\left( \theta \right) \right] \right) \text{.}  \label{cch4}
\end{equation}%
(\ref{cch1}) and (\ref{cch3})\ imply%
\begin{equation}
U_{i}^{\theta }\left[ \beta \times g\left( m_{i}^{\prime },\lambda
_{-i}\right) +\left( 1-\beta \right) \times \mu \right] \leq U_{i}^{\theta }%
\left[ \beta \times g\left( \lambda \right) +\left( 1-\beta \right) \times
\mu \right] =U_{i}^{\theta }\left( \text{UNIF}\left[ F\left( \theta \right) %
\right] \right) \text{.}  \label{cch4a}
\end{equation}

We now consider two cases. First, suppose%
\begin{equation}
\left( 
\begin{tabular}{l}
$F\left( \theta \right) \subset \arg \min_{z\in Z^{\ast }}u_{i}^{\theta
}\left( z\right) $, \\ 
$\Xi _{i}\left( \theta \right) \neq \varnothing $, \\ 
and $Z^{\ast }\cap \mathcal{L}_{i}^{Z}\left( F\left( \theta \right) ,\theta
\right) $ is an $i\text{-}Z^{\ast }\text{-max set}$%
\end{tabular}%
\right) \text{,}  \label{tth12}
\end{equation}%
holds. Thus, by (\ref{ddtt}), we have%
\begin{equation*}
\widehat{\mathcal{L}}_{i}^{Y\text{-}A\text{-}B}\left( \text{UNIF}\left[
F\left( \theta \right) \right] ,\theta \right) =\triangle \left[ Z^{\ast
}\cap \mathcal{L}_{i}^{Z}\left( F\left( \theta \right) ,\theta \right) \cap
\left( \dbigcup\limits_{K\in \Xi _{i}\left( \theta \right) }\dbigcap\limits_{%
\widetilde{\theta }\in K}F\left( \widetilde{\theta }\right) \right) \right] 
\text{,}
\end{equation*}%
and by Lemma \ref{lem:mixed:deviation:SCC}, we have 
\begin{equation*}
g\left( m_{i}^{\prime },\lambda _{-i}\right) \in \triangle \left[ Z^{\ast
}\cap \mathcal{L}_{i}^{Z}\left( F\left( \theta \right) ,\theta \right) \cap
\left( \dbigcup\limits_{K\in \Xi _{i}\left( \theta \right) }\dbigcap\limits_{%
\widetilde{\theta }\in K}F\left( \widetilde{\theta }\right) \right) \right] 
\text{,}
\end{equation*}%
which, together with $\mu \in \triangle \left[ F\left( \theta \right) \right]
$, implies\footnote{$F\left( \theta \right) \subset \arg \min_{z\in Z^{\ast
}}u_{i}^{\theta }\left( z\right) $ implies $\triangle \left[ F\left( \theta
\right) \right] \subset \widehat{\mathcal{L}}_{i}^{Y\text{-}A\text{-}%
B}\left( \text{UNIF}\left[ F\left( \theta \right) \right] ,\theta \right) $.}%
\begin{equation}
\beta \times g\left( m_{i}^{\prime },\lambda _{-i}\right) +\left( 1-\beta
\right) \times \mu \in \widehat{\mathcal{L}}_{i}^{Y\text{-}A\text{-}B}\left( 
\text{UNIF}\left[ F\left( \theta \right) \right] ,\theta \right) \text{.}
\label{ddtt2}
\end{equation}%
(\ref{tth7}) and (\ref{ddtt2}) imply%
\begin{equation*}
\beta \times g\left( m_{i}^{\prime },\lambda _{-i}\right) +\left( 1-\beta
\right) \times \mu \in \mathcal{L}_{i}^{Y}\left( \text{UNIF}\left[ F\left(
\theta \right) \right] ,\theta ^{\prime }\right) \text{.}
\end{equation*}%
contradicting (\ref{cch4}).

Second, suppose that (\ref{tth12}) does not hold, and by (\ref{ddtt}), we
have%
\begin{equation*}
\widehat{\mathcal{L}}_{i}^{Y\text{-}A\text{-}B}\left( \text{UNIF}\left[
F\left( \theta \right) \right] ,\theta \right) =\left[ \triangle \left(
Z^{\ast }\right) \right] \cap \mathcal{L}_{i}^{Y}\left( \text{UNIF}\left[
F\left( \theta \right) \right] ,\theta \right) \text{,}
\end{equation*}%
which, together with (\ref{cch4a}) and Lemma \ref{lem:Z-A}, implies%
\begin{eqnarray}
\beta \times g\left( m_{i}^{\prime },\lambda _{-i}\right) +\left( 1-\beta
\right) \times \mu &\subset &\left[ \triangle \left( Z^{\ast }\right) \right]
\cap \mathcal{L}_{i}^{Y}\left( \text{UNIF}\left[ F\left( \theta \right) %
\right] ,\theta \right)  \notag \\
&=&\widehat{\mathcal{L}}_{i}^{Y\text{-}A\text{-}B}\left( \text{UNIF}\left[
F\left( \theta \right) \right] ,\theta \right) \text{.}  \label{ddtt3}
\end{eqnarray}%
(\ref{tth7}) and (\ref{ddtt3}) imply%
\begin{equation*}
\beta \times g\left( m_{i}^{\prime },\lambda _{-i}\right) +\left( 1-\beta
\right) \times \mu \in \mathcal{L}_{i}^{Y}\left( \text{UNIF}\left[ F\left(
\theta \right) \right] ,\theta ^{\prime }\right) \text{.}
\end{equation*}%
contradicting (\ref{cch4}).$\blacksquare $

\subsubsection{The proof of (iii)$\Longrightarrow $(ii)}

\label{sec:theorem:full:mix:SCC-A:3->2}

\paragraph{Preliminary construction}

In order to build our canonical mechanism to implement $F$, we need to take
two preliminary constructions. First, for each $\left( \theta ,j\right) \in
\Theta \times \mathcal{I}$, fix any function $\psi _{j}^{\theta }:\Theta
\longrightarrow Y$ such that 
\begin{equation}
\psi _{j}^{\theta }\left( \theta ^{\prime }\right) \in \left( \arg
\max_{y\in \widehat{\mathcal{L}}_{j}^{Y\text{-}A\text{-}B}\left( \text{UNIF}%
\left[ F\left( \theta \right) \right] ,\theta \right) }U_{j}^{\theta
^{\prime }}\left[ y\right] \right) \text{, }\forall \theta ^{\prime }\in
\Theta \text{,}  \label{yuy1a-A-B}
\end{equation}%
and by (\ref{ddtt4}), we have%
\begin{equation}
\psi _{j}^{\theta }\left( \theta ^{\prime }\right) \in \left( \arg
\max_{y\in \widehat{\mathcal{L}}_{j}^{Y\text{-}A\text{-}B}\left( \text{UNIF}%
\left[ F\left( \theta \right) \right] ,\theta \right) }U_{j}^{\theta
^{\prime }}\left[ y\right] \right) \cap \triangle \left( \widehat{\Gamma }%
_{j}^{A\text{-}B}\left( \theta \right) \right) \text{, }\forall \theta
^{\prime }\in \Theta \text{.}  \label{yuy1-A-B}
\end{equation}%
The following lemma completes our second construction.

\begin{lemma}
\label{lem:if:A-B}For each $\left( \theta ,j\right) \in \Theta \times 
\mathcal{I}$, there exist%
\begin{equation*}
\varepsilon _{j}^{\theta }>0\text{ and }y_{j}^{\theta }\in \widehat{\mathcal{%
L}}_{j}^{Y\text{-}A\text{-}B}\left( \text{UNIF}\left[ F\left( \theta \right) %
\right] ,\theta \right) \text{,}
\end{equation*}%
such that%
\begin{equation}
\left[ \varepsilon _{j}^{\theta }\times y+\left( 1-\varepsilon _{j}^{\theta
}\right) \times y_{j}^{\theta }\right] \in \widehat{\mathcal{L}}_{j}^{Y\text{%
-}A\text{-}B}\left( \text{UNIF}\left[ F\left( \theta \right) \right] ,\theta
\right) \text{, }\forall y\in \triangle \left( \widehat{\Gamma }_{j}^{A\text{%
-}B}\left( \theta \right) \right) \text{.}  \label{yuy2-A-B}
\end{equation}
\end{lemma}

The proof of Lemma \ref{lem:if:A-B} is similar to that of Lemma \ref{lem:if}%
, and we omit it.

\paragraph{A canonical mechanism}

\label{sec:canonic:A-B}

Let $\mathbb{N}$ denote the set of positive integers. We use the mechanism $%
\mathcal{M}^{A\text{-}B}=\left\langle M^{A\text{-}B}\equiv \times _{i\in 
\mathcal{I}}M_{i}^{A\text{-}B}\text{, \ }g:M^{A\text{-}B}\longrightarrow
\triangle \left( Z^{\ast }\right) \right\rangle $ defined below to implement 
$F$. In particular, we have 
\begin{equation*}
M_{i}^{A\text{-}B}=\left\{ \left( \theta _{i},k_{i}^{2},k_{i}^{3},\gamma
_{i},b_{i}\right) \in \Theta \times \mathbb{N}\times \mathbb{N}\times \left(
Z^{\ast }\right) ^{\left[ 2^{Z^{\ast }}\diagdown \left\{ \varnothing
\right\} \right] }\times Z^{\ast }:%
\begin{tabular}{l}
$\gamma _{i}\left( E\right) \in E$, \\ 
$\forall E\in \left[ 2^{Z^{\ast }}\diagdown \left\{ \varnothing \right\} %
\right] $%
\end{tabular}%
\right\} \text{, }\forall i\in \mathcal{I}\text{,}
\end{equation*}%
and $g\left[ m=\left( m_{i}\right) _{i\in \mathcal{I}}=\left( \theta
_{i},k_{i}^{2},k_{i}^{3},\gamma _{i},b_{i}\right) _{i\in \mathcal{I}}\right] 
$ is defined in three cases.

\begin{description}
\item[Case (1): consensus] if there exists $\theta \in \Theta $ such that%
\begin{equation*}
\left( \theta _{i},k_{i}^{2}\right) =\left( \theta ,1\right) \text{, }%
\forall i\in \mathcal{I}\text{,}
\end{equation*}%
then $g\left[ m\right] =$UNIF$\left[ F\left( \theta \right) \right] $;

\item[Case (2), unilateral deviation: ] if there exists $\left( \theta
,j\right) \in \Theta \times \mathcal{I}$ such that%
\begin{equation*}
\left( \theta _{i},k_{i}^{2}\right) =\left( \theta ,1\right) \text{ if and
only if }i\in \mathcal{I}\diagdown \left\{ j\right\} \text{,}
\end{equation*}%
then 
\begin{eqnarray}
g\left[ m\right] &=&\left( 1-\frac{1}{k_{j}^{2}}\right) \times \psi
_{j}^{\theta }\left( \theta _{j}\right)  \label{uue1-A-B} \\
&&+\frac{1}{k_{j}^{2}}\times \left( 
\begin{tabular}{l}
$\varepsilon _{j}^{\theta }\times \left[ \left( 1-\frac{1}{k_{j}^{3}}\right)
\times \gamma _{j}\left( \widehat{\Gamma }_{j}^{A\text{-}B}\left( \theta
\right) \right) +\frac{1}{k_{j}^{3}}\times \text{UNIF}\left( \widehat{\Gamma 
}_{j}^{A\text{-}B}\left( \theta \right) \right) \right] $ \\ 
$+\left( 1-\varepsilon _{j}^{\theta }\right) \times y_{j}^{\theta }$%
\end{tabular}%
\right) \text{,}  \notag
\end{eqnarray}%
where $\left( \varepsilon _{j}^{\theta },y_{j}^{\theta }\right) $ are chosen
for each $\left( \theta ,j\right) \in \Theta \times \mathcal{I}$ according
to Lemma \ref{lem:if:A-B};

\item[Case (3), multi-lateral deviation: ] otherwise, 
\begin{equation}
g\left[ m\right] =\left( 1-\frac{1}{k_{j^{\ast }}^{2}}\right) \times
b_{j^{\ast }}+\frac{1}{k_{j^{\ast }}^{2}}\times \text{UNIF}\left( Z^{\ast
}\right) \text{,}  \label{uue2-A-B}
\end{equation}%
where $j^{\ast }=\max \left( \arg \max_{i\in \mathcal{I}}k_{i}^{2}\right) $,
i.e., $j^{\ast }$ is the largest-numbered agent who submits the highest
number on the second dimension of the message.
\end{description}

\begin{lemma}
\label{lem:mixed:canonical:pure:A-B}Consider the canonical mechanism $%
\mathcal{M}^{A\text{-}B}$ above. For any $\theta \in \Theta $ and any $%
\lambda \in MNE^{\left( \mathcal{M}^{A\text{-}B},\text{ }\theta \right) }$,
we have SUPP$\left[ \lambda \right] \subset PNE^{\left( \mathcal{M}^{A\text{-%
}B},\text{ }\theta \right) }$.
\end{lemma}

The proof of Lemma \ref{lem:mixed:canonical:pure:A-B} is similar to that of
Lemma \ref{lem:mixed:canonical:pure}, and we omit it.

\paragraph{(iii)$\Longrightarrow $(ii) in Theorem \protect\ref%
{theorem:full:mix:SCC-A}: a proof}

Suppose that $\widehat{\mathcal{L}}^{Y\text{-}A\text{-}B}$-monotonicity
holds. Fix any true state $\theta ^{\ast }\in \Theta $. We aim to prove%
\begin{equation*}
\dbigcup\limits_{\lambda \in MNE^{\left( \mathcal{M}^{A\text{-}B},\text{ }%
\theta ^{\ast }\right) }}\text{SUPP}\left( g\left[ \lambda \right] \right)
=F\left( \theta ^{\ast }\right) \text{.}
\end{equation*}%
First, truth revealing is a Nash equilibrium, i.e., any pure strategy profile%
\begin{equation*}
m^{\ast }=\left( \theta _{i}=\theta ^{\ast },k_{i}^{2}=1,\ast ,\ast ,\ast
\right) _{i\in \mathcal{I}}
\end{equation*}%
is a Nash equilibrium, which triggers Case (1) and $g\left[ m^{\ast }\right]
=$UNIF$\left[ F\left( \theta ^{\ast }\right) \right] $. Any unilateral
deviation $\overline{m}_{j}\in M_{j}^{A\text{-}B}$ of agent $j\in \mathcal{I}
$ would either still trigger Case (1) and induce UNIF$\left[ F\left( \theta
^{\ast }\right) \right] $, or trigger Case (2) and induce%
\begin{equation*}
g\left[ \overline{m}_{j},m_{-j}^{\ast }\right] \in \widehat{\mathcal{L}}%
_{j}^{Y\text{-}A\text{-}B}\left( \text{UNIF}\left[ F\left( \theta ^{\ast
}\right) \right] ,\theta ^{\ast }\right) \subset \mathcal{L}_{j}^{Y}\left( 
\text{UNIF}\left[ F\left( \theta ^{\ast }\right) \right] ,\theta ^{\ast
}\right) \text{, }\forall \overline{m}_{j}\in M_{j}\text{.}
\end{equation*}%
Therefore, any $\overline{m}_{j}\in M_{j}$ is not a profitable deviation.

Second, by Lemma \ref{lem:mixed:canonical:pure:A-B}, it suffers no loss of
generality to focus on pure-strategy equilibria. Fix any%
\begin{equation*}
\widetilde{m}=\left( \widetilde{\theta _{i}},\widetilde{k_{i}^{2}},%
\widetilde{k_{i}^{3}},\widetilde{\gamma _{i}},\widetilde{b_{i}}\right)
_{i\in \mathcal{I}}\in PNE^{\left( \mathcal{M}^{A\text{-}B},\text{ }\theta
^{\ast }\right) }\text{,}
\end{equation*}%
and we aim to prove $g\left[ \widetilde{m}\right] \in \triangle \left[
F\left( \theta ^{\ast }\right) \right] $.

$\widetilde{m}$ may trigger either Case (1) or Case (2) or Case (3). We
first consider the scenarios in which $\widetilde{m}$ triggers Case (1),
i.e.,%
\begin{equation*}
\widetilde{m}=\left( \widetilde{\theta _{i}}=\widetilde{\theta },\widetilde{%
k_{i}^{2}}=1,\widetilde{k_{i}^{3}},\widetilde{\gamma _{i}},\widetilde{b_{i}}%
\right) _{i\in \mathcal{I}}\text{ for some }\widetilde{\theta }\in \Theta 
\text{,}
\end{equation*}%
and $g\left[ \widetilde{m}\right] =$UNIF$\left[ F\left( \widetilde{\theta }%
\right) \right] $. We now show $F\left( \widetilde{\theta }\right) \subset
F\left( \theta ^{\ast }\right) $ by contradiction. Suppose otherwise. By $%
\widehat{\mathcal{L}}^{Y\text{-}A\text{-}B}$-monotonicity, there exists $%
j\in \mathcal{I}$ such that%
\begin{equation*}
\exists y^{\ast }\in \widehat{\mathcal{L}}_{j}^{Y\text{-}A\text{-}B}\left( 
\text{UNIF}\left[ F\left( \widetilde{\theta }\right) \right] ,\widetilde{%
\theta }\right) \diagdown \mathcal{L}_{j}^{Y}\left( \text{UNIF}\left[
F\left( \widetilde{\theta }\right) \right] ,\theta ^{\ast }\right) \text{,}
\end{equation*}%
which, together with (\ref{yuy1-A-B}), implies%
\begin{equation*}
U_{j}^{\theta ^{\ast }}\left[ \psi _{j}^{\widetilde{\theta }}\left( \theta
^{\ast }\right) \right] \geq U_{j}^{\theta ^{\ast }}\left[ y^{\ast }\right]
>U_{j}^{\theta ^{\ast }}\left( \text{UNIF}\left[ F\left( \widetilde{\theta }%
\right) \right] \right) =U_{j}^{\theta ^{\ast }}\left( g\left[ \widetilde{m}%
\right] \right) \text{.}
\end{equation*}%
Therefore, it is strictly profitable for agent $j$ to deviate to%
\begin{equation*}
m_{j}=\left( \theta ^{\ast },k_{j}^{2},\widetilde{k_{j}^{3}},\widetilde{%
\gamma _{j}},\widetilde{b_{j}}\right) \text{ for sufficiently large }%
k_{j}^{2}\text{,}
\end{equation*}%
contradicting $\widetilde{m}\in PNE^{\left( \mathcal{M}^{A\text{-}B},\text{ }%
\theta ^{\ast }\right) }$.

Consider the scenarios in which $\widetilde{m}$ triggers Case (2), i.e.,
there exists $j\in \mathcal{I}$ such that 
\begin{equation*}
\exists \widetilde{\theta }\in \Theta \text{, }\widetilde{m}_{i}=\left( 
\widetilde{\theta _{i}}=\widetilde{\theta },\widetilde{k_{i}^{2}}=1,%
\widetilde{k_{i}^{3}},\widetilde{\gamma _{i}},\widetilde{b_{i}}\right) \text{
, }\forall i\in \mathcal{I\diagdown }\left\{ j\right\} \text{,}
\end{equation*}%
and%
\begin{eqnarray}
g\left[ \widetilde{m}\right] &=&\left( 1-\frac{1}{\widetilde{k_{j}^{2}}}%
\right) \times \psi _{j}^{\widetilde{\theta }}\left( \widetilde{\theta _{j}}%
\right)  \label{yuy4-A-B} \\
&&+\frac{1}{\widetilde{k_{j}^{2}}}\times \left( 
\begin{tabular}{l}
$\varepsilon _{j}^{\widetilde{\theta }}\times \left[ \left( 1-\frac{1}{%
\widetilde{k_{j}^{3}}}\right) \times \gamma _{j}\left( \widehat{\Gamma }%
_{j}^{A\text{-}B}\left( \theta \right) \right) +\frac{1}{\widetilde{k_{j}^{3}%
}}\times \text{UNIF}\left( \widehat{\Gamma }_{j}^{A\text{-}B}\left( \theta
\right) \right) \right] $ \\ 
$+\left( 1-\varepsilon _{j}^{\widetilde{\theta }}\right) \times y_{j}^{%
\widetilde{\theta }}$%
\end{tabular}%
\right) \text{.}  \notag
\end{eqnarray}%
We now prove $g\left[ \widetilde{m}\right] \in \triangle \left[ F\left(
\theta ^{\ast }\right) \right] $. By our construction,%
\begin{equation}
g\left[ \widetilde{m}\right] \in \triangle \left[ \widehat{\Gamma }_{j}^{A%
\text{-}B}\left( \widetilde{\theta }\right) \right] \text{.}
\label{uiu1-A-B}
\end{equation}%
Since every $i\in \mathcal{I\diagdown }\left\{ j\right\} $ can deviate to
trigger Case (3), and dictate her top outcome in $Z^{\ast }$ with
arbitrarily high probability, $\widetilde{m}\in PNE^{\left( \mathcal{M}^{A%
\text{-}B},\text{ }\theta ^{\ast }\right) }$ implies%
\begin{equation}
\widehat{\Gamma }_{j}^{A\text{-}B}\left( \widetilde{\theta }\right) \subset
\arg \max_{z\in Z^{\ast }}u_{i}^{\theta ^{\ast }}\left( z\right) \text{, }%
\forall i\in \mathcal{I}\diagdown \left\{ j\right\} \text{.}
\label{uiu2-A-B}
\end{equation}%
Inside the the compound lottery $g\left[ \widetilde{m}\right] $ in (\ref%
{yuy4-A-B}), conditional on an event with probability $\frac{1}{\widetilde{%
k_{j}^{2}}}\times \varepsilon _{j}^{\widetilde{\theta }}$, we have the
compound lottery 
\begin{equation*}
\left[ \left( 1-\frac{1}{\widetilde{k_{j}^{3}}}\right) \times \widetilde{%
\gamma _{j}}\left( \widehat{\Gamma }_{j}^{A\text{-}B}\left( \widetilde{%
\theta }\right) \right) +\frac{1}{\widetilde{k_{j}^{3}}}\times \text{UNIF}%
\left( \widehat{\Gamma }_{j}^{A\text{-}B}\left( \widetilde{\theta }\right)
\right) \right] \text{,}
\end{equation*}%
and hence, agent $j$ can always deviate to%
\begin{equation*}
m_{j}=\left( \widetilde{\theta _{j}},\widetilde{k_{j}^{2}},k_{j}^{3},\gamma
_{j},\widetilde{b_{j}}\right) _{i\in \mathcal{I\diagdown }\left\{ j\right\} }%
\text{ with }\gamma _{j}\left( \widehat{\Gamma }_{j}^{A\text{-}B}\left( 
\widetilde{\theta }\right) \right) \in \arg \max_{z\in \widehat{\Gamma }%
_{j}^{A\text{-}B}\left( \widetilde{\theta }\right) }u_{j}^{\theta ^{\ast
}}\left( z\right)
\end{equation*}%
for sufficiently large $k_{j}^{3}$. Thus, $\widetilde{m}\in PNE^{\left( 
\mathcal{M}^{A\text{-}B},\text{ }\theta ^{\ast }\right) }$ implies%
\begin{equation}
\widehat{\Gamma }_{j}^{A\text{-}B}\left( \widetilde{\theta }\right) \subset
\arg \max_{z\in \widehat{\Gamma }_{j}^{A\text{-}B}\left( \widetilde{\theta }%
\right) }u_{j}^{\theta ^{\ast }}\left( z\right) \text{.}  \label{uiu3-A-B}
\end{equation}%
(\ref{uiu2-A-B}) and (\ref{uiu3-A-B}) imply that $\widehat{\Gamma }_{j}^{A%
\text{-}B}\left( \widetilde{\theta }\right) $ is a $j$-$Z^{\ast }$-$\theta
^{\ast }$-max set, which together Lemma \ref{lem:no-veto:generalized:SCC},
further implies%
\begin{equation}
\widehat{\Gamma }_{j}^{A\text{-}B}\left( \widetilde{\theta }\right) \subset
F\left( \theta ^{\ast }\right) \text{.}  \label{uiu3a-A-B}
\end{equation}%
(\ref{uiu1-A-B}) and (\ref{uiu3a-A-B}) imply $g\left[ \widetilde{m}\right]
\in \triangle \left[ F\left( \theta ^{\ast }\right) \right] $.

Finally, consider the scenarios in which $\widetilde{m}$ triggers Case (3),
i.e., 
\begin{equation*}
g\left[ \widetilde{m}\right] =\left( 1-\frac{1}{\widetilde{k_{j^{\ast }}^{2}}%
}\right) \times \widetilde{b_{j^{\ast }}}+\frac{1}{\widetilde{k_{j^{\ast
}}^{2}}}\times \text{UNIF}\left( Z^{\ast }\right) \text{,}
\end{equation*}%
where $j^{\ast }=\max \left( \arg \max_{i\in \mathcal{I}}\widetilde{k_{i}^{2}%
}\right) $. Since every $i\in \mathcal{I}$ can increase their integer in the
second dimension and dictate her top outcome in $Z^{\ast }$ with arbitrarily
high probability, $\widetilde{m}\in PNE^{\left( \mathcal{M}^{A\text{-}B},%
\text{ }\theta ^{\ast }\right) }$ implies%
\begin{equation*}
Z^{\ast }\subset \arg \max_{z\in Z^{\ast }}u_{i}^{\theta ^{\ast }}\left(
z\right) \text{, }\forall i\in \mathcal{I}\text{,}
\end{equation*}%
i.e., $Z^{\ast }$ is a is a $j$-$Z^{\ast }$-$\theta ^{\ast }$-max set, which
together Lemma \ref{lem:no-veto:generalized:SCC}, further implies%
\begin{equation*}
Z^{\ast }\subset F\left( \theta ^{\ast }\right) \text{.}
\end{equation*}%
Therefore, $g\left[ \widetilde{m}\right] \in \triangle \left( Z^{\ast
}\right) \subset \triangle \left[ F\left( \theta ^{\ast }\right) \right] $.$%
\blacksquare $

\subsection{Proof of Theorem \protect\ref{thm:pure:SCC}}

\label{sec:thm:pure:SCC}

By their definitions, we have (iii)$\Longrightarrow $(i). Theorem \ref%
{theorem:full:mix:SCC-A} implies (ii)$\Longrightarrow $(iii). We will show
(i)$\Longrightarrow $(ii) to complete the proof. We need two additional
lemmas, before proving "(i)$\Longrightarrow $(ii)."

\begin{lemma}
\label{lem:mixed:Z*:pure}Suppose that an SCC $F$ is pure-Nash-implementable
by $\mathcal{M}=\left\langle M\text{, \ }g:M\longrightarrow Y\right\rangle $%
. We have $g\left( M\right) \subset \triangle \left( Z^{\ast }\right) $.
\end{lemma}

\noindent \textbf{Proof of Lemma \ref{lem:mixed:Z*:pure}:} Suppose that an
SCC $F$ is pure-Nash-implemented by $\mathcal{M}=\left\langle M\text{, \ }%
g:M\longrightarrow Y\right\rangle $. We consider two scenarios: (I) $Z$ is
not an $i$-max set for any $i\in \mathcal{I}$ and (II) $Z$ is an $i$-max set
for some $i\in \mathcal{I}$. In scenario (I), we have $Z^{\ast }=Z$ by (\ref%
{tth2}), and hence $g\left( M\right) \subset \triangle \left( Z\right)
=\triangle \left( Z^{\ast }\right) $. In scenario (II), we have $Z^{\ast
}=\cup _{\theta \in \Theta }F\left( \theta \right) $ by (\ref{tth2}). $Z$
being an $i$-max set implies existence of some state $\theta ^{\prime }\in
\Theta $ such that all agents are indifferent between any two elements in $Z$
at $\theta ^{\prime }$. Therefore, every $m\in M$ is a pure-strategy Nash
equilibrium at $\theta ^{\prime }$, and hence%
\begin{equation*}
\dbigcup\limits_{m\in M}\text{SUPP}\left[ g\left( m\right) \right] \subset
F\left( \theta ^{\prime }\right) \subset \cup _{\theta \in \Theta }F\left(
\theta \right) =Z^{\ast }\text{,}
\end{equation*}%
i.e., $g\left( M\right) \subset \triangle \left( Z^{\ast }\right) $.$%
\blacksquare $

\begin{lemma}
\label{lem:mixed:unilateral-deviation:pure}Suppose that an SCC $F$ is
pure-Nash-implemented by $\mathcal{M}=\left\langle M\text{, \ }%
g:M\longrightarrow Y\right\rangle $. For any $\left( i,\theta \right) \in 
\mathcal{I}\times \Theta $ and any $\lambda \in PNE^{\left( \mathcal{M},%
\text{ }\theta \right) }$, we have%
\begin{multline*}
\left( 
\begin{array}{c}
F\left( \theta \right) \subset \arg \min_{z\in Z^{\ast }}u_{i}^{\theta
}\left( z\right) \text{,} \\ 
\Xi _{i}\left( \theta \right) \neq \varnothing \text{ and} \\ 
\text{and }Z^{\ast }\cap \mathcal{L}_{i}^{Z}\left( F\left( \theta \right)
,\theta \right) \text{ is an }i\text{-}Z^{\ast }\text{-max set}%
\end{array}%
\right) \\
\Longrightarrow \dbigcup\limits_{m_{i}\in M_{i}}\text{SUPP}\left[ g\left(
m_{i},\lambda _{-i}\right) \right] \subset \left[ Z^{\ast }\cap \mathcal{L}%
_{i}^{Z}\left( F\left( \theta \right) ,\theta \right) \cap \left(
\dbigcup\limits_{E\in \Xi _{i}\left( \theta \right) }\dbigcap\limits_{\theta
^{\prime }\in E}F\left( \theta ^{\prime }\right) \right) \right] \text{.}
\end{multline*}
\end{lemma}

\noindent \textbf{Proof of Lemma \ref{lem:mixed:unilateral-deviation:pure}:}
Suppose that an SCC $F$ is pure-Nash-implemented by $\mathcal{M}%
=\left\langle M\text{, \ }g:M\longrightarrow Y\right\rangle $. Fix any $%
\left( i,\theta \right) \in \mathcal{I}\times \Theta $ and any $\lambda \in
PNE^{\left( \mathcal{M},\text{ }\theta \right) }$ such that%
\begin{gather*}
F\left( \theta \right) \subset \arg \min_{z\in Z^{\ast }}u_{i}^{\theta
}\left( z\right) \text{,} \\
\Xi _{i}\left( \theta \right) \neq \varnothing \text{,} \\
\text{and }Z^{\ast }\cap \mathcal{L}_{i}^{Z}\left( F\left( \theta \right)
,\theta \right) \text{ is an }i\text{-}Z^{\ast }\text{-max set.}
\end{gather*}

First, with $Z^{\ast }\cap \mathcal{L}_{i}^{Z}\left( F\left( \theta \right)
,\theta \right) $ being an $i$-$Z^{\ast }$-max set, there exists $\widetilde{%
\theta }\in \Theta $ such that $Z^{\ast }\cap \mathcal{L}_{i}^{Z}\left(
F\left( \theta \right) ,\theta \right) $ is an $i$-$Z^{\ast }$-$\widetilde{%
\theta }$-max set, i.e.,%
\begin{eqnarray*}
F\left( \theta \right) &\subset &Z^{\ast }\cap \mathcal{L}_{i}^{Z}\left(
F\left( \theta \right) ,\theta \right) \subset \arg \max_{z\in Z^{\ast
}}u_{j}^{\widetilde{\theta }}\left( z\right) \text{, }\forall j\in \mathcal{I%
}\diagdown \left\{ i\right\} \text{,} \\
F\left( \theta \right) &\subset &Z^{\ast }\cap \mathcal{L}_{i}^{Z}\left(
F\left( \theta \right) ,\theta \right) \subset \arg \max_{z\in Z^{\ast }\cap 
\mathcal{L}_{i}^{Z}\left( F\left( \theta \right) ,\theta \right) }u_{i}^{%
\widetilde{\theta }}\left( z\right) \text{,}
\end{eqnarray*}%
which immediately implies $PNE^{\left( \mathcal{M},\text{ }\theta \right)
}\subset PNE^{\left( \mathcal{M},\text{ }\widetilde{\theta }\right) }$, and
hence, $F\left( \theta \right) \subset F\left( \widetilde{\theta }\right) $.
Therefore, we have%
\begin{equation}
\left( 
\begin{array}{c}
\widetilde{\theta }\in \Theta _{i}^{\theta }\neq \varnothing \text{,} \\ 
\text{and }Z^{\ast }\cap \mathcal{L}_{i}^{Z}\left( F\left( \theta \right)
,\theta \right) \text{ is an }i\text{-}Z^{\ast }\text{-}\widetilde{\theta }%
\text{-max set}%
\end{array}%
\right) \text{.}  \label{hhty3}
\end{equation}

Second, we define an$\text{ algorithm.}$%
\begin{equation}
\left( 
\begin{array}{c}
\text{Step }1\text{: let }K^{1}\equiv \left\{ \theta ^{\prime }\in \Theta
_{i}^{\theta }:Z^{\ast }\cap \mathcal{L}_{i}^{Z}\left( F\left( \theta
\right) ,\theta \right) \text{ is an }i\text{-}Z^{\ast }\text{-}\theta
^{\prime }\text{-max set}\right\} \text{,} \\ 
\text{Step }2\text{: let }K^{2}\equiv \left\{ \theta ^{\prime }\in \Theta
_{i}^{\theta }:Z^{\ast }\cap \mathcal{L}_{i}^{Z}\left( F\left( \theta
\right) ,\theta \right) \cap \left[ \dbigcap\limits_{\theta ^{\prime }\in
K^{1}}F\left( \theta ^{\prime }\right) \right] \text{ is an }i\text{-}%
Z^{\ast }\text{-}\theta ^{\prime }\text{-max set}\right\} \text{,} \\ 
... \\ 
\text{Step }n\text{: let }K^{n}\equiv \left\{ \theta ^{\prime }\in \Theta
_{i}^{\theta }:Z^{\ast }\cap \mathcal{L}_{i}^{Z}\left( F\left( \theta
\right) ,\theta \right) \cap \left[ \dbigcap\limits_{\theta ^{\prime }\in
K^{n-1}}F\left( \theta ^{\prime }\right) \right] \text{ is an }i\text{-}%
Z^{\ast }\text{-}\theta ^{\prime }\text{-max set}\right\} \text{,} \\ 
...%
\end{array}%
\right)  \label{tpt3a}
\end{equation}%
(\ref{hhty3}) implies $\widetilde{\theta }\in K^{1}\neq \varnothing $.
Furthermore, inductively, it is easy to show 
\begin{eqnarray}
\forall n &\geq &1\text{,}  \notag \\
K^{n} &\subset &K^{n+1}\text{,}  \notag \\
Z^{\ast }\cap \mathcal{L}_{i}^{Z}\left( F\left( \theta \right) ,\theta
\right) \cap \left[ \dbigcap\limits_{\theta ^{\prime }\in K^{n}}F\left(
\theta ^{\prime }\right) \right] &\supset &Z^{\ast }\cap \mathcal{L}%
_{i}^{Z}\left( F\left( \theta \right) ,\theta \right) \cap \left[
\dbigcap\limits_{\theta ^{\prime }\in K^{n+1}}F\left( \theta ^{\prime
}\right) \right] \text{.}  \label{tpt2a}
\end{eqnarray}%
$\text{Since }Z$ is finite, (\ref{tpt2a}) implies that there exists $n\geq 1$
such that%
\begin{equation}
Z^{\ast }\cap \mathcal{L}_{i}^{Z}\left( F\left( \theta \right) ,\theta
\right) \cap \left[ \dbigcap\limits_{\theta ^{\prime }\in K^{n}}F\left(
\theta ^{\prime }\right) \right] =Z^{\ast }\cap \mathcal{L}_{i}^{Z}\left(
F\left( \theta \right) ,\theta \right) \cap \left[ \dbigcap\limits_{\theta
^{\prime }\in K^{n+1}}F\left( \theta ^{\prime }\right) \right] \text{.}
\label{tpt1a}
\end{equation}%
and as a result,%
\begin{eqnarray*}
K^{n+1} &\equiv &\left\{ \theta ^{\prime }\in \Theta _{i}^{\theta }:Z^{\ast
}\cap \mathcal{L}_{i}^{Z}\left( F\left( \theta \right) ,\theta \right) \cap %
\left[ \dbigcap\limits_{\theta ^{\prime }\in K^{n}}F\left( \theta ^{\prime
}\right) \right] \text{ is an }i\text{-}Z^{\ast }\text{-}\theta ^{\prime }%
\text{-max set}\right\} \\
&=&\left\{ \theta ^{\prime }\in \Theta _{i}^{\theta }:Z^{\ast }\cap \mathcal{%
L}_{i}^{Z}\left( F\left( \theta \right) ,\theta \right) \cap \left[
\dbigcap\limits_{\theta ^{\prime }\in K^{n+1}}F\left( \theta ^{\prime
}\right) \right] \text{ is an }i\text{-}Z^{\ast }\text{-}\theta ^{\prime }%
\text{-max set}\right\} \text{,}
\end{eqnarray*}%
where the second inequality follows from (\ref{tpt1a}). Therefore, $%
\varnothing \neq K^{1}\subset K^{n+1}$ and%
\begin{equation}
K^{n+1}\in \Xi _{i}\left( \theta \right) \neq \varnothing \text{.}
\label{tptt1aa}
\end{equation}

Third, we aim to prove%
\begin{equation}
\dbigcup\limits_{m_{i}\in M_{i}}\text{SUPP}\left[ g\left( m_{i},\lambda
_{-i}\right) \right] \subset \left[ Z^{\ast }\cap \mathcal{L}_{i}^{Z}\left(
F\left( \theta \right) ,\theta \right) \cap \left( \dbigcup\limits_{K\in \Xi
_{i}\left( \theta \right) }\dbigcap\limits_{\theta ^{\prime }\in K}F\left(
\theta ^{\prime }\right) \right) \right] \text{.}  \label{tpt7a}
\end{equation}%
(\ref{tptt1aa}) and (\ref{tpt7a}) implies that it suffices to prove%
\begin{equation}
\dbigcup\limits_{m_{i}\in M_{i}}\text{SUPP}\left[ g\left( m_{i},\lambda
_{-i}\right) \right] \subset \left[ Z^{\ast }\cap \mathcal{L}_{i}^{Z}\left(
F\left( \theta \right) ,\theta \right) \cap \left( \dbigcap\limits_{\theta
^{\prime }\in K^{n+1}}F\left( \theta ^{\prime }\right) \right) \right] \text{%
.}  \label{tptt1b}
\end{equation}%
We prove (\ref{tptt1b}) inductively. First, 
\begin{equation*}
K^{1}\equiv \left\{ \theta ^{\prime }\in \Theta _{i}^{\theta }:Z^{\ast }\cap 
\mathcal{L}_{i}^{Z}\left( F\left( \theta \right) ,\theta \right) \text{ is
an }i\text{-}Z^{\ast }\text{-}\theta ^{\prime }\text{-max set}\right\} \text{%
,}
\end{equation*}%
and as a result,%
\begin{equation*}
g\left( m_{i},\lambda _{-i}\right) \in PNE^{\left( \mathcal{M},\text{ }%
\theta ^{\prime }\right) }\text{, }\forall m_{i}\in M_{i}\text{, }\forall
\theta ^{\prime }\in K^{1}\text{,}
\end{equation*}%
and hence,%
\begin{equation*}
g\left( m_{i},\lambda _{-i}\right) \in \triangle \left[ F\left( \theta
^{\prime }\right) \right] \text{, }\forall m_{i}\in M_{i}\text{, }\forall
\theta ^{\prime }\in K^{1}\text{,}
\end{equation*}%
which, together with Lemma \ref{lem:Z-A}, implies%
\begin{equation*}
\dbigcup\limits_{m_{i}\in M_{i}}\text{SUPP}\left[ g\left( m_{i},\lambda
_{-i}\right) \right] \subset \left[ Z^{\ast }\cap \mathcal{L}_{i}^{Z}\left(
F\left( \theta \right) ,\theta \right) \cap \left( \dbigcap\limits_{\theta
^{\prime }\in K^{1}}F\left( \theta ^{\prime }\right) \right) \right] \text{.}
\end{equation*}%
This completes the first step of the induction.

Suppose that for some $t<k+1$, we have proved%
\begin{equation*}
\dbigcup\limits_{m_{i}\in M_{i}}\text{SUPP}\left[ g\left( m_{i},\lambda
_{-i}\right) \right] \subset \left[ Z^{\ast }\cap \mathcal{L}_{i}^{Z}\left(
F\left( \theta \right) ,\theta \right) \cap \left( \dbigcap\limits_{\theta
^{\prime }\in K^{t}}F\left( \theta ^{\prime }\right) \right) \right] \text{.}
\end{equation*}%
Consider 
\begin{equation*}
K^{t+1}\equiv \left\{ \theta ^{\prime }\in \Theta _{i}^{\theta }:Z^{\ast
}\cap \mathcal{L}_{i}^{Z}\left( F\left( \theta \right) ,\theta \right) \cap %
\left[ \dbigcap\limits_{\theta ^{\prime }\in K^{t}}F\left( \theta ^{\prime
}\right) \right] \text{ is an }i\text{-}Z^{\ast }\text{-}\theta ^{\prime }%
\text{-max set}\right\} \text{,}
\end{equation*}%
and as a result,%
\begin{equation*}
g\left( m_{i},\lambda _{-i}\right) \in PNE^{\left( \mathcal{M},\text{ }%
\theta ^{\prime }\right) }\text{, }\forall m_{i}\in M_{i}\text{, }\forall
\theta ^{\prime }\in K^{t+1}\text{,}
\end{equation*}%
and hence,%
\begin{equation*}
g\left( m_{i},\lambda _{-i}\right) \in \triangle \left[ F\left( \theta
^{\prime }\right) \right] \text{, }\forall m_{i}\in M_{i}\text{, }\forall
\theta ^{\prime }\in K^{t+1}\text{,}
\end{equation*}%
which, together with Lemma \ref{lem:Z-A}, implies%
\begin{equation*}
\dbigcup\limits_{m_{i}\in M_{i}}\text{SUPP}\left[ g\left( m_{i},\lambda
_{-i}\right) \right] \subset \left[ Z^{\ast }\cap \mathcal{L}_{i}^{Z}\left(
F\left( \theta \right) ,\theta \right) \cap \left( \dbigcap\limits_{\theta
^{\prime }\in K^{t+1}}F\left( \theta ^{\prime }\right) \right) \right] \text{%
.}
\end{equation*}%
This completes the induction process, and proves (\ref{tptt1b}).$%
\blacksquare $

\noindent \textbf{Proof of Theorem "(i)}$\Longrightarrow $\textbf{(ii)" in
Theorem \ref{thm:pure:SCC}:} Suppose that an SCC $F$ is
pure-Nash-implemented by $\mathcal{M}=\left\langle M\text{, \ }%
g:M\longrightarrow Y\right\rangle $. By Theorem \ref{theorem:full:mix:SCC-A}%
, it suffices to show $\widehat{\mathcal{L}}^{Y\text{-}A\text{-}B}$%
-uniform-monotonicity. Fix any $\left( \theta ,\theta ^{\prime }\right) \in
\Theta \times \Theta $ such that%
\begin{equation}
\left[ 
\begin{array}{c}
\widehat{\mathcal{L}}_{i}^{Y\text{-}A\text{-}B}\left( \text{UNIF}\left[
F\left( \theta \right) \right] ,\theta \right) \subset \mathcal{L}%
_{i}^{Y}\left( \text{UNIF}\left[ F\left( \theta \right) \right] ,\theta
^{\prime }\right) \text{, } \\ 
\forall i\in \mathcal{I}%
\end{array}%
\right] \text{ ,}  \label{tth7-p}
\end{equation}%
and we aim to show $F\left( \theta \right) \subset F\left( \theta ^{\prime
}\right) $, i.e., $\widehat{\mathcal{L}}^{Y\text{-}A\text{-}B}$%
-uniform-monotonicity holds. In particular, fix any $a\in F\left( \theta
\right) $, and we aim to show $a\in F\left( \theta ^{\prime }\right) $.
Since $a\in F\left( \theta \right) $, there exists some $\lambda \in
PNE^{\left( \mathcal{M},\text{ }\theta \right) }$, such that $a$ is induced
with positive probability by $\lambda $. Thus, SUPP$\left[ g\left( \lambda
\right) \right] \subset F\left( \theta \right) $. We now prove $\lambda \in
PNE^{\left( \mathcal{M},\text{ }\theta ^{\prime }\right) }$ by
contradiction, which further implies $a\in F\left( \theta ^{\prime }\right) $%
. Suppose $\lambda \notin PNE^{\left( \mathcal{M},\text{ }\theta ^{\prime
}\right) }$, i.e., there exist $i\in \mathcal{I}$, and $m_{i}^{\prime }\in
M_{i}$ such that%
\begin{eqnarray}
U_{i}^{\theta }\left[ g\left( m_{i}^{\prime },\lambda _{-i}\right) \right]
&\leq &U_{i}^{\theta }\left[ g\left( \lambda \right) \right] \text{,}
\label{ddy1} \\
U_{i}^{\theta ^{\prime }}\left[ g\left( m_{i}^{\prime },\lambda _{-i}\right) %
\right] &>&U_{i}^{\theta ^{\prime }}\left[ g\left( \lambda \right) \right] 
\text{.}  \label{ddy2}
\end{eqnarray}%
Lemma \ref{lem:mixed-lottery} and SUPP$\left[ g\left( \lambda \right) \right]
\subset F\left( \theta \right) $ imply existence of $\mu \in \triangle \left[
F\left( \theta \right) \right] $ and $\beta \in \left[ 0,1\right] $ such that%
\begin{equation}
\text{UNIF}\left[ F\left( \theta \right) \right] =\beta \times g\left(
\lambda \right) +\left( 1-\beta \right) \times \mu \text{.}  \label{ddy3}
\end{equation}%
Thus, (\ref{ddy2}) and (\ref{ddy3})\ imply%
\begin{equation}
U_{i}^{\theta ^{\prime }}\left[ \beta \times g\left( m_{i}^{\prime },\lambda
_{-i}\right) +\left( 1-\beta \right) \times \mu \right] >U_{i}^{\theta
^{\prime }}\left[ \beta \times g\left( \lambda \right) +\left( 1-\beta
\right) \times \mu \right] =U_{i}^{\theta ^{\prime }}\left( \text{UNIF}\left[
F\left( \theta \right) \right] \right) \text{.}  \label{ddy4}
\end{equation}%
(\ref{ddy1}) and (\ref{ddy3})\ imply%
\begin{equation}
U_{i}^{\theta }\left[ \beta \times g\left( m_{i}^{\prime },\lambda
_{-i}\right) +\left( 1-\beta \right) \times \mu \right] \leq U_{i}^{\theta }%
\left[ \beta \times g\left( \lambda \right) +\left( 1-\beta \right) \times
\mu \right] =U_{i}^{\theta }\left( \text{UNIF}\left[ F\left( \theta \right) %
\right] \right) \text{.}  \label{ddy5}
\end{equation}

We now consider two cases. First, suppose%
\begin{equation}
\left( 
\begin{tabular}{l}
$F\left( \theta \right) \subset \arg \min_{z\in Z^{\ast }}u_{i}^{\theta
}\left( z\right) $, \\ 
$\Xi _{i}\left( \theta \right) \neq \varnothing $, \\ 
and $Z^{\ast }\cap \mathcal{L}_{i}^{Z}\left( F\left( \theta \right) ,\theta
\right) $ is an $i\text{-}Z^{\ast }\text{-max set}$%
\end{tabular}%
\right) \text{,}  \label{ddy6}
\end{equation}%
holds. Thus, by (\ref{ddtt}), we have%
\begin{equation*}
\widehat{\mathcal{L}}_{i}^{Y\text{-}A\text{-}B}\left( \text{UNIF}\left[
F\left( \theta \right) \right] ,\theta \right) =\triangle \left[ Z^{\ast
}\cap \mathcal{L}_{i}^{Z}\left( F\left( \theta \right) ,\theta \right) \cap
\left( \dbigcup\limits_{E\in \Xi _{i}\left( \theta \right) }\dbigcap\limits_{%
\widetilde{\theta }\in E}F\left( \widetilde{\theta }\right) \right) \right] 
\text{,}
\end{equation*}%
and by Lemma \ref{lem:mixed:unilateral-deviation:pure}, we have 
\begin{equation*}
g\left( m_{i}^{\prime },\lambda _{-i}\right) \in \triangle \left[ Z^{\ast
}\cap \mathcal{L}_{i}^{Z}\left( F\left( \theta \right) ,\theta \right) \cap
\left( \dbigcup\limits_{E\in \Xi _{i}\left( \theta \right) }\dbigcap\limits_{%
\widetilde{\theta }\in E}F\left( \widetilde{\theta }\right) \right) \right] 
\text{,}
\end{equation*}%
which, together with $\mu \in \triangle \left[ F\left( \theta \right) \right]
$, implies%
\begin{equation}
\beta \times g\left( m_{i}^{\prime },\lambda _{-i}\right) +\left( 1-\beta
\right) \times \mu \in \widehat{\mathcal{L}}_{i}^{Y\text{-}A\text{-}B}\left( 
\text{UNIF}\left[ F\left( \theta \right) \right] ,\theta \right) \text{.}
\label{ddy7}
\end{equation}%
(\ref{tth7-p}) and (\ref{ddy7}) imply%
\begin{equation*}
\beta \times g\left( m_{i}^{\prime },\lambda _{-i}\right) +\left( 1-\beta
\right) \times \mu \in \mathcal{L}_{i}^{Y}\left( \text{UNIF}\left[ F\left(
\theta \right) \right] ,\theta ^{\prime }\right) \text{.}
\end{equation*}%
contradicting (\ref{ddy4}).

Second, suppose that (\ref{ddy6}) does not hold, and by (\ref{ddtt}), we have%
\begin{equation*}
\widehat{\mathcal{L}}_{i}^{Y\text{-}A\text{-}B}\left( \text{UNIF}\left[
F\left( \theta \right) \right] ,\theta \right) =\left[ \triangle \left(
Z^{\ast }\right) \right] \cap \mathcal{L}_{i}^{Y}\left( \text{UNIF}\left[
F\left( \theta \right) \right] ,\theta \right) \text{,}
\end{equation*}%
which, together with (\ref{ddy5}), implies%
\begin{eqnarray}
\beta \times g\left( m_{i}^{\prime },\lambda _{-i}\right) +\left( 1-\beta
\right) \times \mu &\subset &\left[ \triangle \left( Z^{\ast }\right) \right]
\cap \mathcal{L}_{i}^{Y}\left( \text{UNIF}\left[ F\left( \theta \right) %
\right] ,\theta \right)  \notag \\
&=&\widehat{\mathcal{L}}_{i}^{Y\text{-}A\text{-}B}\left( \text{UNIF}\left[
F\left( \theta \right) \right] ,\theta \right) \text{.}  \label{ddy8}
\end{eqnarray}%
(\ref{tth7-p}) and (\ref{ddy8}) imply%
\begin{equation*}
\beta \times g\left( m_{i}^{\prime },\lambda _{-i}\right) +\left( 1-\beta
\right) \times \mu \in \mathcal{L}_{i}^{Y}\left( \text{UNIF}\left[ F\left(
\theta \right) \right] ,\theta ^{\prime }\right) \text{.}
\end{equation*}%
contradicting (\ref{ddy4}).$\blacksquare $

\subsection{Proof of Lemma \protect\ref{lem:mixed-lottery:lower-contour}}

\label{sec:lem:mixed-lottery:lower-contour}

The "$\Rightarrow $" direction in (\ref{kkg1a}) is trivial, and we prove the
"$\Leftarrow $" direction by contradiction. Suppose $\mathcal{L}%
_{i}^{Y}\left( \gamma ,\theta \right) \subset \mathcal{L}_{i}^{Y}\left(
\gamma ,\theta ^{\prime }\right) $, and for some $\eta \in \triangle \left[ E%
\right] $,%
\begin{equation*}
\exists y^{\ast }\in \mathcal{L}_{i}^{Y}\left( \eta ,\theta \right)
\diagdown \mathcal{L}_{i}^{Y}\left( \eta ,\theta ^{\prime }\right) \text{,}
\end{equation*}%
or equivalently,%
\begin{eqnarray}
U_{i}^{\theta }\left[ y^{\ast }\right] &\leq &U_{i}^{\theta }\left[ \eta %
\right] \text{,}  \label{eey1} \\
U_{i}^{\theta ^{\prime }}\left[ y^{\ast }\right] &>&U_{i}^{\theta ^{\prime }}%
\left[ \eta \right] \text{.}  \label{eey2}
\end{eqnarray}%
By Lemma \ref{lem:mixed-lottery}, there exist $\beta \in \left( 0,1\right) $
and $\mu \in \triangle \left[ E\right] $ such that%
\begin{equation}
\gamma =\beta \times \eta +\left( 1-\beta \right) \times \mu \text{.}
\label{eey3}
\end{equation}%
Thus, (\ref{eey1}), (\ref{eey2}) and (\ref{eey3}) imply%
\begin{eqnarray*}
U_{i}^{\theta }\left[ \beta \times y^{\ast }+\left( 1-\beta \right) \times
\mu \right] &\leq &U_{i}^{\theta }\left[ \beta \times \eta +\left( 1-\beta
\right) \times \mu \right] =U_{i}^{\theta }\left[ \gamma \right] \text{,} \\
U_{i}^{\theta ^{\prime }}\left[ \beta \times y^{\ast }+\left( 1-\beta
\right) \times \mu \right] &>&U_{i}^{\theta ^{\prime }}\left[ \beta \times
\eta +\left( 1-\beta \right) \times \mu \right] =U_{i}^{\theta ^{\prime }}%
\left[ \gamma \right] \text{,}
\end{eqnarray*}%
or equivalently,%
\begin{equation*}
\left[ \beta \times y^{\ast }+\left( 1-\beta \right) \times \mu \right] \in 
\mathcal{L}_{i}^{Y}\left( \gamma ,\theta \right) \diagdown \mathcal{L}%
_{i}^{Y}\left( \gamma ,\theta ^{\prime }\right) \text{,}
\end{equation*}%
contradicting $\mathcal{L}_{i}^{Y}\left( \gamma ,\theta \right) \subset 
\mathcal{L}_{i}^{Y}\left( \gamma ,\theta ^{\prime }\right) $.$\blacksquare $

\subsection{Proof of Theorem \protect\ref{theorem:MR:iff}}

\label{sec:theorem:MR:iff}

Clearly, "(i)$\Longleftrightarrow $(ii)" is implied by Theorem \ref%
{theorem:full:mix:SCC-A} and Lemma \ref{lem:set-monotone:strong}. "(iii)$%
\Longrightarrow $(ii)" is implied by $\cup _{\theta \in \Theta ^{\ast
}}\Omega ^{\left[ \succeq ^{\theta },\text{ }\mathbb{Q}\right] }\subset \cup
_{\theta \in \Theta ^{\ast }}\Omega ^{\left[ \succeq ^{\theta },\text{ }%
\mathbb{R}
\right] }$. We prove "(ii)$\Longrightarrow $(iii)" by combining the
techniques developed in both \cite{cmlr} and this paper. Suppose $F\ $is
mixed-Nash-A-implementable on $\Theta \equiv \cup _{\theta \in \Theta ^{\ast
}}\Omega ^{\left[ \succeq ^{\theta },\text{ }\mathbb{Q}\right] }$. We aim to
show $F\ $is mixed-Nash-A-implementable on $\widetilde{\Theta }\equiv \cup
_{\theta \in \Theta ^{\ast }}\Omega ^{\left[ \succeq ^{\theta },\text{ }%
\mathbb{R}
\right] }$.

We need four constructions before we can define a canonical mechanism to
implement $F$. First, by Theorem \ref{theorem:full:mix:SCC-A}, $\widehat{%
\mathcal{L}}^{Y\text{-}A\text{-}B}$-uniform-monotonicity holds on $\Theta
\equiv \cup _{\theta \in \Theta ^{\ast }}\Omega ^{\left[ \succeq ^{\theta },%
\text{ }\mathbb{Q}\right] }$. The following result is adapted from Lemma \ref%
{lem:no-veto:generalized:SCC}.

\begin{lemma}
\label{lem:no-veto:generalized:SCC:ordinal}Suppose that $\widehat{\mathcal{L}%
}^{Y\text{-}A\text{-}B}$-monotonicity holds on $\Theta \equiv \cup _{\theta
\in \Theta ^{\ast }}\Omega ^{\left[ \succeq ^{\theta },\text{ }\mathbb{Q}%
\right] }$. We have%
\begin{equation*}
\left[ Z^{\ast }\text{ is a }j\text{-}Z^{\ast }\text{-}\theta ^{\prime }%
\text{-max set}\right] \Longrightarrow Z^{\ast }\subset F\left( \theta
^{\prime }\right) \text{, }\forall \left( j,\theta ^{\prime }\right) \in 
\mathcal{I}\times \Theta ^{\ast }\text{,}
\end{equation*}%
\begin{equation*}
\text{and }\left[ \widehat{\Gamma }_{j}^{A\text{-}B}\left( \theta \right) 
\text{ is a }j\text{-}Z^{\ast }\text{-}\theta ^{\prime }\text{-max set}%
\right] \Longrightarrow \widehat{\Gamma }_{j}^{A\text{-}B}\left( \theta
\right) \subset F\left( \theta ^{\prime }\right) \text{, }\forall \left(
j,\theta ,\theta ^{\prime }\right) \in \mathcal{I}\times \Theta ^{\ast
}\times \Theta ^{\ast }\text{,}
\end{equation*}%
\begin{equation*}
\text{where }\widehat{\Gamma }_{j}^{A\text{-}B}\left( \theta \right) \equiv
\left( \dbigcup\limits_{y\in \widehat{\mathcal{L}}_{j}^{Y\text{-}A\text{-}%
B}\left( \text{UNIF}\left[ F\left( \theta \right) \right] ,u^{\theta
}\right) }\text{SUPP}\left[ y\right] \right) \text{ for any }u^{\theta }\in
\Omega ^{\left[ \succeq ^{\theta },\text{ }\mathbb{Q}\right] }\text{.}
\end{equation*}
\end{lemma}

The proof of Lemma \ref{lem:no-veto:generalized:SCC:ordinal} is the same as
that of Lemma \ref{lem:no-veto:generalized:SCC}, and we omit it. The
intuition is that "$j$-$Z^{\ast }$-$\theta ^{\prime }$-max set" and "$%
\widehat{\Gamma }_{j}^{A\text{-}B}\left( \theta \right) $" are ordinal
notions (i.e., they denpend on ordinal states only).

Second, we define a matric on $\widetilde{\Theta }\equiv \cup _{\theta \in
\Theta ^{\ast }}\Omega ^{\left[ \succeq ^{\theta },\text{ }%
\mathbb{R}
\right] }$:%
\begin{equation*}
\rho \left( u,\widehat{u}\right) =\max_{\left( i,z\right) \in \mathcal{I}%
\times Z^{\ast }}\left\vert u_{i}\left( z\right) -\widehat{u}_{i}\left(
z\right) \right\vert \text{, }\forall \left( u,\widehat{u}\right) \in 
\widetilde{\Theta }\times \widetilde{\Theta }\text{.}
\end{equation*}%
Furthermore, consider any $u\in \cup _{\theta \in \Theta ^{\ast }}\Omega ^{%
\left[ \succeq ^{\theta },\text{ }%
\mathbb{R}
\right] }$ such that 
\begin{equation*}
\max_{i\in \mathcal{I}}\left[ \max_{z\in Z^{\ast }}u_{i}^{\theta }\left(
z\right) -\min_{z\in Z^{\ast }}u_{i}^{\theta }\left( z\right) \right] >0%
\text{,}
\end{equation*}%
denote%
\begin{equation*}
\gamma ^{u}\equiv \min_{\left\{ \left( i,a,b\right) \in \mathcal{I}\times
Z^{\ast }\times Z^{\ast }:u_{i}\left( a\right) \neq u_{i}\left( b\right)
\right\} }\left\vert u_{i}\left( a\right) -u_{i}\left( b\right) \right\vert
>0\text{, }\forall u\in \cup _{\theta \in \Theta ^{\ast }}\Omega ^{\left[
\succeq ^{\theta },\text{ }%
\mathbb{R}
\right] }\text{.}
\end{equation*}%
We need the following two lemmas to complete the proof, which describes how
to use cardinal states in $\Omega ^{\left[ \succeq ^{\theta },\text{ }%
\mathbb{Q}\right] }$ to approximate cardinal states in $\Omega ^{\left[
\succeq ^{\theta },\text{ }%
\mathbb{R}
\right] }$.

\begin{lemma}
\label{lem:ordinal:approximate1}For each $\theta \in \Theta ^{\ast }$ and
each $\widehat{u}^{\theta }\in \Omega ^{\left[ \succeq ^{\theta },\text{ }%
\mathbb{Q}\right] }$, there exists $\varsigma ^{\widehat{u}^{\theta }}\in
\left( 0,1\right) $ such that for any $\left( i,x,y\right) \in \mathcal{I}%
\times Z^{\ast }\times Z^{\ast }$ and any $u^{\theta }\in \Omega ^{\left[
\succeq ^{\theta },\text{ }%
\mathbb{R}
\right] }$, 
\begin{eqnarray}
&&\left( 
\begin{array}{c}
\widehat{u}_{i}^{\theta }\left( x\right) -\widehat{u}_{i}^{\theta }\left(
y\right) >0\text{,} \\ 
\rho \left( \widehat{u}^{\theta },u^{\theta }\right) <\frac{1}{3}\times
\gamma ^{\widehat{u}^{\theta }}%
\end{array}%
\right)  \label{nnb1a} \\
&\Longrightarrow &u_{i}^{\theta }\left( x\right) -\left[ \left( 1-\varsigma
^{\widehat{u}^{\theta }}\right) u_{i}^{\theta }\left( y\right) +\varsigma ^{%
\widehat{u}^{\theta }}\max_{z\in Z^{\ast }}u_{i}^{\theta }\left( z\right)
+\varsigma ^{\widehat{u}^{\theta }}\left( \left\vert Z^{\ast }\right\vert
-1\right) \left( \max_{z\in Z^{\ast }}u_{i}^{\theta }\left( z\right)
-\min_{z\in Z^{\ast }}u_{i}^{\theta }\left( z\right) \right) \right] >0\text{%
.}  \notag
\end{eqnarray}
\end{lemma}

\begin{lemma}
\label{lem:perturb}For each $\left( \theta ,j\right) \in \Theta ^{\ast
}\times \mathcal{I}$ and each $\widehat{u}^{\theta }\in \Omega ^{\left[
\succeq ^{\theta },\text{ }\mathbb{Q}\right] }$, there exist%
\begin{equation*}
\varepsilon _{j}^{\widehat{u}^{\theta }}>0\text{ and }y_{j}^{\widehat{u}%
^{\theta }}\in \widehat{\mathcal{L}}_{j}^{Y\text{-}A\text{-}B}\left( \text{%
UNIF}\left[ F\left( \theta \right) \right] ,\widehat{u}^{\theta }\right) 
\text{,}
\end{equation*}%
such that for any $u^{\theta }\in \Omega ^{\left[ \succeq ^{\theta },\text{ }%
\mathbb{R}
\right] }$, 
\begin{eqnarray}
&&\left( 
\begin{array}{c}
\left[ \max_{z\in Z^{\ast }}\widehat{u}_{j}^{\theta }\left( z\right)
-\min_{z\in Z^{\ast }}\widehat{u}_{j}^{\theta }\left( z\right) \right] >0%
\text{,} \\ 
\rho \left( \widehat{u}^{\theta },u^{\theta }\right) <\frac{1}{3\times
\left\vert Z^{\ast }\right\vert }\times \gamma ^{\widehat{u}^{\theta }}%
\end{array}%
\right)  \notag \\
&\Longrightarrow &\left( 
\begin{array}{c}
\left[ \varepsilon _{j}^{\widehat{u}^{\theta }}\times y+\left( 1-\varepsilon
_{j}^{\widehat{u}^{\theta }}\right) \times y_{j}^{\widehat{u}^{\theta }}%
\right] \in \mathcal{L}_{j}^{Y}\left( \text{UNIF}\left[ F\left( \theta
\right) \right] ,u^{\theta }\right) \text{, } \\ 
\forall y\in \triangle \left( \widehat{\Gamma }_{j}^{A\text{-}B}\left(
\theta \right) \right)%
\end{array}%
\right) \text{.}  \label{tyu1a}
\end{eqnarray}
\end{lemma}

Third, we need a modified version of the lottery proposed in \cite{cmlr},
which is described as follows.

\paragraph{The (modified) Mezzetti-Renou lottery}

Following \cite{cmlr}, for each $\left( \theta ,j\right) \in \Theta ^{\ast
}\times \mathcal{I}$, each $u^{\theta }\in \Omega ^{\left[ \succeq ^{\theta
},\text{ }\mathbb{Q}\right] }$, each $z\in Z^{\ast }$ and each $\tau
:Z^{\ast }\rightarrow Z^{\ast }$, consider the following lottery:%
\begin{eqnarray*}
T^{\left( \theta ,u^{\theta },j,\tau ,z\right) } &=&\frac{1}{\left\vert
F\left( \theta \right) \right\vert }\dsum\limits_{x\in F\left( x\right)
}\left( 
\begin{array}{c}
\delta \left( x,\theta ,j,\tau \right) \times \left[ 
\begin{array}{c}
\left( 1-\varsigma \left( \theta ,u^{\theta },j,\tau \right) \right) \times
\tau \left( x\right) \\ 
+\varsigma \left( \theta ,u^{\theta },j,\tau \right) \times z%
\end{array}%
\right] \\ 
+\left( 1-\delta \left( x,\theta ,j,\tau \right) \right) \times x%
\end{array}%
\right) \text{,} \\
\text{with }\delta \left( x,\theta ,j,\tau \right) &\equiv &\left\{ 
\begin{tabular}{ll}
$\frac{1}{2}\text{,}$ & if $\tau \left( x\right) \in \widehat{\mathcal{L}}%
_{j}^{Z^{\ast }\text{-}A\text{-}B}\left( x,\theta \right) $, \\ 
$0$, & if $\tau \left( x\right) \notin \widehat{\mathcal{L}}_{j}^{Z^{\ast }%
\text{-}A\text{-}B}\left( x,\theta \right) $,%
\end{tabular}%
\right. \\
\text{and }\varsigma \left( \theta ,u^{\theta },j,\tau \right) &\equiv
&\left\{ 
\begin{tabular}{ll}
$\varsigma ^{u^{\theta }}\text{,}$ & if $\tau \left( x\right) \in \widehat{S%
\mathcal{L}}_{j}^{Z^{\ast }\text{-}A\text{-}B}\left( x,\theta \right) $ for
some $x\in F\left( \theta \right) $, \\ 
$0$, & if $\tau \left( x\right) \notin \widehat{S\mathcal{L}}_{j}^{Z^{\ast }%
\text{-}A\text{-}B}\left( x,\theta \right) $ for any $x\in F\left( \theta
\right) $,%
\end{tabular}%
\right.
\end{eqnarray*}%
where $\varsigma ^{u^{\theta }}$ is chosen according to Lemma \ref%
{lem:ordinal:approximate1}.

Given $\left( \theta ,u^{\theta },j\right) $, we define $\psi
_{j}^{u^{\theta }}\left( u^{\theta ^{\prime }}\right) $ for each $u^{\theta
^{\prime }}\in \cup _{\theta \in \Theta ^{\ast }}\Omega ^{\left[ \succeq
^{\theta },\text{ }\mathbb{Q}\right] }$ as follows. Fix any%
\begin{equation}
\left( \tau ^{\left( u^{\theta ^{\prime }}|\left( \theta ,u^{\theta
},j\right) \right) },z^{\left( u^{\theta ^{\prime }}|\left( \theta
,u^{\theta },j\right) \right) }\right) \in \arg \max_{\left( \tau ^{\prime
},z^{\prime }\right) \in \left( Z^{\ast }\right) ^{Z^{\ast }}\times Z^{\ast
}}u_{j}^{\theta ^{\prime }}\left[ T^{\left( \theta ,u^{\theta },j,\tau
^{\prime },z^{\prime }\right) }\right] \text{.}  \label{hhr1}
\end{equation}%
By finiteness of $Z^{\ast }$, $\left( \tau ^{\left( u^{\theta ^{\prime
}}|\left( \theta ,u^{\theta },j\right) \right) },z^{\left( u^{\theta
^{\prime }}|\left( \theta ,u^{\theta },j\right) \right) }\right) $ is
well-defined. Then, define%
\begin{equation}
\psi _{j}^{u^{\theta }}\left( u^{\theta ^{\prime }}\right) \equiv T^{\left(
\theta ,u^{\theta },j,\left( \tau ^{\left( u^{\theta ^{\prime }}|\left(
\theta ,u^{\theta },j\right) \right) },z^{\left( u^{\theta ^{\prime
}}|\left( \theta ,u^{\theta },j\right) \right) }\right) \right) }\text{.}
\label{hhr2}
\end{equation}%
The interpretation is that agent $j$ is the whistle-blower, and agents $-j$
report $\left( \theta ,u^{\theta }\right) $. Our canonical mechanism would
pick the lottery $T^{\left( \theta ,u^{\theta },j,\tau ,z\right) }$ with
some positive probability, while $j$ is allowed to choose any $\left( \tau
,z\right) $. Suppose the true cardinal state is $u^{\theta ^{\prime }}\in
\cup _{\theta \in \Theta ^{\ast }}\Omega ^{\left[ \succeq ^{\theta },\text{ }%
\mathbb{Q}\right] }$. Then, $\left( \tau ^{\left( u^{\theta ^{\prime
}}|\left( \theta ,u^{\theta },j\right) \right) },z^{\left( u^{\theta
^{\prime }}|\left( \theta ,u^{\theta },j\right) \right) }\right) $ is an
optimal choice for $j$ by (\ref{hhr1}). Therefore, by (\ref{hhr2}), $\psi
_{j}^{u^{\theta }}\left( u^{\theta ^{\prime }}\right) $ is an optimal
blocking scheme for $j$ in this scenario.

Fourth, we define a blocking selector as follows.

\paragraph{A optimal blocking-plan selector}

For each $\left( \theta ,\theta ^{\prime },j\right) \in \Theta ^{\ast
}\times \Theta ^{\ast }\times \mathcal{I}$, each $\left( u^{\theta
},u^{\theta ^{\prime }}\right) \in \Omega ^{\left[ \succeq ^{\theta },\text{ 
}\mathbb{Q}\right] }\times \Omega ^{\left[ \succeq ^{\theta ^{\prime }},%
\text{ }\mathbb{Q}\right] }$, consider the following condition:

\begin{equation}
u_{j}^{\theta ^{\prime }}\left[ \psi _{j}^{u^{\theta }}\left( u^{\theta
^{\prime }}\right) \right] \geq \max_{\gamma _{j}\in \left( Z^{\ast }\right)
^{\left[ 2^{Z^{\ast }}\diagdown \left\{ \varnothing \right\} \right]
}}\left( u_{j}^{\theta ^{\prime }}\left[ \varepsilon _{j}^{u^{\theta
}}\times \gamma _{j}\left( \widehat{\Gamma }_{j}^{A\text{-}B}\left( \theta
\right) \right) +\left( 1-\varepsilon _{j}^{u^{\theta }}\right) \times
y_{j}^{u^{\theta }}\right] \right) \text{,}  \label{gtt5}
\end{equation}%
where $\psi _{j}^{u^{\theta }}\left( u_{j}^{\theta _{j}}\right) $ is defined
in (\ref{hhr2}), and $\left( \varepsilon _{j}^{u^{\theta }},y_{j}^{u^{\theta
}}\right) $ are chosen for each $\left( \theta ,j\right) \in \Theta \times 
\mathcal{I}$ according to Lemma \ref{lem:perturb}. Define%
\begin{equation}
\eta _{j}^{u^{\theta }}\left( u^{\theta ^{\prime }}\right) \equiv \left\{ 
\begin{tabular}{ll}
$1\text{,}$ & if (\ref{gtt5})\ holds; \\ 
$0$, & otherwise.%
\end{tabular}%
\right.  \label{gtt6}
\end{equation}%
The interpretation of $\eta _{j}^{u^{\theta }}\left( u^{\theta ^{\prime
}}\right) $ is provided in the next section, when we describe Case (2) of
the canonical mechanism.

\paragraph{A canonical mechanism}

\label{sec:ordinal:canonical}

Let $\mathbb{N}$ denote the set of positive integers. We use the mechanism $%
\mathcal{M}^{\mathbb{Q}}=\left\langle M^{\mathbb{Q}}\equiv \times _{i\in 
\mathcal{I}}M_{i}^{\mathbb{Q}}\text{, \ }g:M^{\mathbb{Q}}\longrightarrow
\triangle \left( Z^{\ast }\right) \right\rangle $ defined below to implement 
$F$. In particular, for each $i\in \mathcal{I}$, define $M_{i}^{\mathbb{Q}}$
as:%
\begin{equation*}
\left\{ \left( \theta _{i},u_{i}^{\theta _{i}},k_{i}^{3},k_{i}^{4},\gamma
_{i},b_{i}\right) \in \Theta ^{\ast }\times \Theta \times \mathbb{N}\times 
\mathbb{N}\times \left( Z^{\ast }\right) ^{\left[ 2^{Z^{\ast }}\diagdown
\left\{ \varnothing \right\} \right] }\times Z^{\ast }:%
\begin{tabular}{l}
$u^{\theta _{i}}\in \Omega ^{\left[ \succeq ^{\theta _{i}},\text{ }\mathbb{Q}%
\right] }$ \\ 
$\gamma _{i}\left( E\right) \in E$, \\ 
$\forall E\in \left[ 2^{Z^{\ast }}\diagdown \left\{ \varnothing \right\} %
\right] $%
\end{tabular}%
\right\} \text{,}
\end{equation*}%
and $g\left[ m=\left( m_{i}\right) _{i\in \mathcal{I}}=\left( \theta
_{i},u_{i}^{\theta _{i}},k_{i}^{3},k_{i}^{4},\gamma _{i},b_{i}\right) _{i\in 
\mathcal{I}}\right] $ is defined in three cases.

\begin{description}
\item[Case (1): consensus] if there exists $\left( \theta ,u^{\theta
}\right) \in \Theta ^{\ast }\times \Theta $ such that%
\begin{eqnarray*}
u^{\theta } &\in &\Omega ^{\left[ \succeq ^{\theta },\text{ }\mathbb{Q}%
\right] }\text{,} \\
\left( \theta _{i},u_{i}^{\theta _{i}},k_{i}^{3}\right) &=&\left( \theta
,u^{\theta },1\right) \text{, }\forall i\in \mathcal{I}\text{,}
\end{eqnarray*}%
then $g\left[ m\right] =$UNIF$\left[ F\left( \theta \right) \right] $;

\item[Case (2), unilateral deviation: ] if there exists $\left( \theta
,u^{\theta },j\right) \in \Theta ^{\ast }\times \Theta \times \mathcal{I}$
such that%
\begin{eqnarray*}
u^{\theta } &\in &\Omega ^{\left[ \succeq ^{\theta },\text{ }\mathbb{Q}%
\right] }\text{,} \\
\text{and }\left( \theta _{i},u_{i}^{\theta _{i}},k_{i}^{3}\right) &=&\left(
\theta ,u^{\theta },1\right) \text{ if and only if }i\in \mathcal{I}%
\diagdown \left\{ j\right\} \text{,}
\end{eqnarray*}%
then 
\begin{eqnarray*}
g\left[ m\right] &=&\iota ^{\ast }\times \psi _{j}^{u^{\theta }}\left(
u_{j}^{\theta _{j}}\right) + \\
&&\left[ 1-\iota ^{\ast }\right] \times \left( 
\begin{tabular}{l}
$\varepsilon _{j}^{u^{\theta }}\times \left[ \left( 1-\frac{1}{k_{j}^{4}}%
\right) \times \gamma _{j}\left( \widehat{\Gamma }_{j}^{A\text{-}B}\left(
\theta \right) \right) +\frac{1}{k_{j}^{4}}\times \text{UNIF}\left( \widehat{%
\Gamma }_{j}^{A\text{-}B}\left( \theta \right) \right) \right] $ \\ 
$+\left( 1-\varepsilon _{j}^{u^{\theta }}\right) \times y_{j}^{u^{\theta }}$%
\end{tabular}%
\right) \text{,}
\end{eqnarray*}%
where%
\begin{equation}
\iota ^{\ast }\equiv \left[ \left( 1-\frac{1}{k_{j}^{3}+1}\right) \times
\eta _{j}^{u^{\theta }}\left( u^{\theta ^{\prime }}\right) +\left( \frac{1}{%
k_{j}^{3}+1}\right) \times \left( 1-\eta _{j}^{u^{\theta }}\left( u^{\theta
^{\prime }}\right) \right) \right] \in \left( 0,1\right) \text{,}
\label{gtt3}
\end{equation}%
$\psi _{j}^{u^{\theta }}\left( u_{j}^{\theta _{j}}\right) $ is defined in (%
\ref{hhr2}), and $\left( \varepsilon _{j}^{u^{\theta }},y_{j}^{u^{\theta
}}\right) $ are chosen for each $\left( \theta ,j\right) \in \Theta \times 
\mathcal{I}$ according to Lemma \ref{lem:perturb}, and $\eta _{j}^{u^{\theta
}}\left( u^{\theta ^{\prime }}\right) $ is defined in (\ref{gtt6}). Two
points are worthy of mentioning. First, by our construction,%
\begin{equation*}
g\left[ m\right] \in \left( \triangle \left[ \widehat{\Gamma }_{j}^{A\text{-}%
B}\left( \theta \right) \right] \right) \dbigcap \mathcal{L}_{j}^{Y}\left( 
\text{UNIF}\left[ F\left( \theta \right) \right] ,u^{\theta }\right) \text{.}
\end{equation*}%
Second, the Mezzetti-Renou technique is described by the compound lottery $%
\psi _{j}^{u^{\theta }}\left( u_{j}^{\theta _{j}}\right) $ (with probability 
$\iota ^{\ast }$), while our technique is described by the compound lottery
with probability $\left[ 1-\iota ^{\ast }\right] $, which ensures that all
equilibria triggering Case (2) deliver good outcomes. However, it is not
clear which of the two compound lotteries is better for the whistle-blower,%
\footnote{%
In the canonical mechanism which we use to prove Theorem \ref%
{theorem:full:mix:SCC-A}, we define $\psi _{j}^{u^{\theta }}\left(
u_{j}^{\theta _{j}}\right) $ as a best blocking plan in $\widehat{\mathcal{L}%
}_{j}^{Y\text{-}A\text{-}B}\left( \text{UNIF}\left[ F\left( \theta \right) %
\right] ,u^{\theta }\right) $ at $u^{\theta _{j}}$. Here, we must adopt the
the Mezzetti-Renou lottery, which may not be a best blocking plan in $%
\widehat{\mathcal{L}}_{j}^{Y\text{-}A\text{-}B}\left( \text{UNIF}\left[
F\left( \theta \right) \right] ,u^{\theta }\right) $ at $u^{\theta _{j}}$.}
and hence, we need $\iota ^{\ast }$ defined in (\ref{gtt3}), so that the
whistle-blower may use $\eta _{j}^{u^{\theta }}\left( u^{\theta ^{\prime
}}\right) $ to choose a better one between the two.\footnote{%
More precisely, when the true cardinal state is $u^{\theta ^{\prime }}$ and (%
\ref{gtt3})\ holds, the whistle-blower $j$ finds the Mezzetti-Renou lottery
better and use $\eta _{j}^{u^{\theta }}\left( u^{\theta ^{\prime }}\right)
=1 $ to choose it. Otherwise, $j$ uses $\eta _{j}^{u^{\theta }}\left(
u^{\theta ^{\prime }}\right) =0$ to choose our compound lottery.} ---This is
crucial to prove Lemma \ref{lem:mixed:canonical:pure:ordinal} below.

\item[Case (3), multi-lateral deviation: ] otherwise, 
\begin{equation*}
g\left[ m\right] =\left( 1-\frac{1}{k_{j^{\ast }}^{3}}\right) \times
b_{j^{\ast }}+\frac{1}{k_{j^{\ast }}^{3}}\times \text{UNIF}\left( Z^{\ast
}\right) \text{,}
\end{equation*}%
where $j^{\ast }=\max \left( \arg \max_{i\in \mathcal{I}}k_{i}^{2}\right) $,
i.e., $j^{\ast }$ is the largest-numbered agent who submits the highest
number on the second dimension of the message.
\end{description}

\begin{lemma}
\label{lem:mixed:canonical:pure:ordinal}Consider the canonical mechanism $%
\mathcal{M}^{\mathbb{Q}}$ above. For any $u\in \cup _{\theta \in \Theta
^{\ast }}\Omega ^{\left[ \succeq ^{\theta },\text{ }%
\mathbb{R}
\right] }$ and any $\lambda \in MNE^{\left( \mathcal{M}^{\mathbb{Q}},\text{ }%
u\right) }$, we have SUPP$\left[ \lambda \right] \subset PNE^{\left( 
\mathcal{M}^{\mathbb{Q}},\text{ }u\right) }$.
\end{lemma}

The proof of Lemma \ref{lem:mixed:canonical:pure:ordinal} is relegated to
Appendix \ref{sec:lem:mixed:canonical:pure:ordinal}.

\paragraph{Theorem \protect\ref{theorem:MR:iff}: a proof}

We are now ready to prove Theorem \ref{theorem:MR:iff}. Fix any true ordinal
state $\theta ^{\ast }\in \Theta ^{\ast }$ and any true cardinal state $%
u^{\theta ^{\ast }}\in \Omega ^{\left[ \succeq ^{\theta ^{\ast }},\text{ }%
\mathbb{R}
\right] }$. By Lemma \ref{lem:mixed:canonical:pure:ordinal}, it suffers no
loss of generality to focus on pure-strategy Nash equilibrium, and thus, we
aim to prove%
\begin{equation*}
\dbigcup\limits_{\lambda \in PNE^{\left( \mathcal{M}^{\mathbb{Q}},\text{ }%
u^{\theta ^{\ast }}\right) }}\text{SUPP}\left( g\left[ \lambda \right]
\right) =F\left( \theta ^{\ast }\right) \text{.}
\end{equation*}

First, suppose 
\begin{equation*}
\max_{i\in \mathcal{I}}\left[ \max_{z\in Z^{\ast }}u_{i}^{\theta ^{\ast
}}\left( z\right) -\min_{z\in Z^{\ast }}u_{i}^{\theta ^{\ast }}\left(
z\right) \right] =0\text{,}
\end{equation*}%
i.e., all agents are indifferent between any two outcomes in $Z^{\ast }$.
Pick any $\widehat{u^{\theta ^{\ast }}}\in \Omega ^{\left[ \succeq ^{\theta
^{\ast }},\text{ }\mathbb{Q}\right] }$, and any pure strategy profile%
\begin{equation*}
m^{\ast }=\left( \theta _{i}=\theta ^{\ast },u_{i}^{\theta _{i}}=\widehat{%
u^{\theta ^{\ast }}},k_{i}^{3}=1,\ast ,\ast ,\ast ,\ast \right) _{i\in 
\mathcal{I}}
\end{equation*}%
is a Nash equilibrium, which triggers Case (1) and $g\left[ m^{\ast }\right]
=$UNIF$\left[ F\left( \theta ^{\ast }\right) \right] $.

Second, suppose 
\begin{equation*}
\max_{i\in \mathcal{I}}\left[ \max_{z\in Z^{\ast }}u_{i}^{\theta ^{\ast
}}\left( z\right) -\min_{z\in Z^{\ast }}u_{i}^{\theta ^{\ast }}\left(
z\right) \right] >0\text{,}
\end{equation*}%
and denote%
\begin{equation*}
\gamma ^{u^{\theta ^{\ast }}}\equiv \min_{\left\{ \left( i,a,b\right) \in 
\mathcal{I}\times Z^{\ast }\times Z^{\ast }:u_{i}\left( a\right) \neq
u_{i}\left( b\right) \right\} }\left\vert u_{i}^{\theta ^{\ast }}\left(
a\right) -u_{i}^{\theta ^{\ast }}\left( b\right) \right\vert >0\text{.}
\end{equation*}%
Pick any $\widehat{u^{\theta ^{\ast }}}\in \Omega ^{\left[ \succeq ^{\theta
^{\ast }},\text{ }\mathbb{Q}\right] }$ such that%
\begin{equation}
\rho \left( u^{\theta ^{\ast }},\widehat{u^{\theta ^{\ast }}}\right) <\frac{1%
}{2}\times \frac{1}{\left( 3\times \left\vert Z^{\ast }\right\vert +1\right) 
}\gamma ^{u^{\theta ^{\ast }}}\text{,}  \label{hhr3a}
\end{equation}%
and as a result,%
\begin{equation}
\gamma ^{\widehat{u^{\theta ^{\ast }}}}\equiv \min_{\left\{ \left(
i,a,b\right) \in \mathcal{I}\times Z^{\ast }\times Z^{\ast }:u_{i}\left(
a\right) \neq u_{i}\left( b\right) \right\} }\left\vert \widehat{u^{\theta
^{\ast }}}_{i}\left( a\right) -\widehat{u^{\theta ^{\ast }}}_{i}\left(
b\right) \right\vert >\left[ 1-\frac{1}{\left( 3\times \left\vert Z^{\ast
}\right\vert +1\right) }\right] \gamma ^{u^{\theta ^{\ast }}}>0\text{,}
\label{hhr3b}
\end{equation}%
and hence,%
\begin{equation}
\frac{1}{3\times \left\vert Z^{\ast }\right\vert }\gamma ^{\widehat{%
u^{\theta ^{\ast }}}}>\frac{1}{3\times \left\vert Z^{\ast }\right\vert }%
\left[ 1-\frac{1}{\left( 3\times \left\vert Z^{\ast }\right\vert +1\right) }%
\right] \gamma ^{u^{\theta ^{\ast }}}=\frac{1}{\left( 3\times \left\vert
Z^{\ast }\right\vert +1\right) }\gamma ^{u^{\theta ^{\ast }}}>\rho \left(
u^{\theta ^{\ast }},\widehat{u^{\theta ^{\ast }}}\right) \text{,}
\label{hhr3}
\end{equation}%
where the first inequality follows from (\ref{hhr3b}), and the second
inequality from (\ref{hhr3a}). Then, any pure strategy profile%
\begin{equation*}
m^{\ast }=\left( \theta _{i}=\theta ^{\ast },u_{i}^{\theta _{i}}=\widehat{%
u^{\theta ^{\ast }}},k_{i}^{3}=1,\ast ,\ast ,\ast \right) _{i\in \mathcal{I}}
\end{equation*}%
is a Nash equilibrium, which triggers Case (1) and $g\left[ m^{\ast }\right]
=$UNIF$\left[ F\left( \theta ^{\ast }\right) \right] $. Any unilateral
deviation $\overline{m}_{j}\in M_{j}^{\mathbb{Q}}$ of agent $j\in \mathcal{I}
$ would either still trigger Case (1) and induce UNIF$\left[ F\left( \theta
^{\ast }\right) \right] $, or trigger Case (2) and induce a mixture of%
\begin{equation*}
\psi _{j}^{\widehat{u^{\theta ^{\ast }}}}\left( u_{j}^{\theta _{j}}\right) 
\text{ and }\left( 
\begin{tabular}{l}
$\varepsilon _{j}^{\widehat{u^{\theta ^{\ast }}}}\times \left[ \left( 1-%
\frac{1}{k_{j}^{4}}\right) \times \gamma _{j}\left( \widehat{\Gamma }_{j}^{A%
\text{-}B}\left( \theta \right) \right) +\frac{1}{k_{j}^{4}}\times \text{UNIF%
}\left( \widehat{\Gamma }_{j}^{A\text{-}B}\left( \theta \right) \right) %
\right] $ \\ 
$+\left( 1-\varepsilon _{j}^{\widehat{u^{\theta ^{\ast }}}}\right) \times
y_{j}^{u^{\theta }}$%
\end{tabular}%
\right) \text{.}
\end{equation*}%
(\ref{hhr3}) and Lemma \ref{lem:perturb} imply%
\begin{equation*}
\left( 
\begin{tabular}{l}
$\varepsilon _{j}^{\widehat{u^{\theta ^{\ast }}}}\times \left[ \left( 1-%
\frac{1}{k_{j}^{4}}\right) \times \gamma _{j}\left( \widehat{\Gamma }_{j}^{A%
\text{-}B}\left( \theta \right) \right) +\frac{1}{k_{j}^{4}}\times \text{UNIF%
}\left( \widehat{\Gamma }_{j}^{A\text{-}B}\left( \theta \right) \right) %
\right] $ \\ 
$+\left( 1-\varepsilon _{j}^{\widehat{u^{\theta ^{\ast }}}}\right) \times
y_{j}^{u^{\theta }}$%
\end{tabular}%
\right) \in \mathcal{L}_{j}^{Y}\left( \text{UNIF}\left[ F\left( \theta
^{\ast }\right) \right] ,u^{\theta ^{\ast }}\right) \text{.}
\end{equation*}%
We now show%
\begin{equation}
\psi _{j}^{\widehat{u^{\theta ^{\ast }}}}\left( u_{j}^{\theta _{j}}\right)
\in \mathcal{L}_{j}^{Y}\left( \text{UNIF}\left[ F\left( \theta ^{\ast
}\right) \right] ,u^{\theta ^{\ast }}\right) \text{.}  \label{hhr4}
\end{equation}%
We consider two scenarios:%
\begin{equation*}
\left( 
\begin{array}{c}
\text{(i) }\tau ^{\left( u_{j}^{\theta _{j}}|\left( \theta ^{\ast },\widehat{%
u^{\theta ^{\ast }}},j\right) \right) }\left( x\right) \notin \widehat{S%
\mathcal{L}}_{j}^{Z^{\ast }\text{-}A\text{-}B}\left( x,\theta ^{\ast
}\right) \text{ for any }x\in F\left( \theta ^{\ast }\right) \text{,} \\ 
\text{\ (ii) }\tau ^{\left( u_{j}^{\theta _{j}}|\left( \theta ^{\ast },%
\widehat{u^{\theta ^{\ast }}},j\right) \right) }\left( x\right) \in \widehat{%
S\mathcal{L}}_{j}^{Z^{\ast }\text{-}A\text{-}B}\left( x,\theta ^{\ast
}\right) \text{ for some }x\in F\left( \theta ^{\ast }\right) \text{,}%
\end{array}%
\right)
\end{equation*}%
where $\tau ^{\left( u_{j}^{\theta _{j}}|\left( \theta ^{\ast },\widehat{%
u^{\theta ^{\ast }}},j\right) \right) }\left( x\right) $ is defined in (\ref%
{hhr1}). In scenario (i) we have%
\begin{gather*}
\varsigma \left( \theta ^{\ast },\widehat{u^{\theta ^{\ast }}},j,\tau
^{\left( u_{j}^{\theta _{j}}|\left( \theta ^{\ast },\widehat{u^{\theta
^{\ast }}},j\right) \right) }\right) =0\text{ for any }x\in F\left( \theta
^{\ast }\right) \text{ and} \\
\widehat{u^{\theta ^{\ast }}}_{j}\left( x\right) =\widehat{u^{\theta ^{\ast
}}}_{j}\left( \tau ^{\left( u_{j}^{\theta _{j}}|\left( \theta ^{\ast },%
\widehat{u^{\theta ^{\ast }}},j\right) \right) }\left( x\right) \right) 
\text{ for any }x\in F\left( \theta ^{\ast }\right) \text{,}
\end{gather*}%
which, together with $\left\{ u^{\theta ^{\ast }},\widehat{u^{\theta ^{\ast
}}}\right\} \subset \Omega ^{\left[ \succeq ^{\theta ^{\ast }},\text{ }%
\mathbb{R}
\right] }$, immediately implies%
\begin{equation*}
u_{j}^{\theta ^{\ast }}\left( x\right) =u_{j}^{\theta ^{\ast }}\left( \tau
^{\left( u_{j}^{\theta _{j}}|\left( \theta ^{\ast },\widehat{u^{\theta
^{\ast }}},j\right) \right) }\left( x\right) \right) \text{ for any }x\in
F\left( \theta ^{\ast }\right) \text{.}
\end{equation*}%
Therefore, (\ref{hhr4}) holds.

In scenario (ii), (\ref{hhr3}) and Lemma \ref{lem:ordinal:approximate1}
imply (\ref{hhr4}). Therefore, any $\overline{m}_{j}\in M_{j}^{\mathbb{Q}}$
is not a profitable deviation.

Third, fix any%
\begin{equation*}
\widetilde{m}=\left( \widetilde{\theta _{i}},\widetilde{u_{i}^{\theta _{i}}},%
\widetilde{k_{i}^{3}},\widetilde{k_{i}^{4}},\widetilde{\gamma _{i}},%
\widetilde{b_{i}}\right) _{i\in \mathcal{I}}\in PNE^{\left( \mathcal{M}^{%
\mathbb{Q}},\text{ }u^{\theta ^{\ast }}\right) }\text{,}
\end{equation*}%
and we aim to prove $g\left[ \widetilde{m}\right] \in \triangle \left[
F\left( \theta ^{\ast }\right) \right] $. If $\widetilde{m}$ triggers Case
(1), by a similar argument as in \cite{cmlr}, strong set-monotonicity
implies $g\left[ \widetilde{m}\right] \in \triangle \left[ F\left( \theta
^{\ast }\right) \right] $. If $\widetilde{m}$ triggers Case (2) and $j$ is
the whistle-blower, i.e., 
\begin{equation*}
\exists \theta \in \Theta ^{\ast }\text{, }\widetilde{\theta _{i}}=\theta 
\text{, }\forall i\in \mathcal{I}\diagdown \left\{ j\right\} \text{.}
\end{equation*}%
As argued above, $\widehat{\Gamma }_{j}^{A\text{-}B}\left( \theta \right) $
must be a $j$-$Z^{\ast }$-$u^{\theta ^{\ast }}$-max set, which, together
Lemma \ref{lem:no-veto:generalized:SCC:ordinal}, implies $g\left[ \widetilde{%
m}\right] \in \triangle \left[ \widehat{\Gamma }_{j}^{A\text{-}B}\left(
\theta \right) \right] \subset \triangle \left[ F\left( \theta ^{\ast
}\right) \right] $. Similarly, if $\widetilde{m}$ triggers Case (3), $%
Z^{\ast }$ must be a $j$-$Z^{\ast }$-$u^{\theta ^{\ast }}$-max set, which,
together Lemma \ref{lem:no-veto:generalized:SCC:ordinal}, implies $g\left[ 
\widetilde{m}\right] \in \triangle \left[ Z^{\ast }\right] \subset \triangle %
\left[ F\left( \theta ^{\ast }\right) \right] $.$\blacksquare $

\subsubsection{Proof of Lemma \protect\ref{lem:ordinal:approximate1}}

\label{sec:lem:ordinal:approximate1}

Fix any $\theta \in \Theta ^{\ast }$ and any $\widehat{u}^{\theta }\in
\Omega ^{\left[ \succeq ^{\theta },\text{ }\mathbb{Q}\right] }$. Consider
any $\left( i,x,y\right) \in \mathcal{I}\times Z^{\ast }\times Z^{\ast }$
and any $u^{\theta }\in \Omega ^{\left[ \succeq ^{\theta },\text{ }%
\mathbb{R}
\right] }$ such that%
\begin{equation*}
\left( 
\begin{array}{c}
\widehat{u}_{i}^{\theta }\left( x\right) -\widehat{u}_{i}^{\theta }\left(
y\right) >0\text{,} \\ 
\rho \left( \widehat{u}^{\theta },u^{\theta }\right) <\frac{1}{3}\times
\gamma ^{\widehat{u}^{\theta }}%
\end{array}%
\right) \text{.}
\end{equation*}%
We have%
\begin{eqnarray}
&&u_{i}^{\theta }\left( x\right) -\left[ \left( 1-\varsigma ^{\widehat{u}%
^{\theta }}\right) u_{i}^{\theta }\left( y\right) +\varsigma ^{\widehat{u}%
^{\theta }}\max_{z\in Z^{\ast }}u_{i}^{\theta }\left( z\right) +\varsigma ^{%
\widehat{u}^{\theta }}\left( \left\vert Z^{\ast }\right\vert -1\right)
\left( \max_{z\in Z^{\ast }}u_{i}^{\theta }\left( z\right) -\min_{z\in
Z^{\ast }}u_{i}^{\theta }\left( z\right) \right) \right]  \notag \\
&\geq &\widehat{u}_{i}^{\theta }\left( x\right) -\left[ \left( 1-\varsigma ^{%
\widehat{u}^{\theta }}\right) \widehat{u}_{i}^{\theta }\left( y\right)
+\varsigma ^{\widehat{u}^{\theta }}\max_{z\in Z^{\ast }}\widehat{u}%
_{i}^{\theta }\left( z\right) +\varsigma ^{\widehat{u}^{\theta }}\left(
\left\vert Z^{\ast }\right\vert -1\right) \left( \max_{z\in Z^{\ast }}%
\widehat{u}_{i}^{\theta }\left( z\right) -\min_{z\in Z^{\ast }}\widehat{u}%
_{i}^{\theta }\left( z\right) \right) \right]  \notag \\
&&-\rho \left( \widehat{u}^{\theta },u^{\theta }\right) \times \left( 1+%
\left[ \left( 1-\varsigma ^{\widehat{u}^{\theta }}\right) +\varsigma ^{%
\widehat{u}^{\theta }}+2\varsigma ^{\widehat{u}^{\theta }}\left( \left\vert
Z^{\ast }\right\vert -1\right) \right] \right) \text{.}  \label{gtt1}
\end{eqnarray}%
Furthermore, we have%
\begin{eqnarray}
&&\lim_{\varsigma ^{\widehat{u}^{\theta }}\rightarrow 0}\left( 
\begin{array}{c}
\widehat{u}_{i}^{\theta }\left( x\right) \\ 
-\left[ 
\begin{array}{c}
\left( 1-\varsigma ^{\widehat{u}^{\theta }}\right) \widehat{u}_{i}^{\theta
}\left( y\right) \\ 
+\varsigma ^{\widehat{u}^{\theta }}\max_{z\in Z^{\ast }}\widehat{u}%
_{i}^{\theta }\left( z\right) +\varsigma ^{\widehat{u}^{\theta }}\left(
\left\vert Z^{\ast }\right\vert -1\right) \left( \max_{z\in Z^{\ast }}%
\widehat{u}_{i}^{\theta }\left( z\right) -\min_{z\in Z^{\ast }}\widehat{u}%
_{i}^{\theta }\left( z\right) \right)%
\end{array}%
\right] \\ 
-\rho \left( \widehat{u}^{\theta },u^{\theta }\right) \times \left( 1+\left[
\left( 1-\varsigma ^{\widehat{u}^{\theta }}\right) +\varsigma ^{\widehat{u}%
^{\theta }}+2\varsigma ^{\widehat{u}^{\theta }}\left( \left\vert Z^{\ast
}\right\vert -1\right) \right] \right)%
\end{array}%
\right)  \notag \\
&=&\widehat{u}_{i}^{\theta }\left( x\right) -\widehat{u}_{i}^{\theta }\left(
y\right) -2\times \rho \left( \widehat{u}^{\theta },u^{\theta }\right) 
\notag \\
&>&\widehat{u}_{i}^{\theta }\left( x\right) -\widehat{u}_{i}^{\theta }\left(
y\right) -2\times \frac{1}{3}\times \gamma ^{\widehat{u}^{\theta }}  \notag
\\
&=&\widehat{u}_{i}^{\theta }\left( x\right) -\widehat{u}_{i}^{\theta }\left(
y\right) -2\times \frac{1}{3}\times \min_{\left\{ \left( j,a,b\right) \in 
\mathcal{I}\times Z^{\ast }\times Z^{\ast }:\widehat{u}_{j}^{\theta }\left(
a\right) \neq \widehat{u}_{j}^{\theta }\left( b\right) \right\} }\left\vert 
\widehat{u}_{j}^{\theta }\left( a\right) -\widehat{u}_{j}^{\theta }\left(
b\right) \right\vert >0\text{,}  \label{dff1}
\end{eqnarray}%
(\ref{gtt1}) and (\ref{dff1}) imply existence of $\varsigma ^{\widehat{u}%
^{\theta }}\in \left( 0,1\right) $ such that%
\begin{equation}
u_{i}^{\theta }\left( x\right) -\left[ \left( 1-\varsigma ^{\widehat{u}%
^{\theta }}\right) u_{i}^{\theta }\left( y\right) +\varsigma ^{\widehat{u}%
^{\theta }}\max_{z\in Z^{\ast }}u_{i}^{\theta }\left( z\right) +\varsigma ^{%
\widehat{u}^{\theta }}\left( \left\vert Z^{\ast }\right\vert -1\right)
\left( \max_{z\in Z^{\ast }}u_{i}^{\theta }\left( z\right) -\min_{z\in
Z^{\ast }}u_{i}^{\theta }\left( z\right) \right) \right] >0\text{,}  \notag
\end{equation}%
i.e., (\ref{nnb1a}) holds.$\blacksquare $

\subsubsection{Proof of Lemma \protect\ref{lem:perturb}}

\label{sec:lem:perturb}

Fix any $\left( \theta ,j\right) \in \Theta ^{\ast }\times \mathcal{I}$ and
any $\widehat{u}^{\theta }\in \Omega ^{\left[ \succeq ^{\theta },\text{ }%
\mathbb{Q}\right] }$. If%
\begin{equation*}
\left[ \max_{z\in Z^{\ast }}\widehat{u}_{j}^{\theta }\left( z\right)
-\min_{z\in Z^{\ast }}\widehat{u}_{j}^{\theta }\left( z\right) \right] =0%
\text{,}
\end{equation*}%
choose any $\varepsilon _{j}^{\widehat{u}^{\theta }}\in \left( 0,1\right) $
and any $y_{j}^{\widehat{u}^{\theta }}\in \widehat{\mathcal{L}}_{j}^{Y\text{-%
}A\text{-}B}\left( \text{UNIF}\left[ F\left( \theta \right) \right] ,%
\widehat{u}^{\theta }\right) $, and (\ref{tyu1a}) holds vacuously. Suppose 
\begin{equation*}
\left[ \max_{z\in Z^{\ast }}\widehat{u}_{j}^{\theta }\left( z\right)
-\min_{z\in Z^{\ast }}\widehat{u}_{j}^{\theta }\left( z\right) \right] >0%
\text{,}
\end{equation*}%
and consider any $u^{\theta }\in \Omega ^{\left[ \succeq ^{\theta },\text{ }%
\mathbb{R}
\right] }$ such that 
\begin{equation}
\rho \left( \widehat{u}^{\theta },u^{\theta }\right) <\frac{1}{2}\times 
\frac{1}{3\times \left\vert Z^{\ast }\right\vert }\times \gamma ^{\widehat{u}%
^{\theta }}\text{.}  \label{tyu2}
\end{equation}%
Choose any $y_{j}^{\widehat{u}^{\theta }}\in \arg \min_{z\in Z^{\ast }}%
\widehat{u}_{j}^{\theta }\left( z\right) $. We now consider two scenarios:
(i) $F\left( \theta \right) \subset \arg \min_{z\in Z^{\ast }}\widehat{u}%
_{j}^{\theta }\left( z\right) $, and (ii)\ $F\left( \theta \right) \diagdown
\arg \min_{z\in Z^{\ast }}\widehat{u}_{j}^{\theta }\left( z\right) \neq
\varnothing $. First, in scenario (i), we have 
\begin{equation*}
\widehat{\Gamma }_{j}^{A\text{-}B}\left( \theta \right) \subset Z^{\ast
}\cap \mathcal{L}_{j}^{Z}\left( F\left( \theta \right) ,\widehat{u}^{\theta
}\right) \text{.}
\end{equation*}%
Thus, $\left\{ \widehat{u}^{\theta },u^{\theta }\right\} \in \Omega ^{\left[
\succeq ^{\theta },\text{ }%
\mathbb{R}
\right] }$ implies%
\begin{gather*}
\widehat{\Gamma }_{j}^{A\text{-}B}\left( \theta \right) \subset Z^{\ast
}\cap \mathcal{L}_{j}^{Z}\left( F\left( \theta \right) ,u^{\theta }\right) 
\text{,} \\
y_{j}^{\widehat{u}^{\theta }}\in \arg \min_{z\in Z^{\ast }}\widehat{u}%
_{j}^{\theta }\left( z\right) =\arg \min_{z\in Z^{\ast }}u_{j}^{\theta
}\left( z\right) \text{.}
\end{gather*}%
Choose $\varepsilon _{j}^{\widehat{u}^{\theta }}=\frac{1}{2}$, and we have 
\begin{eqnarray*}
\left[ \varepsilon _{j}^{\widehat{u}^{\theta }}\times y+\left( 1-\varepsilon
_{j}^{\widehat{u}^{\theta }}\right) \times y_{j}^{\widehat{u}^{\theta }}%
\right] &\in &\triangle \left[ Z^{\ast }\cap \mathcal{L}_{j}^{Z}\left(
F\left( \theta \right) ,u^{\theta }\right) \right] \subset \mathcal{L}%
_{j}^{Y}\left( \text{UNIF}\left[ F\left( \theta \right) \right] ,u^{\theta
}\right) \text{, } \\
\forall y &\in &\triangle \left( \widehat{\Gamma }_{j}^{A\text{-}B}\left(
\theta \right) \right) \text{,}
\end{eqnarray*}%
i.e., (\ref{tyu1a})\ holds.

Second, in scenario (ii), we have%
\begin{gather*}
y_{j}^{\widehat{u}^{\theta }}\in \arg \min_{z\in Z^{\ast }}\widehat{u}%
_{j}^{\theta }\left( z\right) \text{,} \\
F\left( \theta \right) \diagdown \arg \min_{z\in Z^{\ast }}\widehat{u}%
_{j}^{\theta }\left( z\right) \neq \varnothing \text{.}
\end{gather*}%
Pick any $z^{\prime }\in Z^{\ast }$, and hence,%
\begin{eqnarray*}
\lim_{\varepsilon _{j}^{\widehat{u}^{\theta }}\longrightarrow 0} &&\left[ 
\widehat{u}_{j}^{\theta }\left( \text{UNIF}\left[ F\left( \theta \right) %
\right] \right) -\widehat{u}_{j}^{\theta }\left( \left[ \varepsilon _{j}^{%
\widehat{u}^{\theta }}\times z^{\prime }+\left( 1-\varepsilon _{j}^{\widehat{%
u}^{\theta }}\right) \times y_{j}^{\widehat{u}^{\theta }}\right] \right) %
\right] \\
&=&\sum_{x\in F\left( \theta \right) }\frac{\widehat{u}_{j}^{\theta }\left(
x\right) }{\left\vert F\left( \theta \right) \right\vert }-\widehat{u}%
_{j}^{\theta }\left( y_{j}^{\widehat{u}^{\theta }}\right) \\
&\geq &\frac{1}{\left\vert F\left( \theta \right) \right\vert }\times
\min_{\left\{ \left( i,a,b\right) \in \mathcal{I}\times Z^{\ast }\times
Z^{\ast }:u_{i}\left( a\right) \neq u_{i}\left( b\right) \right\}
}\left\vert \widehat{u}_{i}^{\theta }\left( a\right) -\widehat{u}%
_{i}^{\theta }\left( b\right) \right\vert \\
&\geq &\frac{1}{\left\vert Z^{\ast }\right\vert }\times \gamma ^{\widehat{u}%
^{\theta }}>0\text{.}
\end{eqnarray*}%
By finiteness of $Z^{\ast }$, there exists $\varepsilon _{j}^{\widehat{u}%
^{\theta }}>0$ such that%
\begin{gather*}
\forall y\in \triangle \left( \widehat{\Gamma }_{j}^{A\text{-}B}\left(
\theta \right) \right) \text{,} \\
\left[ \widehat{u}_{j}^{\theta }\left( \text{UNIF}\left[ F\left( \theta
\right) \right] \right) -\widehat{u}_{j}^{\theta }\left( \left[ \varepsilon
_{j}^{\widehat{u}^{\theta }}\times y+\left( 1-\varepsilon _{j}^{\widehat{u}%
^{\theta }}\right) \times y_{j}^{\widehat{u}^{\theta }}\right] \right) %
\right] >\frac{1}{2\times \left\vert Z^{\ast }\right\vert }\times \gamma ^{%
\widehat{u}^{\theta }}>0\text{,}
\end{gather*}%
which, together with (\ref{tyu2}), implies 
\begin{gather*}
\forall y\in \triangle \left( \widehat{\Gamma }_{j}^{A\text{-}B}\left(
\theta \right) \right) \text{,} \\
\left[ u_{j}^{\theta }\left( \text{UNIF}\left[ F\left( \theta \right) \right]
\right) -u_{j}^{\theta }\left( \left[ \varepsilon _{j}^{\widehat{u}^{\theta
}}\times y+\left( 1-\varepsilon _{j}^{\widehat{u}^{\theta }}\right) \times
y_{j}^{\widehat{u}^{\theta }}\right] \right) \right] >\left( \frac{1}{%
2\times \left\vert Z^{\ast }\right\vert }-\frac{1}{3\times \left\vert
Z^{\ast }\right\vert }\right) \times \gamma ^{\widehat{u}^{\theta }}>0\text{,%
}
\end{gather*}%
i.e., (\ref{tyu1a})\ holds.$\blacksquare $

\subsubsection{Proof of Lemma \protect\ref{lem:mixed:canonical:pure:ordinal}}

\label{sec:lem:mixed:canonical:pure:ordinal}

We need the following result to prove Lemma \ref%
{lem:mixed:canonical:pure:ordinal}.

\begin{lemma}
\label{lem:mixed:canonical:pure:dominant:ordinal}Consider the canonical
mechanism $\mathcal{M}^{\mathbb{Q}}$ in Appendix \ref{sec:ordinal:canonical}%
. For any $\left( \theta ,i\right) \in \Theta ^{\ast }\times \mathcal{I}$,
any $u^{\theta }\in \Omega ^{\left[ \succeq ^{\theta },\text{ }%
\mathbb{R}
\right] }$ and any $\varepsilon >0$, there exists a sequence $\left\{
m_{i}^{u^{\theta },n}\in M_{i}:n\in \mathbb{N}\right\} $ such that%
\begin{equation}
\lim_{n\rightarrow \infty }u_{i}^{\theta }\left[ g\left( m_{i}^{u^{\theta
},n},m_{-i}\right) \right] \geq u_{i}^{\theta }\left[ g\left(
m_{i},m_{-i}\right) \right] -\varepsilon \text{, }\forall \left(
m_{i},m_{-i}\right) \in M_{i}\times M_{-i}\text{.}  \label{gtt4}
\end{equation}
\end{lemma}

\noindent \textbf{Proof of Lemma \ref%
{lem:mixed:canonical:pure:dominant:ordinal}:} Fix any $\left( \theta
,i\right) \in \Theta ^{\ast }\times \mathcal{I}$. First, consider any $%
u^{\theta }\in \Omega ^{\left[ \succeq ^{\theta _{i}},\text{ }\mathbb{Q}%
\right] }$, and we show (\ref{gtt4})\ for $u^{\theta }$. Fix any%
\begin{eqnarray*}
b_{i}^{u^{\theta }} &\in &\arg \max_{z\in Z}u_{i}^{\theta }\left( z\right) 
\text{ and any} \\
\gamma _{i}\left( E\right) &\in &\left( Z^{\ast }\right) ^{\left[ 2^{Z^{\ast
}}\diagdown \left\{ \varnothing \right\} \right] }\text{ such that }\gamma
_{i}\left( E\right) \in \arg \max_{z\in E}u_{i}^{\theta }\left( z\right) 
\text{, }\forall E\in \left[ 2^{Z^{\ast }}\diagdown \left\{ \varnothing
\right\} \right] \text{,}
\end{eqnarray*}%
and define%
\begin{equation*}
m_{i}^{u^{\theta },n}\equiv \left( \theta ,u^{\theta
},k_{i}^{3}=n,k_{i}^{4}=n,\gamma _{i},b_{i}^{u^{\theta }}\right) \in M_{i}%
\text{, }\forall n\in \mathbb{N}\text{.}
\end{equation*}%
It is straightfoward to show%
\begin{equation*}
\lim_{n\rightarrow \infty }u_{i}^{\theta }\left[ g\left( m_{i}^{u^{\theta
},n},m_{-i}\right) \right] \geq u_{i}^{\theta }\left[ g\left(
m_{i},m_{-i}\right) \right] \text{, }\forall \left( m_{i},m_{-i}\right) \in
M_{i}\times M_{-i}\text{.}
\end{equation*}%
i.e., (\ref{gtt4})\ holds.

Second, consider any $\widehat{u}^{\theta }\in \Omega ^{\left[ \succeq
^{\theta },\text{ }%
\mathbb{R}
\right] }$, and we show (\ref{gtt4})\ for $\widehat{u}^{\theta }$. For any $%
\varepsilon >0$, pick any $u^{\theta }\in \Omega ^{\left[ \succeq ^{\theta
_{i}},\text{ }\mathbb{Q}\right] }$ such that%
\begin{equation}
\rho \left( \widehat{u}^{\theta },u^{\theta }\right) <\frac{1}{2}\times
\varepsilon \text{,}  \label{gtt10}
\end{equation}%
and pick any sequence $\left\{ m_{i}^{u^{\theta },n}\in M_{i}:n\in \mathbb{N}%
\right\} $ such that%
\begin{equation}
\lim_{n\rightarrow \infty }u_{i}^{\theta }\left[ g\left( m_{i}^{u^{\theta
},n},m_{-i}\right) \right] \geq u_{i}^{\theta }\left[ g\left(
m_{i},m_{-i}\right) \right] -\frac{1}{2}\times \varepsilon \text{, }\forall
\left( m_{i},m_{-i}\right) \in M_{i}\times M_{-i}\text{.}  \label{gtt8}
\end{equation}%
Define%
\begin{equation}
m_{i}^{\widehat{u}^{\theta },n}\equiv m_{i}^{u^{\theta },n}\text{.}
\label{gtt9}
\end{equation}%
Thus, for any $\left( m_{i},m_{-i}\right) \in M_{i}\times M_{-i}$, we have 
\begin{eqnarray*}
\lim_{n\rightarrow \infty }\widehat{u}_{i}^{\theta }\left[ g\left( m_{i}^{%
\widehat{u}^{\theta },n},m_{-i}\right) \right] &=&\lim_{n\rightarrow \infty }%
\widehat{u}_{i}^{\theta }\left[ g\left( m_{i}^{u^{\theta },n},m_{-i}\right) %
\right] \\
&\geq &\lim_{n\rightarrow \infty }u_{i}^{\theta }\left[ g\left(
m_{i}^{u^{\theta },n},m_{-i}\right) \right] -\frac{1}{2}\times \varepsilon \\
&\geq &u_{i}^{\theta }\left[ g\left( m_{i},m_{-i}\right) \right] -\frac{1}{2}%
\times \varepsilon -\frac{1}{2}\times \varepsilon \text{,}
\end{eqnarray*}%
where the equality follows from (\ref{gtt9}), the\ first inequality follows
from (\ref{gtt10}), the second inequality follows from (\ref{gtt8}). I.e., (%
\ref{gtt4})\ holds.$\blacksquare $

\noindent \textbf{Proof of Lemma \ref{lem:mixed:canonical:pure:ordinal}:}
Fix any $\theta \in \Theta ^{\ast }$, any $u^{\theta }\in \Omega ^{\left[
\succeq ^{\theta },\text{ }%
\mathbb{R}
\right] }$, any $\lambda \in MNE^{\left( \mathcal{M}^{\mathbb{Q}},\text{ }%
u^{\theta }\right) }$ and any $\widehat{m}\in $SUPP$\left[ \lambda \right] $%
, i.e.,%
\begin{equation*}
\Pi _{i\in \mathcal{I}}\lambda _{i}\left( \widehat{m}_{i}\right) >0\text{.}
\end{equation*}%
We aim to prove $\widehat{m}\in PNE^{\left( \mathcal{M}^{\mathbb{Q}},\text{ }%
\theta \right) }$. Suppose $\widehat{m}\notin PNE^{\left( \mathcal{M}^{%
\mathbb{Q}},\text{ }\theta \right) }$, i.e., there exists $j\in \mathcal{I}$
and $m_{j}^{\prime }\in M_{i}$ such that%
\begin{equation*}
u_{j}^{\theta }\left[ g\left( m_{j}^{\prime },\widehat{m}_{-j}\right) \right]
>u_{j}^{\theta }\left[ g\left( \widehat{m}_{j},\widehat{m}_{-j}\right) %
\right] \text{,}
\end{equation*}%
and define%
\begin{equation}
\varepsilon \equiv \left( u_{j}^{\theta }\left[ g\left( m_{j}^{\prime },%
\widehat{m}_{-j}\right) \right] -u_{j}^{\theta }\left[ g\left( \widehat{m}%
_{j},\widehat{m}_{-j}\right) \right] \right) \times \Pi _{i\in \mathcal{I}%
}\lambda _{i}\left( \widehat{m}_{i}\right) >0\text{.}  \label{gttf2}
\end{equation}%
By Lemma \ref{lem:mixed:canonical:pure:dominant:ordinal}, there exists a
sequence $\left\{ m_{j}^{u^{\theta },n}\in M_{j}:n\in \mathbb{N}\right\} $
such that%
\begin{equation}
\lim_{n\rightarrow \infty }u_{j}^{\theta }\left[ g\left( m_{j}^{u^{\theta
},n},m_{-j}\right) \right] \geq u_{j}^{\theta }\left[ g\left(
m_{j},m_{-j}\right) \right] -\frac{1}{2}\times \varepsilon \text{, }\forall
\left( m_{i},m_{-i}\right) \in M_{i}\times M_{-i}\text{.}  \label{gttf1}
\end{equation}%
We thus have%
\begin{eqnarray*}
&&\lim_{n\rightarrow \infty }u_{j}^{\theta }\left[ g\left( m_{j}^{u^{\theta
},n},\lambda _{-j}\right) \right] -u_{j}^{\theta }\left[ g\left( \lambda
_{j},\lambda _{-j}\right) \right] \\
&=&\lim_{n\rightarrow \infty }\left( 
\begin{tabular}{l}
$\Sigma _{m\in M^{\ast }\diagdown \left\{ \widehat{m}\right\} }\left[ \Pi
_{i\in \mathcal{I}}\lambda _{i}\left( m_{i}\right) \times \left(
u_{j}^{\theta }\left[ g\left( m_{j}^{u^{\theta },n},m_{-j}\right) \right]
-u_{j}^{\theta }\left[ g\left( m\right) \right] \right) \right] $ \\ 
$+\Pi _{i\in \mathcal{I}}\lambda _{i}\left( \widehat{m}_{i}\right) \times
\left( u_{j}^{\theta }\left[ g\left( m_{j}^{u^{\theta },n},\widehat{m}%
_{-j}\right) \right] -u_{j}^{\theta }\left[ g\left( \widehat{m}_{j},\widehat{%
m}_{-j}\right) \right] \right) $%
\end{tabular}%
\right) \\
&\geq &-\left( \Sigma _{m\in M^{\ast }\diagdown \left\{ \widehat{m}\right\}
}\Pi _{i\in \mathcal{I}}\lambda _{i}\left( m_{i}\right) \right) \times \frac{%
1}{2}\times \varepsilon \\
&&-\Pi _{i\in \mathcal{I}}\lambda _{i}\left( \widehat{m}_{i}\right) \times 
\frac{1}{2}\times \varepsilon +\left[ 
\begin{tabular}{l}
$\Pi _{i\in \mathcal{I}}\lambda _{i}\left( \widehat{m}_{i}\right) \times
\left( u_{j}^{\theta }\left[ g\left( m_{j}^{\prime },\widehat{m}_{-j}\right) %
\right] -u_{j}^{\theta }\left[ g\left( \widehat{m}_{j},\widehat{m}%
_{-j}\right) \right] \right) $%
\end{tabular}%
\right] \\
&=&-\frac{1}{2}\times \varepsilon +\varepsilon >0\text{,}
\end{eqnarray*}%
where the first inequality follows from (\ref{gttf1}), and the second
inequality follows from (\ref{gttf2}). As a result, there exists $n\in 
\mathbb{N}$ such that%
\begin{equation*}
u_{j}^{\theta }\left[ g\left( m_{j}^{u^{\theta },n},\lambda _{-j}\right) %
\right] >u_{j}^{\theta }\left[ g\left( \lambda _{j},\lambda _{-j}\right) %
\right] \text{,}
\end{equation*}%
contradicting $\lambda \in MNE^{\left( \mathcal{M}^{\mathbb{Q}},\text{ }%
u^{\theta }\right) }$.$\blacksquare $

\bibliographystyle{econometrica}
\bibliography{IC,mixed}

\end{document}